\documentclass[12pt]{article}
\usepackage{hyperref}
\usepackage{cite}
\usepackage{color}
\usepackage{graphicx}
\usepackage{amsmath}
\usepackage{amssymb}
\usepackage{xspace}
\usepackage[font={small}]{caption}

\definecolor{gold}{rgb}{1,0.839844,0}
\definecolor{darkgreen}{rgb}{0,0.55,0}

\makeatletter
\@addtoreset{equation}{section}

\makeatletter
\renewcommand\section{\@startsection {section}{1}{\z@}%
                                   {-3.5ex \@plus -1ex \@minus -.2ex}%nn
                                   {2.3ex \@plus.2ex}%
                                   {\normalfont\large\bfseries}}
\renewcommand\subsection{\@startsection{subsection}{2}{\z@}%
                                     {-3.25ex\@plus -1ex \@minus -.2ex}%
                                     {1.5ex \@plus .2ex}%
                                     {\normalfont\bfseries}}

\def\baselinestretch{1.2}
\newcommand{\be}{\begin{equation}}
\newcommand{\ee}{\end{equation}}

\newcommand{\gone}[1]{{}}

\newcommand{\A}{\mathcal{A}}
\newcommand{\X}{\mathcal{X}}
\newcommand{\Y}{\mathcal{Y}}
\newcommand{\Z}{\mathcal{Z}}

\numberwithin{equation}{section}

\newcommand{\bC}{\ensuremath{\mathbb{C}}}

\newcommand{\bZ}{\ensuremath{\mathbb{Z}}}

\newcommand{\scA}{\ensuremath{\mathcal{A}}}

\newcommand{\scN}{\ensuremath{\mathcal{N}}}

\newcommand{\scZ}{\ensuremath{\mathcal{Z}}}

\newcommand{\bea}{\begin{equation}\begin{aligned}}
\newcommand{\eea}{\end{aligned}\end{equation}}
\newcommand{\beq}{\begin{eqnarray}}
\newcommand{\eeq}{\end{eqnarray}}
\newcommand{\beqs}{\begin{subequations}\begin{align}}
\newcommand{\eeqs}{\end{align}\end{subequations}}

\newcommand{\hyperKahler}{hyperK\"ahler}

\newcommand{\mprime}{\ensuremath{\prime}}

\begin{document}
\begin{titlepage}
\begin{flushright}
MAD-TH-14-04\\
IPMU-14-0135\\
\today\\
%3dft.06.19a.tex
\end{flushright}
%\vspace{12 mm}

\vfil

\begin{center}

{\large{\bf Boundaries and Defects of ${\cal N}=4$ SYM with 4 Supercharges} \\
\bigskip
Part II: Brane Constructions and 3$d$ $\mathcal{N}=2$ Field Theories}

\vfil

Akikazu Hashimoto$^a$, Peter Ouyang$^b$, and Masahito Yamazaki$^{c,d}$

\vfil

$^a$ Department of Physics, University of Wisconsin, Madison, WI 53706

$^b$ Department of Physics, Purdue University, West Lafayette, IN 47907

$^c$ Institute for Advanced Study, Princeton, NJ 08540

$^d$ Kavli IPMU (WPI), University of Tokyo, Kashiwa 277-8583, Chiba, Japan

\vfil

\end{center}

%%%%%%%%%%%%%%%%%%%%%%%%%%%%%%%%%%%%%%%%%%%%%%%%%%%%%%%%%%%%%%%%%%%%%%%%%%%%%%%%%%%%%%%
\begin{abstract}
\noindent

We study the vacuum moduli spaces of 3$d$ $\scN=2$ supersymmetric 
quantum field theories by
applying the formalism developed in our previous paper \cite{Hashimoto:2014vpa}.  The 3$d$
theories can be realized by branes in type IIB string theory, which in
a decoupling limit reduce to 4$d$ $\mathcal{N}=4$ super-Yang-Mills
theory on an interval with BPS defects inserted.  The moduli space of
a given 3$d$ theory is obtained by solving a generalization of Nahm's
equations with appropriate boundary/junction conditions, along with
help from the S-duality of type IIB string theory.  Our classical
computations reproduce many known results about the quantum-corrected
moduli spaces of 3$d$ theories, e.g.\ $U(N_c)$ theories with $N_f$
flavors with mass and FI parameters turned on.  In particular, our
methods give first-principles derivations of several results in the
literature, including the $s$-rule, quantum splitting of classical
Coulomb branches, the lifting of the Coulomb branch by non-Abelian
instantons, quantum merging of Coulomb and Higgs branches, and phase
transitions from re-ordering 5-branes.

\end{abstract}
%%%%%%%%%%%%%%%%%%%%%%%%%%%%%%%%%%%%%%%%%%%%%%%%%%%%%%%%%%%%%%%%%%%%%%%%%%%%%%%%%%%%%%%%%
\vspace{0.5in}

\end{titlepage}
\renewcommand{\baselinestretch}{1.05}  %Line spacing
%%%%%%%%%%%%%%%%%%%%%%%%%%%%%%%%%%%%%%%%%%%%%%%%%%%%%%%%%%%%%%%%%%%%%%%%%%%%%%%%%%%%%%%%%%%%%

\tableofcontents

\section{Introduction}

String theory enjoys many deep connections with quantum field theory.
For example, the solitonic objects of string theory such as D-branes
support gauge fields living on the brane worldvolume.  One
particularly clear realization of the connection between string theory
and field theory comes about through the Hanany-Witten  brane
construction of field theories \cite{Hanany:1996ie}.

This construction turns out to be an enormously powerful technique for
studying supersymmetric field theories; see \cite{Giveon:1998sr} for a
comprehensive review.  Many properties of the field theory can be
understood pictorially in terms of a brane diagram.  For example, the
vacuum moduli space of the field theory is often simply the space of
allowed geometrical arrangements of the branes; by counting the
degrees of freedom of the brane motions, one counts the dimension of
the moduli space.

Unfortunately, the naive prescription of counting geometric brane
motions is not complete; to correctly determine the dimension of
moduli space, some additional constraints must be imposed.  The most
well-known constraint is the Hanany-Witten $s$-rule
\cite{Hanany:1996ie}.  However, in systems with reduced supersymmetry
there are additional rules, which seem ad hoc from the point of view
of the brane diagrams \cite{Elitzur:1997hc,Giveon:1998sr}; these extra
constraints are necessary to incorporate the effect of
instanton-generated superpotentials\cite{Affleck:1982as}.  Moreover,
the brane diagram is not enough for understanding the detailed
structure of the vacuum moduli space, such as the merging of the
Coulomb and the Higgs branches for non-Abelian theories.  Clearly, it
would be desirable to have a framework where these rules and results
can be derived systematically, while retaining the connection to the
brane diagram.

The point of view we advocate is that one can regard the brane diagram
as a set of rules for constructing a particular system of localized
defects coupled to bulk degrees of freedom.  In
\cite{Hashimoto:2014vpa}, we performed a study of interface conditions
for $\mathcal{N}=4$ super-Yang-Mills theory, and explicitly
constructed UV Lagrangians for such defect systems.  These defect
systems realize 3$d$ $\scN=2$ field theories in the IR, and can be
constructed from type IIB brane configurations with D3-branes (which
support the 4$d$ $\mathcal{N}=4$ theory) suspended between 5-brane
defects.

Our goal in the present paper is to study the vacuum moduli spaces of
$3d$ $\scN=2$ theories in terms of these defect systems.  The 
moduli spaces in question can be identified with the solution spaces of
a generalization of Nahm's monopole equations (a dimensional
reduction of the Donaldson-Uhlenbeck-Yau equations) with a certain set
of boundary conditions which were described explicitly in
\cite{Hashimoto:2014vpa}.

A crucial ingredient in our analysis is the S-duality of type IIB
string theory (and of the $4d$ $\scN=4$ theory on the D3-branes.)  The
classical moduli space computation is potentially subject to quantum
effects, but in many situations of interest, we can find an S-duality
frame where at least some part of the moduli space is free from
quantum corrections.  As we will see, this happens when the gauge
symmetry is completely broken.  In this S-duality frame, our {\it
  classical} computation then gives the {\it quantum--corrected}
moduli space.  For $\mathcal N = 4$ theories in $3d$, this is nothing
but the usual statement of mirror symmetry
\cite{Intriligator:1996ex}. A similar situation holds for $\mathcal N
=2$ theories, although with many subtle differences from the $\mathcal
N=4$ case.

To place our results in context, we might recall the analogous
situation for type IIA brane constructions and their 4$d$ field
theories. In that case, the strong coupling limit of the 4$d$ field
theory can be studied by lifting the IIA configuration to M-theory
\cite{Witten:1997sc}.  The brane configuration becomes a single
M5-brane wrapped on a complex curve which turns out to be the
Seiberg-Witten curve of the 4$d$ theory. For type IIB brane
constructions, we do not have the lift to M-theory as a tool.
Instead, the strong coupling limit of the IIB construction is best
understood by performing an S-duality.  A detailed
description of the defect system, combined with S-duality, will be
sufficient to understand many features of the moduli spaces of the
3$d$ field theories (previously studied in
\cite{deBoer:1997ka,deBoer:1997kr,Aharony:1997bx}.)

The rest of this paper is organized as follows.  We summarize our
methods in Section \ref{sec2}.  We realize 3$d$ $\scN=2$ theories from
brane configurations, and the vacuum moduli space of the 3$d$ theory
can be identified with the moduli space of the bulk BPS equations with
appropriate boundary conditions corresponding to the arrangement of
5-branes.  We then apply our formalism to a variety of 3$d$ field
theories with $U(N_c)$ gauge groups.  First, we work out some examples
with $\mathcal{N}=4$ 3$d$ supersymmetry in detail in Section
\ref{sec:halfbps}; these simple examples are
sufficient to demonstrate most of the technical issues associated with
our methods. We then proceed to apply our techniques to
$\mathcal{N}=2$ Abelian field theories in Section \ref{sec:abelian},
and to non-Abelian $U(N_c)$ gauge theories in Sections \ref{sec:u3}
and \ref{sec:un}.  We will also study in Section \ref{sec:merging} a
few more examples with product gauge groups which exhibit quantum
merging of Higgs and Coulomb branches.

%------------------------------------------
\section{Moduli Space from Generalized Nahm Equations\label{sec2}}

In this section we summarize our general method for studying
the vacuum moduli spaces of $3d$ $\scN=2$ theories. Our analysis will
rely on two crucial ingredients, S-duality and holomorphy. The latter
will be expressed mathematically in terms of a complex gauge
quotient.

We begin with the standard brane configurations for the 3$d$ $\scN=2$
theory (Section \ref{subsec:cartoon}), and point out the subtleties of
3$d$ $\scN=2$ mirror symmetry, as well as the limitations of the
cartoonish brane descriptions (Section \ref{sec:mirror}.)  We will
then write down the generalized BPS equations following our previous
paper \cite{Hashimoto:2014vpa} (Section \ref{sec:BPSeq}), and comment
on technical (but important) issues of gauge symmetry breaking and
stability in the rest of this section.

\subsection{3$d$ $\mathcal{N}=2$ Theory from Type IIB Brane Constructions}\label{subsec:cartoon}

In type IIB string theory, we can engineer 3$d$ $\scN=2$ field
theories by suspending D3-branes between NS5 and D5 branes.  The NS5
and D5 branes may be oriented in directions consistent with the
supercharges preserved by the 3$d$ field theory.  We consider two
allowed types of NS5 brane which we call NS5 and
NS5$^\ensuremath{\prime}$, and two types of D5 brane, which we call D5
and D5$^\ensuremath{\prime}$\cite{deBoer:1997ka,Elitzur:1997hc}.\footnote{We
  can consider D5-branes/NS5-branes oriented at more general angles
  (rotated by the same angle in $47$ and $58$-planes) while preserving
  4 supercharges (see e.g.\ Appendix A3 of \cite{Hashimoto:2014vpa}.)
  In practice this will correspond to adding a finite mass to the
  adjoint chiral multiplets.  While our formalism includes such
  D5-branes, we will not discuss them in this paper since we are
  mostly interested in the IR behavior of 3$d$ theories where the
  precise coefficients of the relevant deformations are not
  important.}  Our convention for orienting the branes is as follows.
\begin{align}
\begin{tabular}{c||ccc|c|ccc|ccc}
       & 0& 1  & 2& 3& 4& 5& 6& 7& 8& 9 \\
       \hline
D3 & $\circ$ &  $\circ$ & $\circ$ & $\circ$ &    &   &    &   &   &     \\
D5 & $\circ$  &  $\circ$ & $\circ$ &  &  $\circ$  & $\circ$  & $\circ$   &   &   &     \\
D5$^\ensuremath{\prime}$ & $\circ$  &  $\circ$ & $\circ$ &  &    &   &  $\circ$  &  $\circ$  & $\circ$  &    \\
NS5 & $\circ$ &  $\circ$ & $\circ$ &  &   &  &  & $\circ$  & $\circ$  & $\circ$    \\
NS5$^\ensuremath{\prime}$ & $\circ$  &  $\circ$ & $\circ$ &   & $\circ$    & $\circ$&    &   & &    $\circ$   \\
\end{tabular}
\label{orient}\end{align}
In general, D3-branes suspended between NS5-type branes contribute
3$d$ vector multiplets while D5-branes intersecting D3-branes
contribute quarks.  The detailed rules for relating the brane
construction to field theory may be found in the review
\cite{Giveon:1998sr}.

\begin{figure}
\centerline{\includegraphics[width=\hsize]{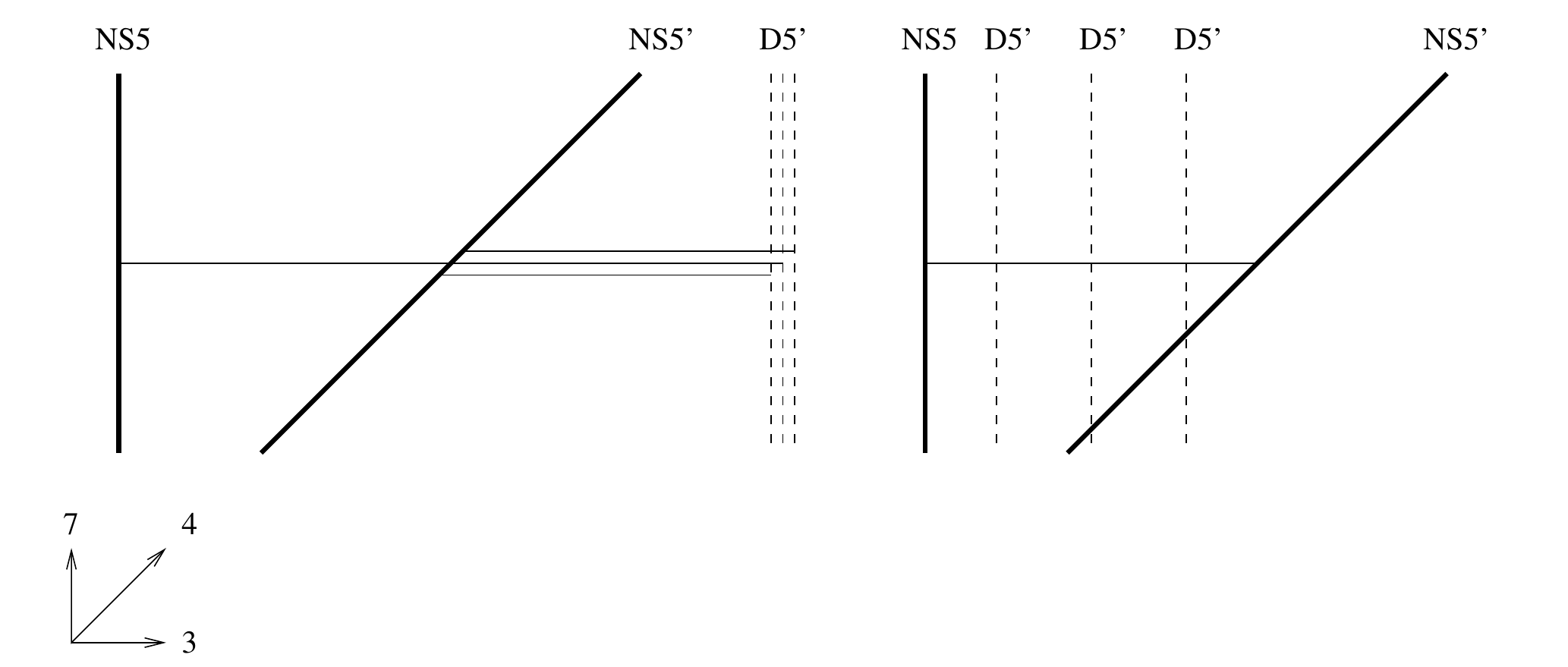}}
\caption{The brane construction of ${\cal N}=2$ $U(1)$ $N_f=3$ theory
  in two different brane orderings, related by Hanany-Witten
  transitions.  Some properties of the moduli space can be understood
  by allowing the D3-brane to break into segments when crossing a
  5-brane, and then counting the possible motions of the D3 segments.
\label{figb}}
\end{figure}

An example of the kind of system we will be analyzing is illustrated
in Figure \ref{figb}.  The figure shows a brane construction which
realizes $3d$ $U(1)$ gauge theory with 3 flavors, in the limit where
the lengths of the D3-brane segments (extended in $x_3$) are taken to
zero.

\subsection{Quantum Corrections, Mirror symmetry and S-duality\label{sec:mirror}}

In many cases, the brane construction gives a recipe for writing down
a supersymmetric field theory, for which the Lagrangian is largely
constrained by symmetry considerations.  Given a Lagrangian, one can
determine the classical moduli space of vacua by solving the F- and
D-flatness conditions.  However, this moduli space can be subject to
corrections at the quantum level, and it is the heart of this paper to
understand these quantum effects.

\subsubsection{3$d$ $\mathcal{N}=4$ Theories}

Let us first recall the case with eight supercharges (i.e.\ 3$d$
$\mathcal{N}=4$ theory), for which the structure of moduli space,
though intricate, is highly constrained.  The $\mathcal{N}=4$ moduli
spaces fall into distinct parts, a Higgs branch, where the gauge
symmetry is broken, and a Coulomb branch, which preserves an unbroken
$U(1)^N$ gauge symmetry, as well as mixed branches which are a direct
product of a Higgs branch and a Coulomb branch with some number of
unbroken $U(1)$'s.  The constraints on the $\mathcal N=4$ moduli space
partly stem from the existence of an $SO(3)_H \times SO(3)_V$
R-symmetry which acts separately on the hypermultiplets and the
vectormultiplets.  The moduli space metric is \hyperKahler, and can
essentially be determined from the asymptotic structure and
singularity structure of the moduli space.  Moreover, (in part) as a
consequence of this global symmetry, the mixed branches are always a
direct product of a Coulomb part and a Higgs part (for a proof, see
the argument in the 4$d$ context in \cite{Argyres:1996eh}.)

The classical field theory computation of the ${\cal N}=4$ moduli
space is not reliable on all the branches of vacua.  On the Higgs
branch, because the vectors are massive, all the remaining fields are
simply free and the classical geometry of the Higgs branch is quantum
mechanically exact.  The Coulomb branch and mixed branches, on the
other hand, have unbroken gauge symmetry and are therefore subject to
both perturbative corrections and nonperturbative instanton
corrections \cite{Affleck:1982as}.

It is often said that $\mathcal{N}=4$ mirror symmetry (which is
understood as S-duality of a IIB brane construction) exchanges the
Higgs and Coulomb branches of two mirror pair theories A and B.  What
is really meant by this statement is that the whole moduli space is
exchanged by mirror symmetry, but the extent of gauge symmetry and its
breaking is different for the branches identified as mirror pairs in
theory A and B.  Because of this, different branches are computable
classically on the two sides of the mirror.  If the mirror pair is
known, then the Higgs branch of theory A (which is classically exact)
determines the Coulomb branch of theory B (which is subject to quantum
corrections), and vice versa.

From the brane construction, it is easy to understand this aspect of
the duality; D5-brane interfaces which break some amount of gauge
symmetry are exchanged with NS5-brane interfaces which do not.  The
Coulomb branch is associated with the movement of D3 segments
stretched between a pair of NS5 branes, and the Higgs branch is
associated with the movement of a D3-brane segment stretched between a
pair of D5 branes. A D3 segment stretched between an NS5 and the D5,
on the other hand, is completely fixed in its position and do not give
rise to moduli.  The $SO(3)\times SO(3)$ global symmetry acts
geometrically by rotating the 456 coordinates and the 789 coordinates
separately.  In this picture, mirror symmetry is a consequence of
S-duality, which exchanges D5 and NS5 branes.  Clearly, in the brane
construction, it exchanges Coulomb and Higgs branch moduli.

In the case of ${\cal N}=4$ supersymmetry, one can also identify the
Coulomb branch as part of the moduli space that is lifted by turning
on Fayet-Iliopoulos terms, and the Higgs branch as the part that is
lifted by turning on masses to the matter fields.  This is a
consequence of the fact that these deformations are charged under the
$SO(3) \times SO(3)$ global symmetry.  From the brane point of view,
the FI terms displace the NS5 branes relative to each other, so that
D3-branes stretched between the NS5 branes do not preserve
supersymmetry, lifting the Coulomb branch.  Similarly, the mass terms
move the D5-branes relative to each other, lifting the Higgs branch.

\subsubsection{3$d$ $\mathcal{N}=2$ Theories}

Much of the structure of ${\cal N}=4$ theories extends to ${\cal N}=2$
theories but with various caveats. Some of these subtleties were
discussed in \cite{deBoer:1997ka,deBoer:1997kr,Aharony:1997bx} but we
choose to take our own perspective on some of these concepts which we
will explain below.  A portion of the moduli space can be considered a
``Higgs branch'' if charged matter fields have expectation values
there, and all the gauge symmetry is broken spontaneously. Similarly
there can be a ``Coulomb branch'' if the scalars in the vector
multiplet are nonvanishing, and some Abelian subgroup of the gauge
group is left unbroken. There can also be mixed branches with both
Coulomb and Higgs components.

However, unlike the $\mathcal N=4$ case, where separate $SO(3)_H$ and
$SO(3)_V$ global symmetries act on the vector multiplets and
hypermultiplets, in $\mathcal N=2$ there is in general only an $SO(2)$
R-symmetry, and the structure of the moduli space is much less
constrained by symmetry.  In this paper, for simplicity we only
consider the case where the global symmetry is $SO(2) \times SO(2)$.
Even in this case, however, the $SO(2) \times SO(2)$ global symmetry
does not sharply distinguish between Coulomb branch and Higgs branch
moduli, and the moduli spaces can exhibit considerable complexity.

Because S-duality still has an action on 1/4 BPS brane configurations,
there is a notion of mirror symmetry mapping an electric theory A to a
magnetic theory B with $\mathcal N=2$ supersymmetry. Some aspects of
this mirror symmetry were also discussed in
\cite{deBoer:1997ka,deBoer:1997kr,Aharony:1997bx}. There are, however,
several important differences in the way mirror symmetry acts in the
${\cal N}=2$ theories as compared with the ${\cal N}=4$ prototype. One
important distinction is that unlike in the ${\cal N}=4$ theories, the
Coulomb branch of model A and the Higgs branch of model B do not map
on to one another one-to-one (and similarly for the Higgs branch of
model A and the Coulomb branch of model B.)

Both the Higgs and the Coulomb branches can receive quantum
corrections. Some components of the Coulomb branch can be lifted by
superpotentials generated by instanton effects.  It is also possible
for some of the branches of moduli space to merge quantum mechanically.
\cite{deBoer:1997ka,deBoer:1997kr,Aharony:1997bx}.  However, the
complex structure of the Higgs branch will be free of instanton
effects\footnote{For instantons to correct the superpotential, they
  need to have two fermionic zero-modes. 1/2 BPS instantons in
  $\mathcal{N}=2$ theories have two such zero-modes, which are the
  goldstini from spontaneously breaking half the supersymmetry.  If
  the gauge symmetry is fully broken, however, the 1/2 BPS instantons
  will not exist, and the superpotential will not be corrected.}.  For the purposes of this
  paper, we will limit our discussion to the complex structure of moduli space, to avoid details of
  the geometry which depend on the moduli space metric.  Some interesting recent work on the moduli space
  metric for closely related intersecting brane systems appeared in \cite{Mintun:2014aka} with implications for 
  $3d$ field theory as studied in \cite{Dorigoni:2014yfa}.

Some of these issues can be made apparent by looking at mirror
symmetry from the point of view of type IIB brane configurations
\cite{deBoer:1997ka,Elitzur:1997hc}. A wide class of ${\cal N}=2$
theories in 2+1 dimensions can be engineered by suspending D3 brane
intervals between NS5, NS5$^\ensuremath{\prime}$, D5, and
D5$^\ensuremath{\prime}$ branes. In this brane construction, mirror
symmetry follows from the S-duality of the type IIB string theory. The
$SO(2)_{45}\times SO(2)_{78}$ global symmetry can be understood as
rotation in the 45 plane and the 78 plane for the branes oriented
according to (\ref{orient}).

It is natural to identify the components of the moduli space
associated with moving the D3 branes stretched between an NS5 and an
NS5$^\ensuremath{\prime}$ (or an NS5) brane as being associated with
the Coulomb branch. Similarly, D3 segments stretched between a D5 and
a D5$^\ensuremath{\prime}$ (or a D5) brane should be interpreted as
being part of the Higgs branch since D3 broken on D5's breaks the
gauge symmetry associated with the D3 brane. These two branches
naturally map onto each other under S-duality of type IIB string
theory.

In the ${\cal N}=2$ construction, however, there are additional moduli
associated with the D3 segments stretched between an
NS5$^\ensuremath{\prime}$ and a D5, or between an NS5 and a
D5$^\ensuremath{\prime}$. These branches, which were not present in
the $\mathcal N =4$ construction, behave somewhat differently than the
two branches we described above.  The $U(1)$ gauge symmetry living on
such a D3-brane segment is broken by the D5/D5' boundary conditions,
so these moduli are naturally Higgs branch moduli.  However, S-duality
maps the NS5$^\ensuremath{\prime}$-D5 configuration to
D5$^\ensuremath{\prime}$-NS5, so the corresponding mirror symmetry
maps Higgs branch moduli to Higgs branch moduli and not to the Coulomb
branch.

It is also instructive to think about how the various branches of
moduli space are affected by mass deformations.  In ${\cal N}=4$
theories, the Coulomb branch is lifted by turning on FI terms but is
not lifted by masses. The Higgs branch, on the other hand, is lifted
by the mass terms but is unlifted by the FI term.  In the case of
${\cal N}=2$ supersymmetry, on the other hand, the Coulomb branch is
still lifted by the FI term, but not all of the Higgs branch is
necessarily lifted by the real mass terms (in this paper we consider
only parity-preserving real mass terms). The reason is that the moduli
associated to the D3 branes stretched between an NS5 brane and a
D5$^\ensuremath{\prime}$ are not lifted either by the FI or the mass
term.

It should be clear from these considerations that one should not think
of Higgs and Coulomb branches as being mapped onto one another under
mirror symmetry for theories with ${\cal N}=2$ supersymmetry.  These
aspects of the action of $\mathcal{N}=2$ mirror symmetry are
summarized in Table \ref{tablebranch}.

As mentioned previously, further complications arise because in some
cases quantum effects can completely blur the distinction between
Higgs branches and Coulomb branches; this was called ``quantum
merging'' of the Higgs and Coulomb branches\footnote{Strictly speaking, it is the Coulomb branch and a mixed
Higgs-Coulomb branch which merge.} in \cite{Aharony:1997bx}.
We will encounter a number of examples where this occurs; in our
analysis it is intrinsically tied to the non-Abelian nature of the
4$d$ $\mathcal{N}=4$ defect system, and is hard to understand from the
point of view of the brane cartoon. We will overcome this limitation
with the help of a more refined analysis of the BPS equations of the
defect system.

\begin{table}
\centerline{\begin{tabular}{l|l|l|l|l|l|l}
brane & phase & gauge symmetry & $SO(2)_{45}$ & $SO(2)_{78}$ & FI & real mass \\ \hline
NS5-NS5$^\ensuremath{\prime}$ & Coulomb & partly unbroken & neutral & neutral & lifted & unlifted  \\
D5-D5$^\ensuremath{\prime}$ & Higgs & broken & neutral & neutral& unlifted & lifted  \\ \hline
NS5-NS5 & Coulomb & partly unbroken &  neutral & charged & lifted & unlifted \\
D5-D5 & Higgs & broken & charged & neutral & unlifted & lifted \\ \hline
NS5$^\ensuremath{\prime}$-NS5$^\ensuremath{\prime}$ & Coulomb & partly unbroken & neutral & charged & lifted & unlifted \\
D5$^\ensuremath{\prime}$-D5$^\ensuremath{\prime}$ & Higgs & broken & charged & neutral& unlifted & lifted \\ \hline
NS5-D5$^\ensuremath{\prime}$ & Higgs &  broken & neutral & charged & unlifted & unlifted \\
NS5$^\ensuremath{\prime}$-D5 & Higgs & broken & charged & neutral & unlifted & unlifted\\
\end{tabular}}
\caption{Classification of phases of moduli-space components associated with the motion of D3 branes in the Hanany-Witten-like brane construction. 
\label{tablebranch}}
\end{table}

\subsection{Bulk BPS Equations} \label{sec:BPSeq}

The brane constructions give rise to defect theories with translation
invariance broken in one direction which we will label as $y$
throughout this paper.  Because the D3-brane has the 3+1-dimensional
$\mathcal{N}=4$ theory living on its worldvolume, this defect theory
consists of the $4d$ $\mathcal N = 4$ theory living on a sequence of
intervals with the defects realized as supersymmetric boundary
conditions for each interval.  In the cases we will consider, the
boundary conditions will correspond to various combinations of NS5 and
D5-branes, possibly oriented in different directions compatible with
preserving 4 supercharges.

In the traditional brane drawing analysis, one studies the moduli
space by allowing the D3-branes to break on the 5-brane defects; in
this picture, the allowed motions of the D3-brane segments are the
moduli of the theory.  In this paper we attempt to analyze the same
systems from the point of view of the $4d$ defect theory.  For
non-Abelian gauge theories, the defect analysis is crucially different
from the brane cartoon.

Let us summarize the bulk BPS equations of the defect system
\cite{Hashimoto:2014vpa}.  The bulk $\mathcal{N}=4$ theory consists of
a gauge field $A_{\mu}$, $\mu =0,1,2,3$, and six adjoint scalars
$X_i$, $i={4,\ldots, 9}$, where the numbering of the indices reflects
the relationship between the 3+1-dimensional $\mathcal{N}=4$ theory
and the 9+1-dimensional $\mathcal{N}=1$ SYM theory.  In addition to
the bosonic fields, the theory contains fermions.  In this paper the
fermions play no role except that we demand that some of their
variations under supersymmetry vanish.  As in \cite{Hashimoto:2014vpa}
we choose to take the fields $A_{\mu}, X_i$ to be anti-Hermitian, to
be consistent with the mathematical literature on Nahm equations.

Away from any boundaries, the scalars of the 3+1-dimensional field
theory must satisfy certain equations to preserve four supersymmetries,
with the assumptions of 2+1-dimensional Lorentz invariance and the
existence of a $U(1) \times U(1)$ global symmetry.  These equations
are conveniently written in terms of three complex equations
\beq
\frac{\mathcal{D}\mathcal{X}}{\mathcal{D}y} &=& 0\ ,\label{qnahm1}\\
\frac{\mathcal{D}\mathcal{Y}}{\mathcal{D}y} &=& 0\ ,\label{qnahm2}\\
\left[\mathcal{X},\mathcal{Y}\right] &=& 0\ ,\label{qnahm3}
\eeq
and one real equation,
\beq
\frac{d}{dy}\left(\mathcal{A}-\bar{\mathcal{A}}\right)-\left[\mathcal{A},\bar{\mathcal{A}}\right] +\left[\mathcal{X}\ ,\bar{\mathcal{X}}\right] + \left[\mathcal{Y},\bar{\mathcal{Y}}\right] &=& 0\label{qnahm4}\ ,
\eeq
where
\beq
\mathcal{X} &\equiv & X^4+iX^5\ ,\\
\mathcal{Y} &\equiv & X^7+i X^8\ ,\\
\mathcal{A} &\equiv & A_3 + i X^6\ .
\eeq
When $\Y=0$, these equations reduce to Nahm's equations
\cite{Nahm:1979yw}, and our analysis here generalizes the relation
between D-branes and Nahm's equations discovered in
\cite{Diaconescu:1996rk}.

When applied to ${\cal N}=4$ SYM on an interval with defects, we must
subject the bulk equation to boundary and junction conditions. The
conditions relevant for our purposes were studied in our earlier paper
\cite{Hashimoto:2014vpa} and are briefly summarized in Appendix
\ref{appB}.

\subsection{Gauge Symmetry Breaking and $X_9$}
\label{sec:x9dual}

In addition to the generalized Nahm equations which involve the
scalars $X_{4,5,6,7,8}$, we have one more adjoint scalar $X_9$, which
combines with the field strength of the 3$d$ gauge field $A_{\mu}$
(the ``dual scalar'' $\varphi$) into a 3$d$ linear multiplet.  The BPS
equation for the $X_9$ reads
\beq
D_3 X_9 &=& 0\ ,\\
\lbrack X_6, X_9\rbrack&=& 0\ ,\\
\lbrack \X, X_9\rbrack&=& 0\ ,\\
\lbrack \Y, X_9 \rbrack &=& 0\ .
\eeq
Although it is not clearly evident in the 4$d$ $\mathcal{N}=4$
analysis, to be consistent with 3$d$ supersymmetry, when the field
$X_9$ is a modulus it must be accompanied by the dual photon
$\varphi$, and so each freely varying component of $X_9$ always gives
rise to two real moduli (or one complex modulus.)

In the analysis we will do later, we will keep track of the dimension
of the moduli space due to the variation of $X_9$ and $\varphi$, but
we will not try to determine more detailed properties of the moduli
space, such as the complex structure, when $X_9$ and $\varphi$ are
involved.  The reason is that when $X_9$ is active there is always
some amount of unbroken gauge symmetry, and in general there will be
quantum corrections.  Hence the complex structure as computed
classically will be wrong anyway.  In some cases, one can overcome
this problem by combining the classical analysis with the S-duality of
4$d$ $\mathcal{N}=4$ theory.  In other cases, we will also find that
despite the existence of quantum corrections, our method still
computes the dimension of moduli space correctly.  In particular, this
will be the case for mixed branches in non-Abelian theories.

Later in the paper, when counting moduli, we will introduce the
variable $\Z$ to keep track of the field $X_9$ and the dual scalar
$\varphi$ -- one can think of it as forming the complex combination
(monopole operator) $\Z \sim e^{X_9+i\varphi}$. One should, however,
always keep in mind the existence of the quantum corrections.

\subsection{$G_{\mathbb{C}}$ Quotient and Stability} 
\label{sec:gc}

The analysis of the moduli space of equations
\eqref{qnahm1}--\eqref{qnahm4} can be drastically simplified by the
method of a complex gauge quotient.  Recall that both the bulk
equations and the boundary conditions can be split into naturally
complex and naturally real equations.  The complex equations are
invariant not just under the gauge symmetry $G$ but rather have a
larger gauge symmetry $G_{\mathbb{C}}$, the complexified version of
$G$.  The real equations are however only invariant under $G$.
Specifically, we may take $\X \rightarrow g^{-1} \X g$, $\Y
\rightarrow g^{-1} \Y g$, $\mathcal{A} \rightarrow g^{-1} \mathcal{A}
g + g^{-1}dg$, where $g$ is valued in the complexified gauge group
$G_{\mathbb{C}}$.  On the other hand, the real equation (\ref{qnahm4})
is only invariant under the real gauge symmetry $G$ and transforms
nontrivially under $G_{\mathbb{C}}$.

There is a beautiful mathematical result that it is possible to simply
ignore the real equations completely (modulo the subtleties to be
mentioned in the next paragraph), and instead solve only the complex
equations, but with a quotient by the complexified gauge group
$G_{\mathbb{C}}$.  The point is that we can find a true solution of
the full system of equations in the closure of the $G_{\mathbb{C}}$
orbit of a point satisfying only the complex equations.

However, there is an important caveat which is that given a point $p$
in the solution space of the complex equations, it is not necessarily
guaranteed that there exists a point in the $G_{\mathbb{C}}$ orbit of
$p$ which actually satisfies the real equation; the points for which
the appropriate gauge transformation does not exist are said to be
``unstable."  The notion of stability was introduced by Mumford
\cite{Mumford,Mumford:1977}, and essentially it amounts to classifying
singular points of moduli spaces.  There are many results in geometric
invariant theory which give various criteria for determining the
stability properties of a point in moduli space.  For algebraic
varieties, the definition of an unstable point is that the closure of
its $G_{\mathbb{C}}$ orbit includes the origin\footnote{The origin for
  our problem is locus where $X_a=c_a \mathbb{I} \,\, (c_a:
  \textrm{constant} , \, a=1, \ldots 6)$.  The shift by such constant
  identity components clearly preserves the existence/non-existence of
  the solution to the real equation.  In string theory language, this
  shift represents the center-of-mass modes of the D3-branes, which
  decouples from the relative positions of the D3-branes.  }.  The
stable points have closed $G_{\mathbb{C}}$ orbits and a finite
stabilizer (that is, the elements of $G_{\mathbb{C}}$ which leave the
point fixed form a finite group.)  The semistable points satisfy the
property that the gauge orbits have a closure which is nonempty but
does not include the origin.

One way of thinking about the stability criterion is that we should
identify points as gauge equivalent if they are related by the closure
of the gauge orbit (not just finite gauge transformations.)  In some
examples, what may appear to be an entire branch of moduli space turns
out to be gauge equivalent to a single point, and therefore should not
be counted as a separate branch.

\subsubsection{Example: The Conifold}

We can illustrate the stability issue with an example familiar to
physicists, the conifold.  The complex structure of the conifold may
be expressed by the algebraic equation
\beq
z_1 z_2 - z_3 z_4 = 0 \ .
\eeq
A second alternative description is given by ``solving'' the algebraic
equation with the variables $A_i, B_j$: \beq z_1 = A_1 B_1\ , \qquad
z_2 = A_2 B_2 \ , \qquad z_3 = A_1 B_2 \ , \qquad z_4 = A_2 B_1 \ ,
\eeq
provided that one mods out by the complex gauge symmetry
$U(1)_{\mathbb{C}} = GL(1,\mathbb{C})=\mathbb{C}^*$
\beq
A_i \rightarrow \lambda A_i \ , \qquad B_j \rightarrow \lambda^{-1} B_j \ .
\eeq
A third description is given by taking the four complex variables $A_i, B_j$
and mod out only by a (real) gauge symmetry $G= U(1)$
\beq
A_i \rightarrow e^{i\theta} A_i \ , \qquad B_j \rightarrow e^{-i\theta} B_j \ ,
\eeq
and also impose a D-term equation,
\beq
|A_1|^2 + |A_2|^2 - |B_1|^2 -|B_2|^2 = 0 \ ,
\eeq
where signs in the D-term equation are determined by the $U(1)$
charges of the $A, B$ fields.

The three constructions are equivalent for semistable points but not
for unstable points.  In the second formulation, there is a family of
solutions with $A_i=0$, $B_j = {\rm arbitrary}$, and it might seem
that this corresponds to a 1$d$ branch of moduli space with the
complex structure of $\mathbb{CP}^1$ (after modding out by
$G_{\mathbb{C}}$.)  But this branch sits at $z_i=0$, the conifold
singularity, which should be a single point in the first description.
The key fact is that this branch consists of unstable points, because
we can take $B_j \rightarrow \lambda^{-1} B_j$ for $\lambda
\rightarrow \infty$, so the closure of the gauge orbit contains the
origin $A_i=B_j=0$.  From the third description, we see that $A_i=0$
and $B_j \neq 0$ violates the D-term equation, unless we deform it by
an appropriate Fayet-Iliopoulos term.
 
On the other hand, the semistable points have at least one nonzero $A$
and one nonzero $B$ -- we see that in this situation a
$G_{\mathbb{C}}$ gauge can always be found to satisfy the D-term
equation of the third description.

Note that the issue of stability does not arise, for the most part, in
the usual construction of monopoles through Nahm's equations.  It
turns out that in the case of purely D5-like boundary conditions, the
corresponding solutions do not have unstable points, and so one can
use the complex gauge quotient freely (see
\cite{Kronheimer:1989zs,Bielawski}.)

However, because we are interested in more general boundary
conditions, in particular those which can include NS5-branes, the
issue of stability will reappear in the analysis we do later.  Namely,
we are allowed to follow the procedure of choosing a convenient gauge
in $G_{\mathbb{C}}$ and to solve only the complex Nahm equation.  This
gauge transformation will not preserve the real Nahm equation or the
$X_6$ boundary condition in general, but this is all right, provided
we restrict attention to the semistable points (which is equivalent to
the statement that the $G_{\mathbb{C}}$ gauge orbit includes a
solution of the full Nahm system.)

Under some circumstances, the unstable points of moduli space are also
important.  In particular, when we deform the real Nahm equation by an
FI term, it is not always possible to gauge-transform an arbitrary
solution of the complex equations to a solution of the full system,
and it becomes necessary to study the real equations explicitly.  In
the conifold example, the corresponding statement is that the
singularity admits a small resolution.

\section{3$d$ $\mathcal{N}=4$ Gauge Theories}
\label{sec:halfbps}

We begin our analysis of defect systems of $\scN=4$ super Yang-Mills
by studying configurations with 8 supercharges \cite{Gaiotto:2008s
  a,Gaiotto:2008ak}.  This will serve as a simple demonstration of our
methods before we proceed to the 1/4 BPS case.  Several important
features arise already in the 1/2 BPS analysis; among these are
subtleties in gauge fixing as well as the role played by stability of
the $G_{\mathbb C}$ quotient.

Of course, the systems with only D3 and D5 branes have already been
studied extensively; they are the standard Nahm system for monopoles
in $U(N)$ gauge theory, where $N$ is the number of D5-branes, and the
number of D3-brane segments is the total monopole charge.

\subsection{$\mathcal{N}=4, N_c=1, N_f=0$}

We begin with a very simple example, the $\mathcal{N}=4$ with $U(1)$
gauge symmetry and no hypermultiplet matter.  From the point of view
of brane configurations, one constructs this theory by suspending a
single D3-brane between two NS5-branes.  Because the NS5-branes are
extended in the 789 directions, the D3-brane has three real moduli
corresponding to the scalar fields $X_{7,8,9}$, and one compact real
modulus from the dual scalar associated with the 2+1-dimensional gauge
field.

\begin{figure}
\centerline{\includegraphics[scale=0.8]{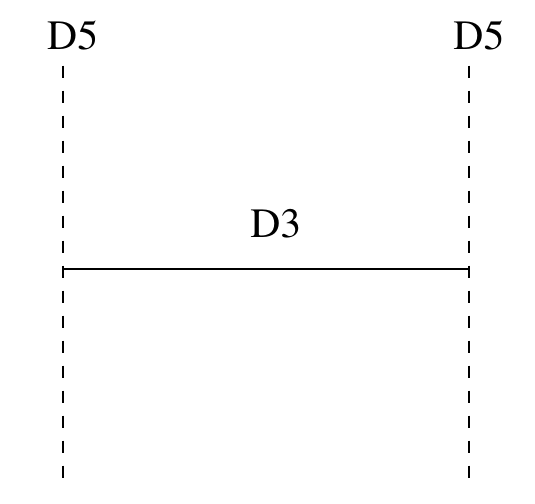}}
\caption{The S-dual of pure $U(1)$ $\mathcal N = 4$ gauge theory.
\label{d5d3d5}}
\end{figure}

The configuration of one D3-brane suspended between two D5-branes,
shown in Figure \ref{d5d3d5}, is the S-dual of pure $\mathcal{N}=4$
$U(1)$ gauge theory.  Its moduli space is simply $\mathbb{R}^3 \times
S^1$, with three noncompact dimensions coming from the motions of the
D3-brane in $X_{4,5,6}$ and one compact modulus from the Wilson line
$\int A_3 dy$.

What might be a little unclear is how to understand this moduli space
from the point of view of $G_{\mathbb C}$, as naively one might think
that one can gauge away $X_6$ and $A_3$.  The point is that this gauge
transformation is possible when the branes on one side of the
configuration are NS5-like, but not when they are D5-like on both
sides.  We are constrained to allow only gauge transformations
satisfying $g=1$ at a D5-type boundary condition.  But the gauge
transformation which sets $A_3=0$, for example, is of the form
\beq
g(y) =\exp\left( \int_0^y dy' A_3(y')\right) \ ,
\eeq
which will only satisfy $g=1$ at $y=0$, where we hereafter (unless
explicitly stated) take the position of the leftmost D5 to be at
$y=0$.

Nevertheless, we can still use the complex gauge transformation
$G_{\mathbb C}$ to choose the gauge where $A_3$ is equal to its
average value over the interval,
\beq
g(y) =\exp\left( \int_0^y dy' (\A(y')-\langle \A \rangle )\right) \ ,
\eeq
which does satisfy $g=1$ at both boundaries.  The same considerations
apply for the complexified gauge transformations.  For this reason we
cannot gauge away $\A$, and instead find that the average value of
$\A$ on the interval is a modulus.  Note that because the gauge
transformation is periodic under shifts of $\langle A_3 \rangle$ by
$2\pi i$, the part of moduli space corresponding to VEVs of $\A$ in
this case has the topology of a cylinder.  It combines with the VEV of
$\X$ to give a two-complex-dimensional moduli space.

\subsection{$\mathcal{N}=4, N_c=1, N_f=2$} 

\begin{figure}
\centerline{\includegraphics[scale=0.8]{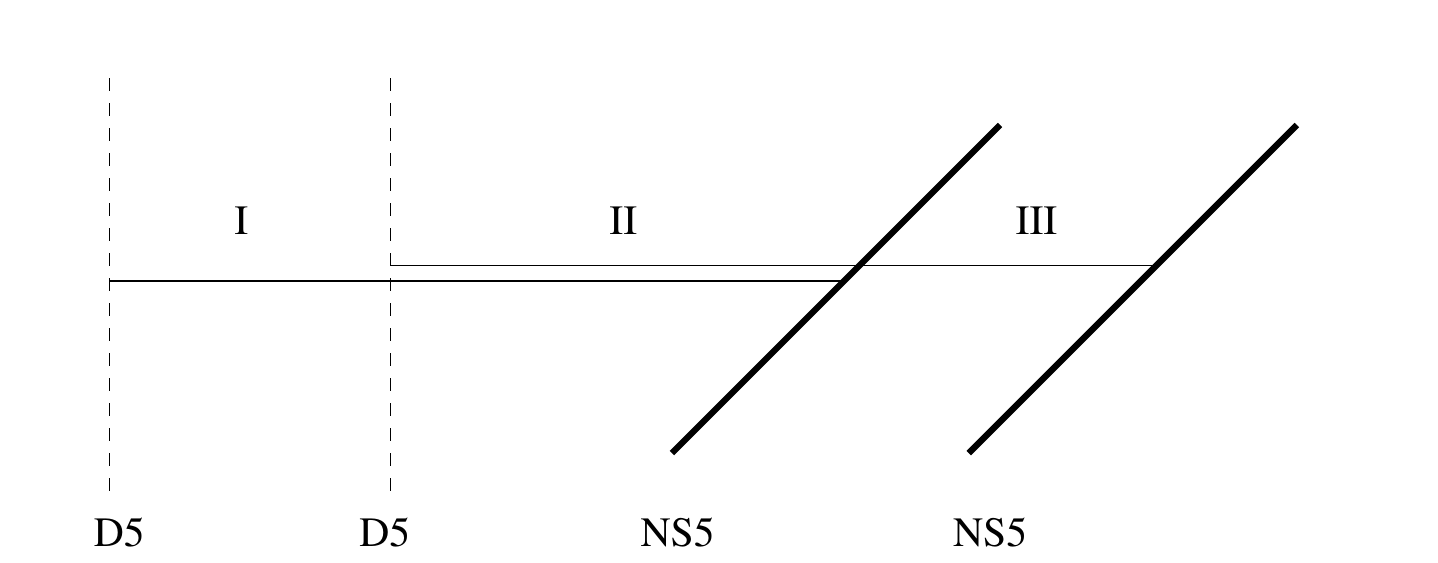}}
\caption{$U(1)$ $\mathcal N =4$ theory with two flavors.\label{nc1nf2N4}}
\end{figure}

Let us next add fundamental hypermultiplets, and consider the Abelian
theory with 2 flavors.  This theory, also called $T[SU(2)]$, is a
well-known example for three-dimensional mirror symmetry.  A brane
realization of this theory is shown in Figure \ref{nc1nf2N4}, which
maps to itself under S-duality, modulo the HW transition.
The defect theory is defined on intervals which we label with Roman
numerals.  In this example, we have three such regions, labeled I, II,
III, such that the gauge group is $U(1)$ in regions I and III and
$U(2)$ in region II.

In our framework, this example is simple because we can use complex
gauge transformations to set $\A=0$ everywhere.  Then the complex part
of the generalized Nahm equations simply imply that $\X$ and $\Y$ are
piecewise constant commuting matrices.

Crossing from region I into region II, we find that
\beq
\X_{II} &=&  \left( \begin{array}{cc} a & b \\ c & d \end{array} \right)  \ , \\
\Y_{II}&=& 0 \ .
\eeq
At the II--III interface we have $\X_{II}= AB$ and $\X_{III}=BA = 0$.
This fixes two eigenvalues of $\X_{II}$ to vanish.  If the
bifundamentals $A,B$ break the gauge symmetry, we have to set
$\Y_{III}=\Z_{III}=0$.  So we have a two-dimensional branch of moduli
space.  In particular, the condition that both eigenvalues of
$\X_{II}$ vanish is equivalent to the conditions $d=-a$, and $ a^2 +
bc =0$.  This Higgs branch moduli space is the orbifold
$\mathbb{C}^2/\bZ_2$, and because the gauge symmetry is completely broken, 
we expect it to be quantum mechanically
exact.

There is another class of solutions where $A=B=0$ and there is an
unbroken $U(1)$ gauge symmetry.  On this branch we have $a=b=c=d=0$
but $\Y_{III}$ and $\Z_{III}$ are free to vary.  So we have a
two-dimensional Coulomb branch with $U(1)$ gauge symmetry, but we do
not trust this branch in detail because it is subject to quantum
corrections.

Because this theory is self-dual under S-duality, a natural conjecture
is that the two 2$d$ branches are exchanged by the duality; therefore,
the quantum corrected Coulomb branch is also the orbifold
$\mathbb{C}^2/\bZ_2$.  This analysis can be generalized to any number of flavors, as 
described in Appendix \ref{appA}.

\subsection{Nahm Pole}

In our previous example (and many other examples below), the fact that
we can take $\A=0$ by a complex gauge transformation simplifies the
analysis dramatically.  The remaining equations are algebraic and thus
the moduli space computation reduces to a problem of linear algebra.

As we move on to more complicated examples, however, it becomes
necessary to properly take into account the singularities of the
complex gauge field.  When Nahm poles are included, it is not possible
to set $\A = 0$ by a non-singular gauge transformation, but it is
often still possible to choose a simple form which makes the problem
essentially algebraic.  

\begin{figure}
\centerline{\includegraphics[scale=0.8]{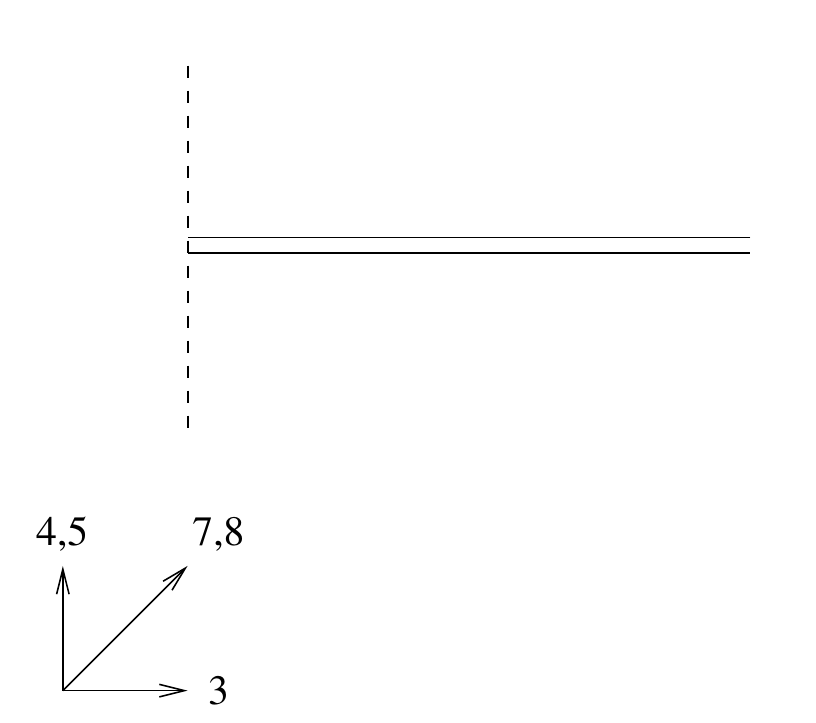}}
\caption{Two D3-branes ending on a D5-brane.
\label{DirichletU2}}
\end{figure}

To explain this, we consider a simple noncompact example where two
semi-infinite D3-branes end on a D5-brane, as shown pictorially in
Figure \ref{DirichletU2}.  There is a $U(2)$ gauge theory with
$\mathcal N = 4$ supersymmetry living on the D3-branes.  The boundary
conditions corresponding to the D5-brane are that there is a Nahm pole
singularity at the location of the D5-brane on the left end.

We can choose the gauge where the complex gauge field has only
singular contribution
\beq
\scA = \left( \begin{array}{cc} \frac{1}{2y} & 0 \\ 0 & -\frac{1}{2y} \end{array} \right) \ .
\label{poleboundary}
\eeq
Then solving the Nahm equations we find
\beq
\X =  \left( \begin{array}{cc} a & \frac{1}{y} \\ b y  & a \end{array} \right)  \ .
\label{poleboundaryX}
\eeq
Note that this solution captures the leading behavior of the fields
for $y\rightarrow 0$.  The subleading terms which satisfy Nahm's
equations can be removed by a $G_{\mathbb C}$ gauge transformation
with $g(0)=1$.\footnote{In particular, one can check that $\X =
  \left( \begin{array}{cc} a & \frac{1}{y} \\ b y & c\end{array}
    \right) $ also solves Nahm's equations, but there is enough
    residual gauge symmetry to set $c=a$.}  In this analysis, we are
allowing $g(\infty)$ to be arbitrary.

For the Nahm pole, the matrix-valued fields $X_{4,5,6}$ do not commute
with each other.  Thus they cannot be simultaneously diagonalized, and
it is not possible to simply interpret their eigenvalues as the
positions of D3-branes.  This is the essential feature of the defect
analysis which is hard to capture in the brane drawing.

\subsection{$\mathcal{N}=4, N_c=2, N_f=0$}

\begin{figure}
\centerline{\includegraphics[scale=0.8]{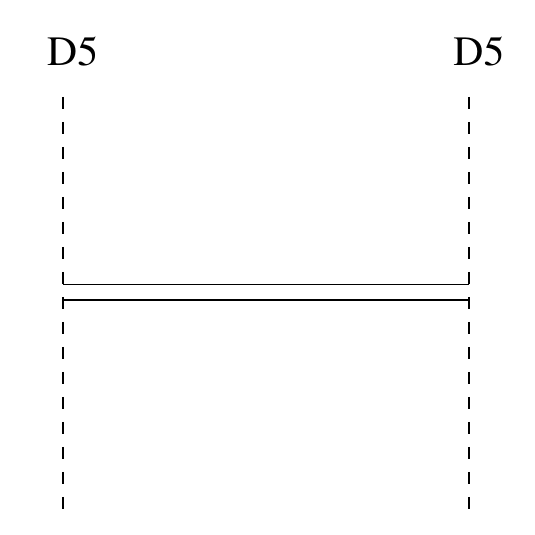}}
\caption{S-dual of pure $U(2)$ $\mathcal N = 4$ gauge theory.
\label{d52d3d5}}
\end{figure}

The standard $SU(2)$ Nahm 2-monopole construction also has an
interesting problem of gauge fixing.  This is equivalent to the brane
construction of pure $U(2)$ gauge theory with $\mathcal N= 4$
supersymmetry, as shown in Figure \ref{d52d3d5}.  At both D5
boundaries, we are only allowed to do gauge transformations where
$g|_{\partial M} = 1$.  This means that we are not allowed to choose
the gauge (\ref{poleboundary}). This is a situation where the complex
gauge formalism is less useful.

Instead, we have to solve the system including the subleading terms,
without the ability to choose a simple gauge.  The $SU(2)$ solution of
Nahm's equations can be written in terms of special functions (the
so-called Euler top functions); see \cite{Weinberg:2006rq} for a
pedagogical review.

The moduli space is 4-complex-dimensional.  When the monopoles are
well-separated, it is possible to understand the dimension intuitively.
There is an $\mathbb R^3$ corresponding to the 3 center-of-mass
coordinates for the 2 monopoles, one (positive) real coordinate
corresponding to the monopole separation, 3 Euler angles rotating the
2-monopole configuration (the monopole solutions are not
axisymmetric), and an $S^1$ from the $U(1)$ gauge framing of the
monopoles.

\subsection{$\mathcal{N}=4, N_c=2, N_f=1$}

\begin{figure}
\centerline{\includegraphics[scale=0.8]{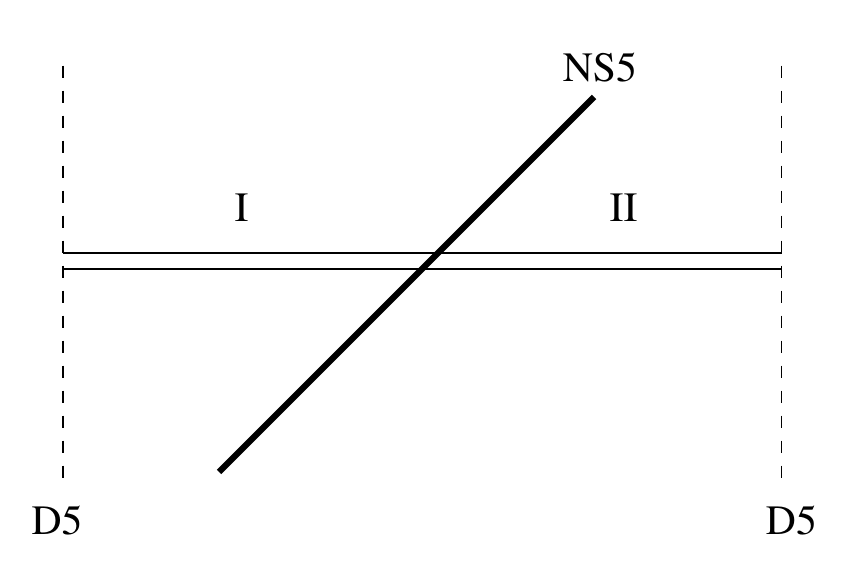}}
\caption{S-dual of $U(2)$ $\mathcal N = 4$ gauge theory with one flavor.
\label{nc2nf1N4}}
\end{figure}

Another interesting system corresponds to adding a flavor to the
$U(2)$ $\scN = 4$ gauge theory by adding an NS5-brane in the interval
(in the S-dual representation.)  This is represented by the brane
configuration shown in Figure \ref{nc2nf1N4}. This theory should also
have a four complex-dimensional Coulomb branch and no Higgs branch.

In this example a crucial simplification occurs compared to the the
previous case, because we are allowed to use the $G_{\mathbb C}$
quotient.  In the real formulation of this problem, the fields
$X_{4,5,6}$ are again described by the complicated Euler top
functions, but if we are only interested in the complex structure, we
{\it can} choose a $G_{\mathbb C}$ gauge where the fields take a
simpler form.

The key difference between this example and the previous one is that
there is an NS5-brane, so we are allowed to do gauge transformations
which are discontinuous at the NS5.  This means that it is easy to
satisfy $g=1$ at both D5-boundaries while choosing a convenient gauge in the bulk.  
We place the NS5 at $y=0$, and we place the D5 and
D5$^\ensuremath{\prime}$ at $ y= \pm 1$.  Then we can use a complex
gauge transformation to set
\beq
\A_I &=& \left( \begin{array}{cc} \frac{1}{2(y+1)} & 0 \\ 0 & - \frac{1}{2(y+1)}  \end{array} \right) \ ,\\
\A_{II} &=& \left( \begin{array}{cc} \frac{1}{2(1-y)} & 0 \\ 0 & - \frac{1}{2(1-y)}  \end{array} \right) \ , 
\eeq
and we can solve Nahm's equations with
\beq
\X_I &=&  \left( \begin{array}{cc} a & \frac{1}{y+1} \\ b (y+1)  & a \end{array} \right) \ , \\
\X_{II} &=&  \left( \begin{array}{cc} a' & \frac{1}{1-y} \\ b' (1-y)  & a' \end{array} \right) \ .
\eeq

At the NS5, we have to satisfy $\X_I = AB$ and $\X_{II} = BA$ where
$A,B$ are $2\times 2$ matrices.  This forces $a'=a$ and $b'=b$, so in
this gauge $\X$ is actually continuous across the boundary, with
\beq
\X_{y=0}=  \left( \begin{array}{cc} a & 1 \\ b   & a \end{array} \right) \ .
\eeq
Note that one solution is given by $B=\mathbb I$ and $A= \X_{y=0}$;
however, this is not the only form of $A$ and $B$ which solves the
constraint.  There is a two-complex-dimensional family of solutions,
given by rigid $GL(2)$ rotations, one of which is generated by
$\mathbb I$ and the other is generated by $\X_{y=0}$.  These transform
$A, B$ nontrivially but do not transform $\X$.

The total moduli space is four complex dimensional.  The point we wish
to emphasize is that in the case with an NS5 brane we were able to
dramatically simplify the problem by using $G_{\mathbb C}$ and in fact
it was unnecessary to consider the Euler top functions.

\subsection{Constraints from the $s$-rule}

\begin{figure}
\centerline{\includegraphics[scale=0.8]{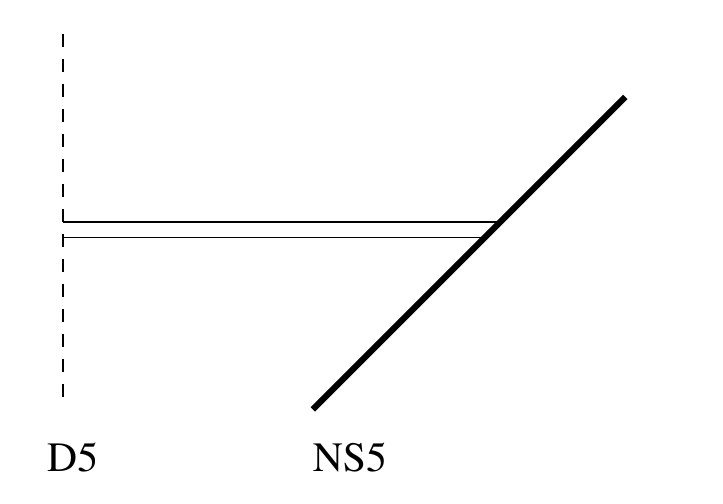}}
\caption{$s$-rule violating configuration. \label{srule}}
\end{figure}

Continuing to more general situations with NS5-branes, we might
consider the case where 2 D3-branes end on a D5 on one side and on an
NS5 on the other side, as shown in Figure \ref{srule}.  Of course,
this is the situation which is well-known to be excluded by the
Hanany-Witten $s$-rule.

The absence of a solution is easy to see using $G_{\mathbb C}$.
Because there is an NS5-brane on the right, we are free to choose the
form (\ref{poleboundary}) for the gauge field.  Then the complex
scalar which solves Nahm's equations takes the form
(\ref{poleboundaryX}).  However, we also have to impose the boundary
conditions at the NS5-brane at a finite distance in $y$ from the Nahm
pole at the D5-brane.  The NS5-brane boundary conditions imply that
$\X=0$ at the NS5.  There is no solution of Nahm's equations with a
pole on the D5 which satisfies this condition at the NS5 on the right.

Because there is no solution to Nahm's equations satisfying the given
boundary conditions, we conclude that supersymmetry must be broken in
this configuration.  Note that we did not have to impose the $s$-rule
as a separate condition; it simply follows as a consequence of our
analysis.  We could also have analyzed this system directly, without
using the complex gauge quotient; such a calculation will reach the
same conclusion.

Our computation gives a new derivation of the $s$-rule, specifically in the D3-D5-NS5 duality frame.  
Previous derivations of the $s$-rule in other duality frames appeared in 
\cite{Ooguri:1997ih,Hori:1997ab,Bachas:1997kn,Bachas:1997sc}.  

\begin{figure}
\centerline{\includegraphics[scale=0.8]{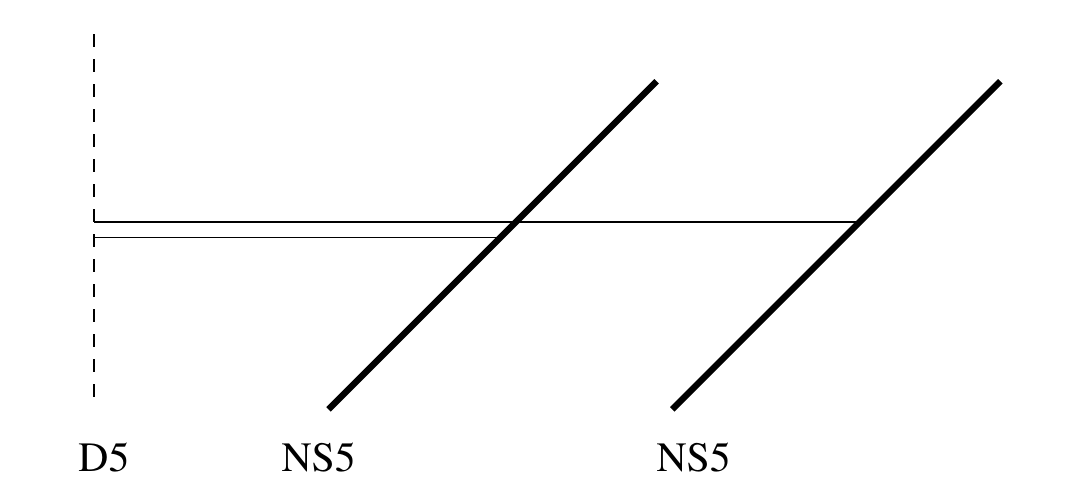}}
\caption{Adding a second NS5-brane satisfies the $s$-rule.  \label{d52d3ns5d3ns5}}
\end{figure}

If the two D3-branes are allowed to end on two different NS5 branes
then we no longer have a restriction from the $s$-rule, as in the
brane configuration in Figure \ref{d52d3ns5d3ns5}.  

Because we have NS5-branes on the right, we have the freedom to choose
a convenient $G_{\mathbb C}$ gauge.  Let us take
\beq
\A =\left( \begin{array}{cc} \frac{1}{2y} & 0 \\ 0 & - \frac{1}{2y}  \end{array} \right) \ ,
\eeq
so that 
\beq
\X = \left( \begin{array}{cc} a & \frac{1}{y} \\ b y  & a \end{array} \right) \ .
\eeq
Without loss of generality, we can locate the leftmost NS5-brane at
$y=1$.  At this NS5-brane, there are bifundamental fields $A, B$.  The
equations they satisfy are
\beq
AB &=& \left( \begin{array}{cc} a & 1 \\ b  & a \end{array} \right)  \ ,\\
BA &=& 0 \ ,
\eeq
which fix
\beq
A &=& \left(\begin{array}{c} v \\ 0 \end{array} \right)  \ , \\
B &=& \left(0, \frac{1}{v}\right) \ .
\eeq
and $a=b=0$.  We may further fix $v$ by a $G_{\mathbb C}$
transformation, or equivalently by solving the equation for $X_6$ and
modding out by a real gauge transformation.  The moduli space is
zero-dimensional, but it is not trivial, which is consistent with the
fact that the configuration in Figure \ref{d52d3ns5d3ns5} satisfies
the $s$-rule.

\section{3$d$ $\mathcal{N}=2$ $U(1)$ Gauge Theories}

\label{sec:abelian}

Now let us come to address theories of our main interest, namely 3$d$
$\mathcal{N}=2$ theories.  In this section we study $U(1)$ gauge
theories with $N_f$ flavors, starting with $N_f=1$ and then $N_f=2$
and more general $N_f$.  The moduli spaces for even these simple
theories have a significant amount of structure, and we will see that
our techniques recover all previously known results
\cite{deBoer:1997ka,deBoer:1997kr,Aharony:1997bx}.

A typical example of a brane construction for this class of theories
is illustrated in Figure \ref{figb}. Both of the brane orderings
illustrated there should give rise to the same theory in the limit
where all but the 2+1 dimensional dynamics is decoupled.  For
concreteness, let us consider the brane ordering illustrated on the
right in Figure \ref{figb}.  As in the $\mathcal{N} =4$ case, the
D3-brane contributes a vector field (which in the classical theory
gives rise to the dual scalar $\varphi$), but because the NS5 and
NS5$^\ensuremath{\prime}$ branes are rotated relative to each other,
the D3 is only free to move in the shared direction $X_9$.  There are
also fundamental quark multiplets localized at the
D5$^\ensuremath{\prime}$-branes.

The $\mathcal{N}=2$ $U(1)$ theories have moduli spaces with a Higgs
branch and a Coulomb branch.  On the Higgs branch, the fundamental
quarks have expectation values.  Because the gauge symmetry is broken
by the quark vevs, this branch can be computed classically and has
dimension $2N_f-1$.  The Coulomb branch, however, is potentially
subject to quantum corrections.  The goal of this section is to show that the
S-dual Nahm analysis captures these quantum effects.

There is also an elaborate structure of deformations one can consider
by giving masses to the quarks and turning on Fayet-Iliopoulos terms,
which correspond in the brane construction to changing the positions
of the 5-branes.  These structures have been studied in earlier work
\cite{deBoer:1997ka,deBoer:1997kr,Aharony:1997bx}, using mirror
symmetry and some degree of educated guesswork.  The new perspective
we aim to present in this work is the observation that the analysis of
supersymmetric field equations of the boundary/defect field theory in
3+1 dimensions provides a complementary {\it systematic} tool to study
features such as the structure of moduli spaces. In the course of this
analysis, we discover some interesting new features with regard to the
action of S-duality on these branches.

\subsection{${\cal N}=2, N_c=N_f=1$}

\begin{figure}
\centerline{\includegraphics[width=\hsize]{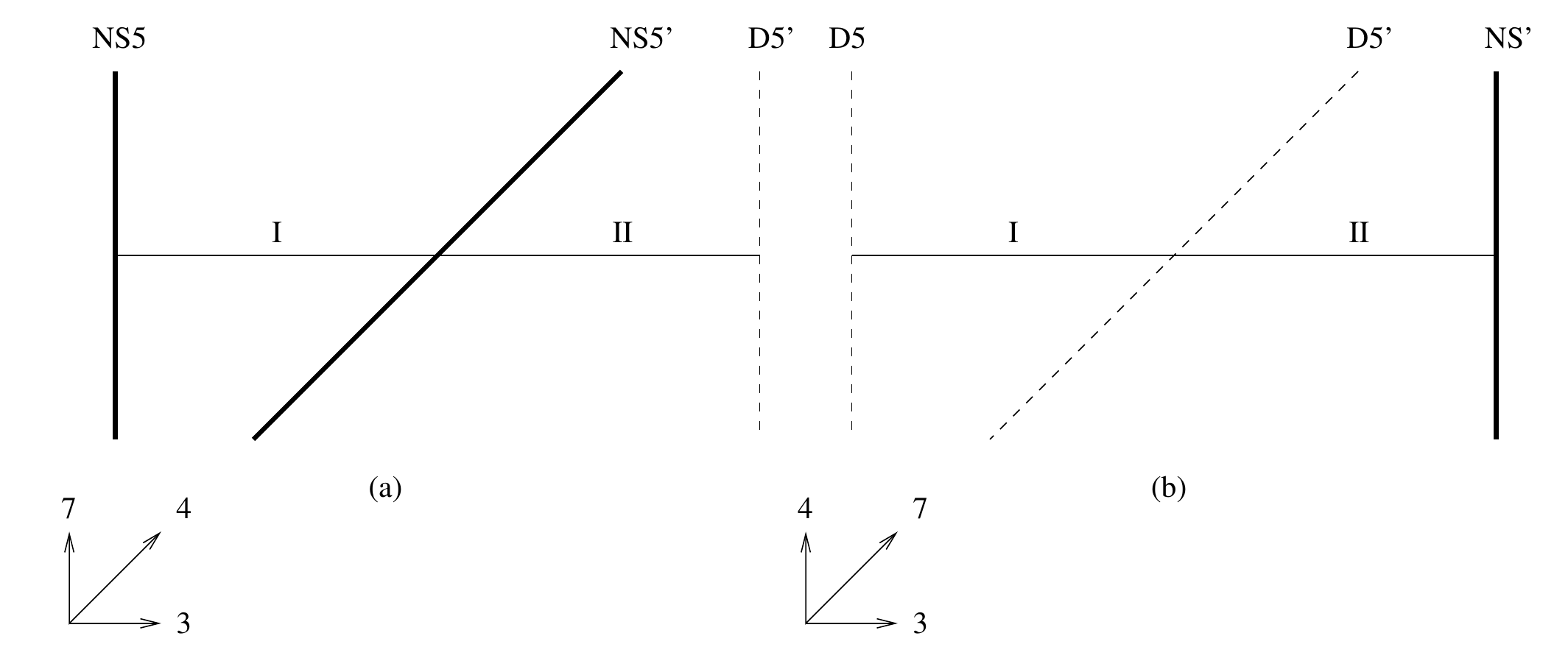}}
\caption{(a) Brane construction of ${\cal N}=2$ $N_c=1$ $N_f=1$ theory
  and (b) its S-dual. We will refer to the original theory
  as the electric theory or the A-model and the S-dual as the magnetic
  theory or the B-model.
\label{figc}}
\end{figure}

In this subsection, we will analyze the simplest non-trivial case of
$U(1)$ theory with $N_f=1$ quarks (the case of $N_f=0$ will be
discussed from the mass deformation of the $N_f=1$ theory.) It turns
out that this example is sufficient to illustrate some of the most
important aspects of $\mathcal{N}=2$ mirror symmetry.  Let us re-draw
Figure \ref{figb} for the specific case of interest in Figure
\ref{figc}.a.  We will refer to the theory as depicted in Figure
\ref{figc}.a as the ``electric'' formulation of the gauge theory.
This is the formulation where the gauge theory interpretation is
simplest; there is a vector multiplet with $U(1)$ gauge symmetry from
the D3 extending between the NS5 and NS5$^\ensuremath{\prime}$ branes,
and the D5$^\ensuremath{\prime}$ brane contributes one flavor.  We
will also consider the S-dual brane configuration, shown in
\ref{figc}.b, which we will sometimes call the ``magnetic''
formulation.  In both cases, we divide the $y$-direction into two
regions, which we label as region I and region II.

Let us analyze the moduli space of the electric configuration shown in
Figure \ref{figc}.a. We start in region I of Figure \ref{figc}.a, and
make the gauge choice $\A=0$; the bulk equations
(\ref{qnahm1})--(\ref{qnahm4}) then imply that all the fields are
constants.  In region I, we have
\beq {\cal Y}_I &=& a \ , \\
{\cal X}_I & = & 0  \ , \\
X_{6I} & = & 0 \ . \label{nc1nf1x6I}
\eeq
At the  NS5$^\ensuremath{\prime}$, we impose the conditions
\beq a & = & {\cal Y}_I = {\cal Y}_II = AB \ , \\
i X_{6I} & = & A A^* - B B^* = 0 \ ,  \label{Nc1Nf1X6}
\eeq
where $A$ and $B$ are the $1 \times 1$ bifundamentals living on the
NS5$^\ensuremath{\prime}$.  We now use the complex gauge formalism to
set $\A=0$ and suppress the real equation (\ref{Nc1Nf1X6}).
Furthermore, we must mod out by the gauge transformation at the
NS5$^\ensuremath{\prime}$ interface,
\beq
A\rightarrow gA \ , \qquad B \rightarrow g^{-1} B\ .
\label{Nc1Nf1gauge}
\eeq 
We see that there is a branch of moduli space which is
one-complex-dimensional (parameterized by $A$, for example, with $B$
fixed by a choice of gauge.)  The quarks have VEVs on this branch,
which one would naturally call the Higgs branch of the gauge theory.

When $A=B=0$, an additional branch of moduli space opens up because
$X_9$ and the dual scalar can have expectation values.  This is the
classical Coulomb branch of the gauge theory, with complex structure
$\mathbb{R} \times S^1$.  Note that there are no branches with $A=0$ and $B\neq 0$ (or vice versa) because they
consist of unstable points of the complex gauge quotient.

Now, let us now analyze the same system from the point of view of the
S-dual ``magnetic'' formulation, as shown in Figure \ref{figc}.b.  In
the S-dual, there are no gauge degrees of freedom in the 2+1
dimensional low energy effective theory.  One of the points of
considering the defect theory of the UV embedding of 2+1 theories is
that the 3+1-dimensional defect theory is a gauge theory with a known
Lagrangian, even if the 2+1-dimensional formulation is not.

To carry out this analysis, we start in region I of Figure
\ref{figc}.b and find
\beq {\cal X}_I &=& a \ ,  \\
i X_{6I} & = &  x_6  \ , \label{x6free}\\
{\cal Y}_I & = & 0 \ .  \label{Y10}
\eeq
Recall that a factor of $i$ appears in front of $X_{6I}$ because we
are using the convention that the bosonic fields are anti-hermitian.

\begin{figure}
\centerline{\includegraphics[scale=0.8]{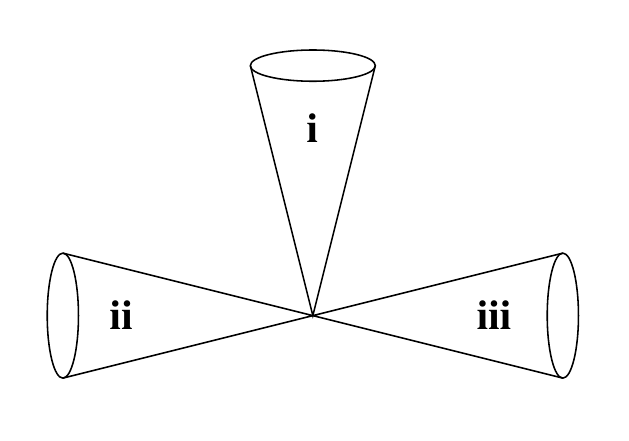}}
  \caption{Quantum-corrected branch structure of the moduli space of
    the ${\cal N}=2$ $N_c=1$ $N_f=1$ theory. The three branches of
    moduli space each have the same complex structure as $\mathbb C$; they are
    drawn as cones to show that the branches intersect at a
    point.  \label{fige}}
\end{figure}

Next, at the D5$^\ensuremath{\prime}$ between regions I and II, we
impose the condition
\beq {\cal Y}_{II} &=& {\cal Y}_{I} + Q \tilde Q  \ ,\label{Y2Y1QQ} \\
{\cal X}_I Q & = & 0\label{XQ1} \ ,\\
{\cal X}_{II} Q & = & 0 \ , \label{XQ2}\\
\tilde Q {\cal X}_I  & = & 0\ ,  \label{XQ3}\\
\tilde Q {\cal X}_{II}  & = & 0\ ,  \label{XQ4} \\
i X_{6 II} & = & i X_{6 I} + (|Q|^2 - |\tilde Q|^2)\ ,  \eeq
where $Q$ and $\tilde Q$ are the quark fields associated with the
D5$^\ensuremath{\prime}$ brane. Finally, the NS5$^\ensuremath{\prime}$
brane imposes the condition
\be {\cal Y}_{II} = 0  \ .\ee
Now, combining  (\ref{Y10}) and (\ref{Y2Y1QQ}), we learn that
\be Q \tilde Q = 0 \ . \ee
Notice that we have no further constraint on the magnitudes of the
scalars in the quark multiplets because the parameter $x_6$ in
(\ref{x6free}) is free to vary.  This means we have as possibilities
that either $Q$ or $\tilde Q$ vanishes, or that both are vanishing. If
either one of $Q$ or $\tilde Q$ is non-vanishing, then one of
(\ref{XQ1})--(\ref{XQ4}) forces $a=0$. On the other hand, should $Q$
and $\tilde Q$ simultaneously vanish, then there is no additional
constraint on $a$. So, we have found three branches
\begin{itemize}
\item[{\bf i}]  $Q=\tilde Q =0$, $a$ arbitrary
\item[{\bf ii}]  $a=\tilde Q =0$, $Q$ arbitrary
\item[{\bf iii}]  $a=Q =0$, $\tilde Q$ arbitrary
\end{itemize}
which we illustrate schematically in Figure \ref{fige}.

In the magnetic formulation, the gauge symmetry is
completely broken on all branches of the moduli space.  This suggests
that the moduli space which we have computed using the S-dual is
actually quantum mechanically exact.  Indeed, this moduli space
structure is consistent with a superpotential of the form
\be W = a Q \tilde Q \ . \ee
This is precisely the superpotential obtained in (3.2) of
\cite{Aharony:1997bx}: $a$ ($Q, \tilde{Q}$) can be identified with the
meson (monopole) operators of the electric theory.  We see that from
purely semi-classical considerations, we have obtained the
quantum-corrected branch structure predicted in Figure 1 of
\cite{Aharony:1997bx} which we reproduce here.  The classical Coulomb
branch which had the complex structure $\mathbb{R} \times S^1$ splits
into two separate branches as shown; under S-duality we see that the
Coulomb branch has been mapped to Higgs branches {\bf ii} and {\bf
  iii} where $Q$ or $\tilde Q$ has an expectation value.  The Higgs
branch ${\bf i}$ parameterized by $a$, on the other hand, is in the
Higgs phase on both sides of the S-duality. See also the bottom of
Figure \ref{figd2} for an illustration of this point.

\subsubsection{Complex Mass Deformation to $\mathcal{N}=2, N_c=1, N_f=0$}

We can add a complex mass deformation to this system (of the electric
theory) to reduce the number of flavors to $N_f=0$.  We do this simply
by following the preceding analysis but shift
\beq 
\Y_{II}=m_c \ ,
\eeq
which gives the complex equation
\beq
Q \tilde{Q} = m_c \ .
\eeq
This requires $\tilde{Q},Q \neq 0$ which in turn forces $a=0$.  The
moduli space is one-dimensional, and is $T^* S^1$ (the cylinder),
which is expected for pure $U(1)$ theory.

\subsubsection{Real Mass and FI Deformations}

There are two naturally real mass deformations we might consider in
the $U(1)$ theory with $N_f=1$.  They are to displace the
D5$^\ensuremath{\prime}$-brane in $X_9$, which we identify as a real
mass deformation, and to separate the NS5 and
NS5$^\ensuremath{\prime}$ branes in the $X_6$ direction.  Because the
NS5 and NS5$^\ensuremath{\prime}$ are both extended in $X_9$, the real
mass deformation is trivial (this will not be the case if $N_f>1$.)

The real FI parameter, however, is not trivial.  We can express this
deformation by modifying (\ref{nc1nf1x6I}) to $iX_{6,I} = \zeta_r $ so
that (\ref{Nc1Nf1X6}) becomes
\beq
A A^* - B B^* = \zeta_r\ .
\eeq
Then the fields $A$, $B$ cannot both vanish.  This lifts the classical
Coulomb branch ($X_9$ and the dual scalar $\varphi$ are both forced to
vanish) and we are left with the Higgs branch.

In the S-dual, the corresponding deformation may be thought of as a
real mass parameter arising from moving the D5-brane in the $X_9$
direction to $iX_{9,I}=m_r$ while keeping the D5$^\ensuremath{\prime}$
fixed at $X_9=0$.  This forces $Q= \tilde Q=0$, and we are left with
only one 1-dimensional branch of moduli space.

\begin{figure}
\centerline{\includegraphics[scale=0.8]{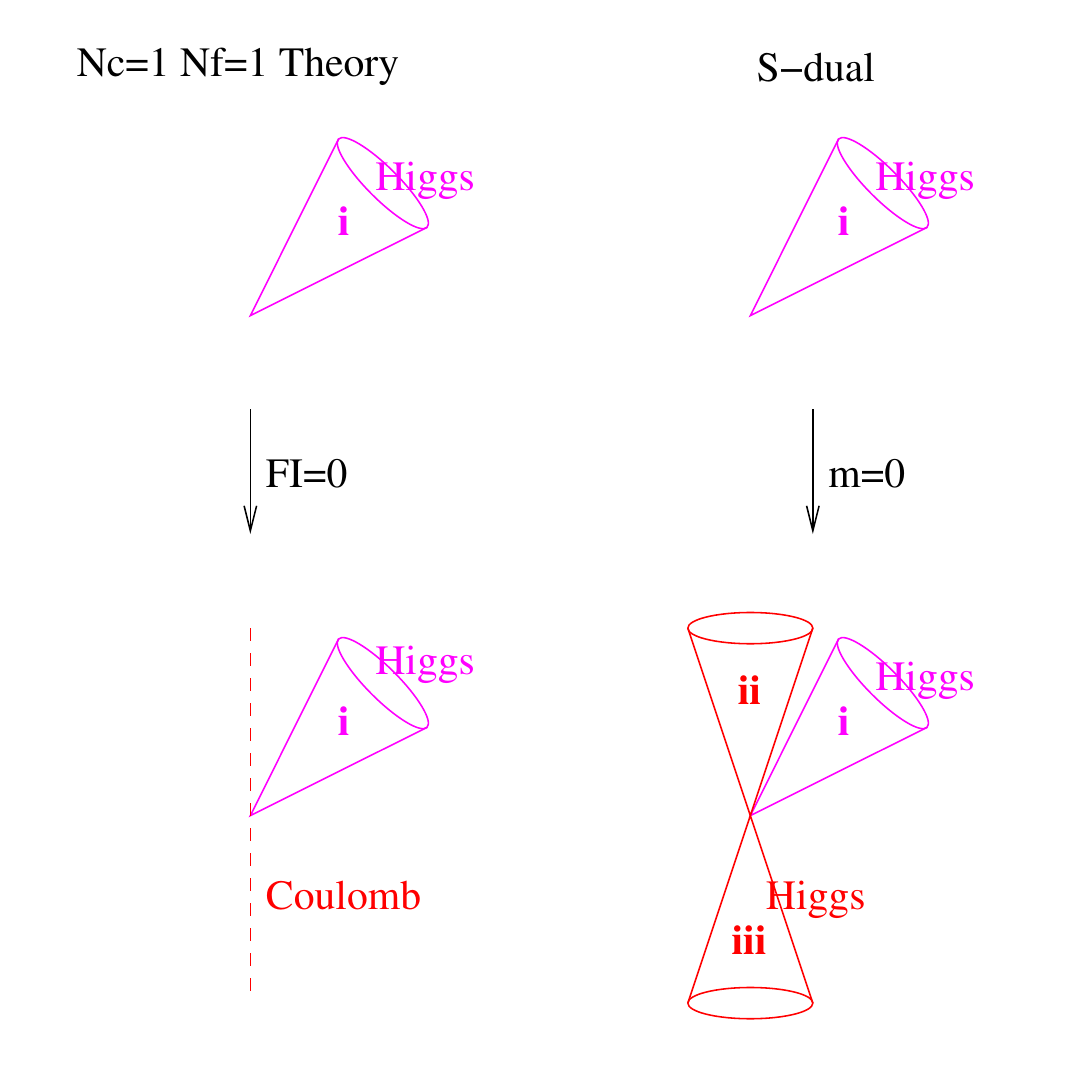}}
\caption{Branch structure of ${\cal N}=2$, $N_c=1$, $N_f=1$ theory,
  deformed by an FI parameter, and its S-dual, deformed by a real mass
  parameter. The FI and mass deformations are S-dual. In the limit of
  vanishing FI parameter, one expects the Coulomb branch to open up,
  but its complex structure and its metric are subject to quantum
  corrections. In a given duality frame , the branches which are
  subject to quantum corrections are illustrated with a dotted
  line. The quantum corrected structure of the Coulomb branch can be
  inferred from the S-dual picture, where the gauge symmetry is
  broken.
\label{figd2}}
\end{figure}

The moduli space we find is in a certain sense ``self-dual'' under
mirror symmetry, with the understanding that we need to trade an FI
parameter for a mass parameter.  The reason why the moduli space
appears self-dual is easy to understand from the brane perspective.
In the representation of Figure \ref{figc}, because of the mass
deformation, the D3-brane is unable to intersect the 5-brane in the
middle of the brane diagram.  Therefore we are just left with a D3
brane stretched between a D5 and NS5$^\ensuremath{\prime}$ or between
an NS5 and D5$^\ensuremath{\prime}$, which are simply exchanged by
S-duality.  The pattern of mass deformation is shown in Figure
\ref{figd2}. This is precisely the class of branches of moduli space
which we highlighted at the end of table \ref{tablebranch}.

\subsection{${\cal N}=2, N_c=1, N_f=2$\label{sec:nc1nf2}}

As our second example, we consider the case of $U(1)$ theory with
$N_f=2$ quarks. This will turn out to be an instructive example
highlighting many interesting features.

\begin{figure}
\centerline{\includegraphics[width=\hsize]{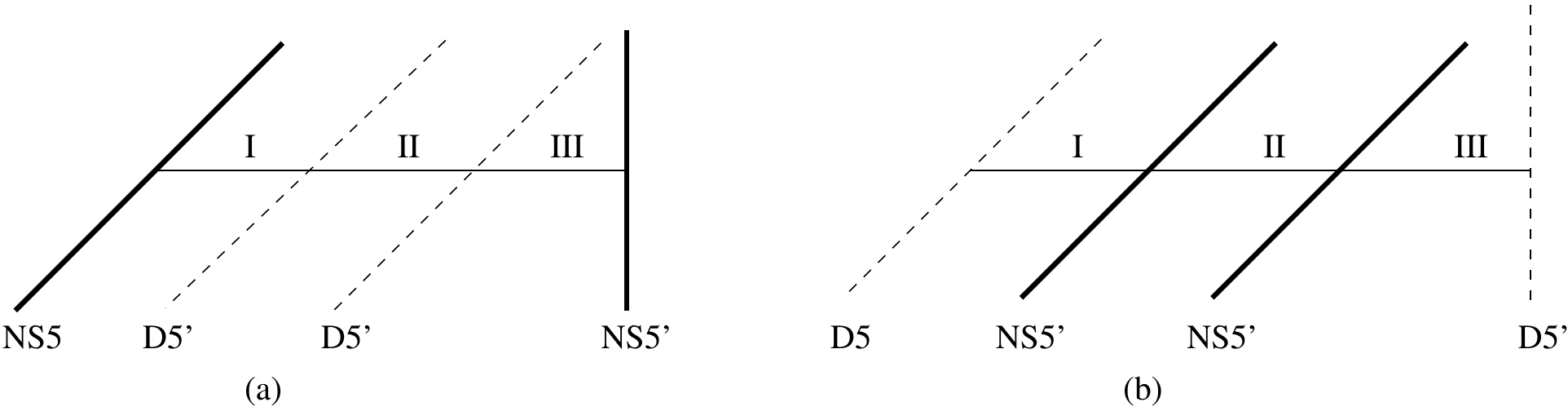}}
\caption{(a) Brane construction for ${\cal N}=2$ $N_c=1$ theory with 2
  flavors and (b) the S-dual. \label{nc1nf2N2}}
\end{figure}

The brane construction realizing this field theory is shown in Figure
\ref{nc1nf2N2}.a.  The two D5$^\ensuremath{\prime}$-branes give rise
to two flavors and the relative orientation of the NS5 and
NS5$^\ensuremath{\prime}$-branes breaks the supersymmetry to
$\mathcal{N}=2$.  We have chosen a convenient ordering of the branes
so that all the fields are Abelian.  We leave it as a simple exercise
to show that the analysis of our Nahm boundary conditions reproduces
the result of the classical field theory computation.  In particular,
there is a $1d$ Coulomb branch (with complex structure $\mathbb R
\times S^1$) and a $3d$ Higgs branch (with the complex structure of
the conifold.)

Because we are interested in the quantum-corrected Coulomb branch, we
consider the S-dual, which is shown in Figure \ref{nc1nf2N2}.b.  From
our experience with the $N_f=1$ example, our expectation is that the
Coulomb branch will be mapped by S-duality to a branch where the gauge
symmetry is completely broken, and then the semiclassical computation
of the moduli space will be reliable.

The scalars may be taken to be
\beq
\mathcal{X}_I &=& a \ ,\\
\mathcal{Y}_I &=& 0\ , \\
\mathcal{X}_{II} &=& b\ , \\
\mathcal{Y}_{II} &=& c\ , \\
\mathcal{X}_{III} &=& 0\ ,\\
\mathcal{Y}_{III} & = & d \ .
\eeq

We also have bifundamental matter $A_1, B_1$ at the I--II interface,
and $A_2, B_2$ at the II--III interface.  Because everything is
Abelian, the constraint equations are rather simple:
\beq
A_1 (\mathcal{X}_I  - \mathcal{X}_{II}) &=& 0\ ,\\
B_1 (\mathcal{X}_I  - \mathcal{X}_{II}) &=& 0\ ,\\
A_2 (\mathcal{X}_{II} - \mathcal{X}_{III}) &=& 0\ ,\\
B_2 (\mathcal{X}_{II}  - \mathcal{X}_{III}) &=& 0\ ,
\eeq
and
\beq
&&\mathcal{Y}_I = A_1 B_1\ ,\\
&&\mathcal{Y}_{II} = B_1 A_1=A_2 B_2 \ ,\label{Nc1Nf2YII}\\
&&\mathcal{Y}_{III}  =  B_2 A_2\ .\label{Nc1Nf2YIII}
\eeq
These conditions actually set $\mathcal{Y}=0$ in all three regions. 

When the real deformations are not turned on, the complex gauge
quotient works straightforwardly.  There are two classes of stable
solutions.  The first class of solutions has $A_i = B_i =0$.  Then the
scalars $a$ and $b$ are not fixed.  In addition to these we have the
scalar $\scZ_{II}$ which is free to vary.  This branch is
3-dimensional.  The second class of solutions has two branches where
all the bulk scalars vanish.  On the first branch the interface fields
$B_1=B_2=0$ but $A_1, A_2$ are nonzero, and on the second branch
$A_1=A_2=0$ while $B_1, B_2$ are nonzero.  After modding out by the
complex gauge symmetry we are left with two one-dimensional branches.

The natural picture for the fully quantum-corrected moduli space comes
from combining the reliable parts of the analysis from both the
electric and magnetic descriptions. In the electric defect system, we
have a 3-dimensional branch with the complex structure of the conifold
and with no gauge symmetry.  There is also a 1-dimensional branch with
unbroken gauge symmetry (so we don't trust the analysis) with complex
structure $\mathbb R \times S^1$.  On the magnetic side, we have a
3-dimensional mixed branch with unbroken $U(1)$ symmetry which we do not
trust, and two 1-dimensional branches where the gauge symmetry is
fully broken, which we do trust.  So to construct the full moduli
space we should take the 3-dimensional branch from the electric
description and the 1-dimensional branches from the magnetic
description.

To describe the case where real FI terms are turned on, we should
restore the equations for $X_6$.  We have (including generic real FI
terms, which correspond to real masses of the original theory)
\beq
iX_{6,I} &=& A_1 A_1^{\dag}- B_1^{\dag}B_1\ ,\\
iX_{6,II} &=& A_1^{\dag} A_1 - B_1 B_1^{\dag}\ ,\\
iX_{6,II} &=& A_2 A_2^{\dag}- B_2^{\dag}B_2-\zeta_r\ ,\\
iX_{6,III} &=& A_2^{\dag} A_2 - B_2 B_2^{\dag}-\zeta_r\ .
\eeq

We immediately identify some branches, where only one of the $A_i,
B_i$ is non-vanishing:
\begin{itemize}
\item[{\bf i}] $A_2=B_2=A_1=0$, $B_1 \neq 0 $.  This requires $\zeta_r>0$, and $\mathcal{X}_I = \mathcal{X}_{II} \neq 0$ for a 1$d$ moduli space.
\item[{\bf ii}] $A_1=B_1=A_2=0$, $B_2\neq 0$. This requires $\zeta_r>0$. Here $\mathcal{X}_I \neq 0$ but $\mathcal{X}_{II,III}=0$.
\item[{\bf iii}] $A_1=B_1=B_2=0$, $A_2 \neq 0$. This requires $\zeta_r<0$. Here also $\mathcal{X}_I \neq 0$ but $\mathcal{X}_{II,III}=0$.
\item[{\bf iv}] $A_2=B_2=B_1=0$, $A_1 \neq 0$. This requires $\zeta_r<0$. Here we have $\mathcal{X}_I = \mathcal{X}_{II} \neq 0$ for a 1$d$ moduli space.
\end{itemize}
These branches of moduli space do not satisfy the real equations for
nonzero $\zeta_r$; we excluded them from the analysis using
$G_{\mathbb C}$ because they consist of unstable points.

The stable branches are those where two of the $A,B$ fields are
non-zero.  This forces us to set $\X=0$ and $\scZ=0$ in all regions.
Explicitly, the four branches are
\begin{itemize}
\item[{\bf v}] $A_1=A_2=0$ and $B_1, B_2 \neq 0$.  This requires $\zeta_r =|B_1|^2-|B_2|^2$.
\item[{\bf vi}] $B_1 = B_2 = 0$ and $ A_1, A_2 \neq 0$.  This requires $\zeta_r =|A_2|^2-|A_1|^2$.
\item[{\bf vii}] $A_1 = B_2=0$ and $A_2, B_1 \neq 0$.  This requires $\zeta_r =|B_1|^2+|A_2|^2>0$.
\item[{\bf viii}] $A_2 = B_1=0$ and $A_1, B_2 \neq 0$.  This requires $\zeta_r =-|A_1|^2-|B_2|^2<0$.
\end{itemize}

\begin{figure}
\centerline{\includegraphics[scale=0.8]{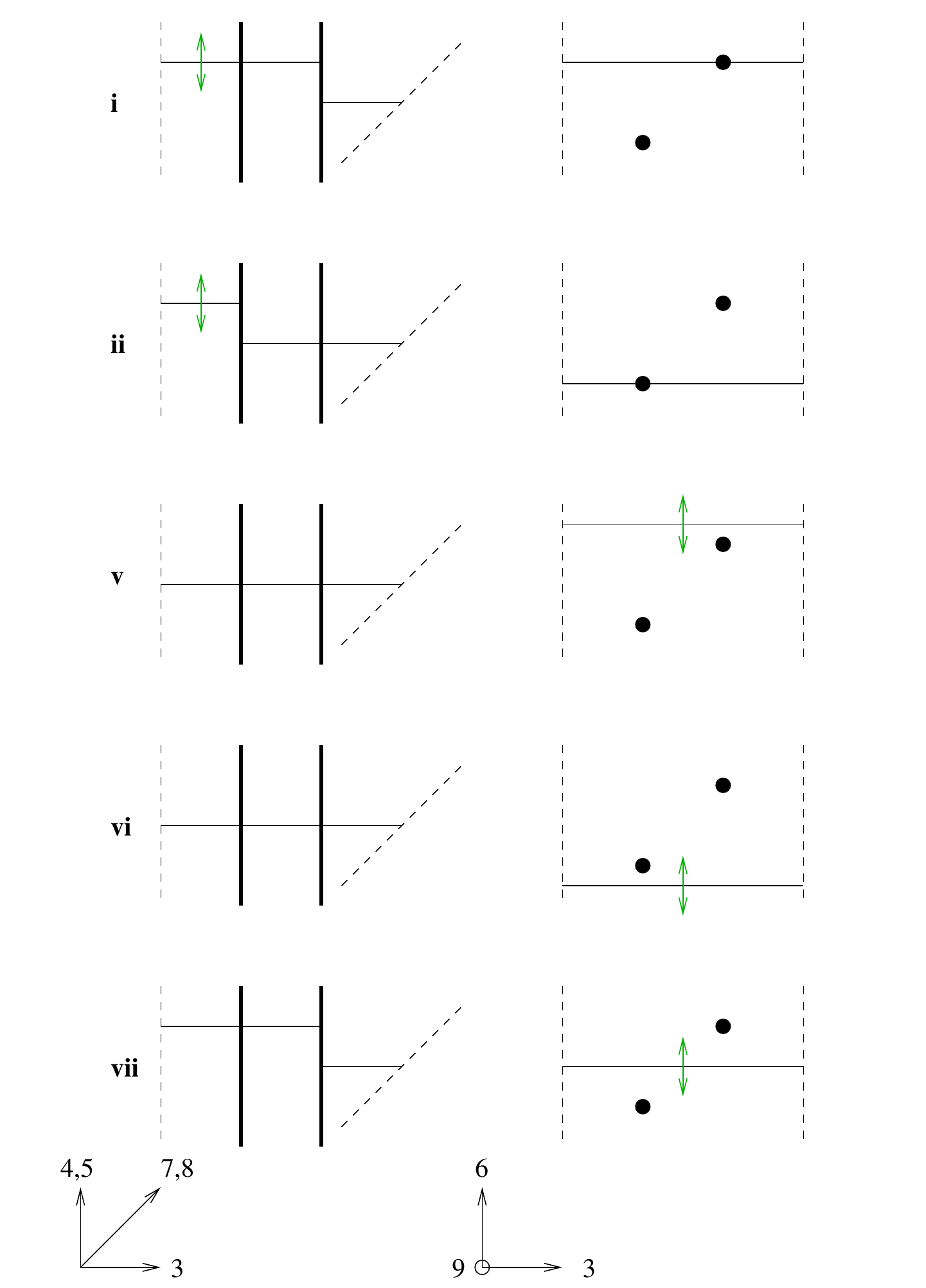}}
\caption{Brane configurations corresponding to stable branches {\bf
    i}, {\bf ii}, {\bf v}, {\bf vi}, and {\bf vii} for the ${\cal
    N}=2$ $N_c=1$ $N_f=2$ theory in the S-dual magnetic description with
  the real FI parameter (of the magnetic theory) $\zeta_r>0$. The
  black dots indicate the NS5 branes. The green arrows indicate the
  unconstrained directions along the moduli space in the brane
  picture.  \label{fign2nc1nf2brane}}
\end{figure}

Some qualitative features of this analysis can be understood from the
brane diagram.  The brane configurations corresponding to the stable
branches for the case with positive $\zeta_r$ are illustrated in
Figure \ref{fign2nc1nf2brane}. The basic picture is that the various
branches intersect when the D3 intersects an NS5-brane. Locally, this
intersection is identical to what was seen in the case of $N_c=1$ and
$N_f=1$. We also see that branch {\bf vii}, which is bounded on both
sides by the NS5-branes, does not have an asymptotic region.

\begin{figure}
\centerline{\includegraphics[scale=0.8]{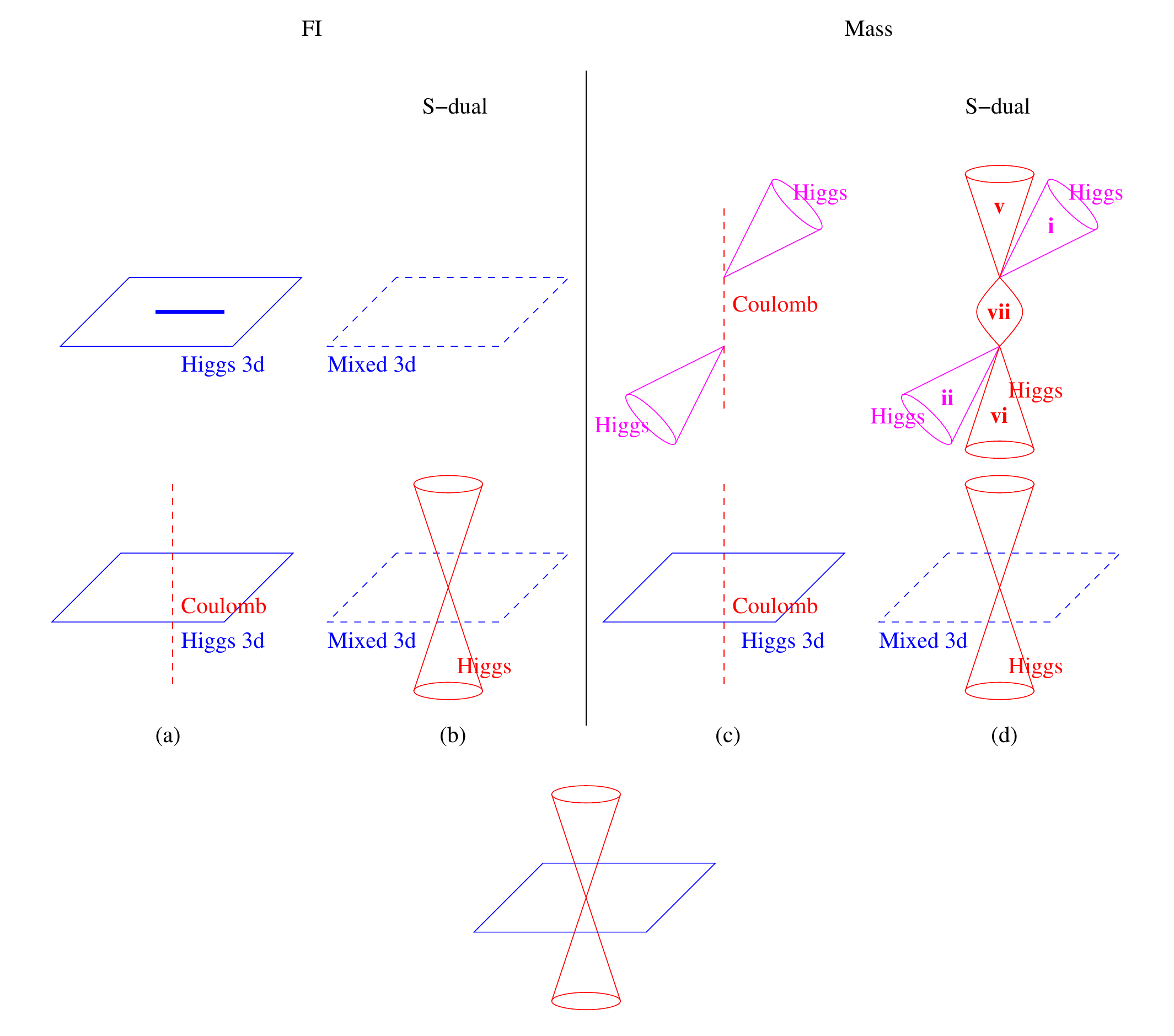}}
\caption{(a) The branch structure of ${\cal N}=2$ $N_c=1$ $N_f=2$
  theory deformed by an FI parameter, (b) its S-dual, (c) ${\cal N}=2$
  $N_c=1$ $N_f=2$ theory deformed by quark mass, and (d) its
  S-dual. The colors reflect the mapping of branches under
  S-duality. The bottom illustrates the fully-quantum-corrected moduli
  space in the undeformed limit. The Higgs branch is the conifold.
\label{figg}}
\end{figure}

We summarize our findings for the moduli space of this system in
figure \ref{figg}.  The analysis of S-dual we carried out first is
summarized in column (d). The dependence on FI parameter $\zeta_r$ in
this dual frame is to be mapped to the dependence on real mass of the
$N_c=1$ $N_f=2$ theory of interest. For non-vanishing $\zeta_r$, we
find Higgs branches {\bf i}, {\bf ii}, {\bf v}, {\bf vi}, and {\bf
  vii}, the complex structure of all of which are protected against
quantum corrections.  The Coulomb branch of the mass deformed $N_c=1$
$N_f=2$ theory, illustrated in column (c), receives quantum
corrections which can be inferred from column (d) using S-duality. In
addition, we studied the $N_c=1$ $N_f=2$ system directly in the
presence of FI term and inferred the structure outlined in column (a).

One can also identify the operators which serves as a gauge invariant
order parameters in each of these phases. On branches {\bf i} and {\bf
  ii}, traces of powers of ${\cal X}$ at any point along regions I and
II will serve that purpose. On the branches {\bf v} and {\bf vi}, one
can consider the following set of chiral operators.
\beq 
{\cal O}_{\bf v} &=&  e^{-\int_{III} {\cal A}} \, B_2 \,  e^{-\int_{II} {\cal A}} \, B_1 \, e^{-\int_I {\cal A} }   \ , \\
{\cal O}_{\bf vi} &=&  e^{\int_I {\cal A}} \, A_1 \,  e^{\int_{II} {\cal A} } \, A_2 \, e^{\int_{III} {\cal A} } \ . \eeq
These operators are invariant under complexified gauge
transformations. The Wilson line factors are necessary in order to
make these operators gauge invariant, but it can be eliminated in the
limit that the size of the regions I, II, III goes to zero. In that
limit, the order parameters are simply
\beq 
{\cal O}_{\bf v} &=&  B_2 B_1\ , \\
{\cal O}_{\bf vi} &=&  A_1 A_2 \ . 
\eeq

The order parameter for the compact branch  {\bf vii} requires some
additional care. Even in the zero interval limit, one can not
construct an invariant combination of $A_2$ and $B_1$ without
involving complex conjugation, e.g.
\be {\cal O}_{\bf vii} =  A_2^\dag B_1 \ . \ee
This operator is not invariant under the complexified gauge
symmetry. 
This is consistent with the fact that the existence and the
stability of this branch depended explicitly on a real datum, namely
the positivity of $\zeta_r$. Here we see one limitation of the complex
gauge formalism. Some observables require partial gauge fixing to the
real formalism where one can construct additional sets of operators
invariant under the smaller gauge group.

This is also a useful point to comment on the status of the
moduli-space metric.  From the explicit solution to Nahm equations at
our disposal, it is straightforward, although tedious, to compute the
Manton metric \cite{Manton:1981mp}. With only ${\cal N}=2$
supersymmetry, one expects generic quantum corrections. In cases such
as the $N_c=1$ and $N_f=2$ examples under consideration, it was
pointed out in \cite{Intriligator:2013lca} that the K{\" a}hler form
is also protected from quantum corrections. Does this mean that one
can extract the quantum exact metric by computing the Manton metric? A
closer look into the argument going into the non-renormalization of
the K{\" a}hler form in \cite{Intriligator:2013lca} relied on
conformal invariance.  In the analysis of the of the 3+1 defect
system, there are features such as the size of the intervals which
introduces a scale and the Manton metric will certainly depend on
these parameters. Perhaps, for certain simple configurations, one can
extract the quantum exact moduli space metric from the zero interval
size limit of the Manton metric. These, however, correspond to the
rich subject of massless monopoles, also known as clouds, reviewed in
\cite{Weinberg:2006rq}. It would be interesting to work out specific
criteria for when such a prescription to compute the metric succeeds
or fails.

By combining the expectations based on $\zeta_r\rightarrow 0$ limit of
columns (a) and (d), we infer the structure of the undeformed moduli
space as intersections of a three dimensional Higgs branch and two
Coulomb branches illustrated at the bottom of Figure \ref{figg}.  We
expect the Coulomb branch, illustrated by red cones, to consist of
complex plane $C$ for branches {\bf v} and {\bf vi}, and have the
structure of $\mathbb{CP}_1$ for {\bf vii}, consistent with the
expectation of taking the subspace of resolved $\mathbb{C}^2/\bZ_2$.

The Higgs branches {\bf i} and {\bf ii}, illustrated in column (d), become
unstable and melt into the Coulomb branch (of theory B) if $\zeta_r =
0$. It is interesting to note, on the other hand, that these Higgs
branches survive when both the FI term and the real mass are turned on
as we alluded to earlier in Section \ref{sec:mirror}.

The FI-deformed moduli space in column (d) of Figure \ref{figg} is
identical in structure to moduli space illustrated in Figure 2 of
\cite{Aharony:1997bx}. The main difference between the our result and
the result of \cite{Aharony:1997bx} is that we arrive at our
conclusion via a strictly {\it classical} analysis of the S-dual
description of Figure \ref{figc}.b.

\begin{figure}
\centerline{\includegraphics[scale=0.8]{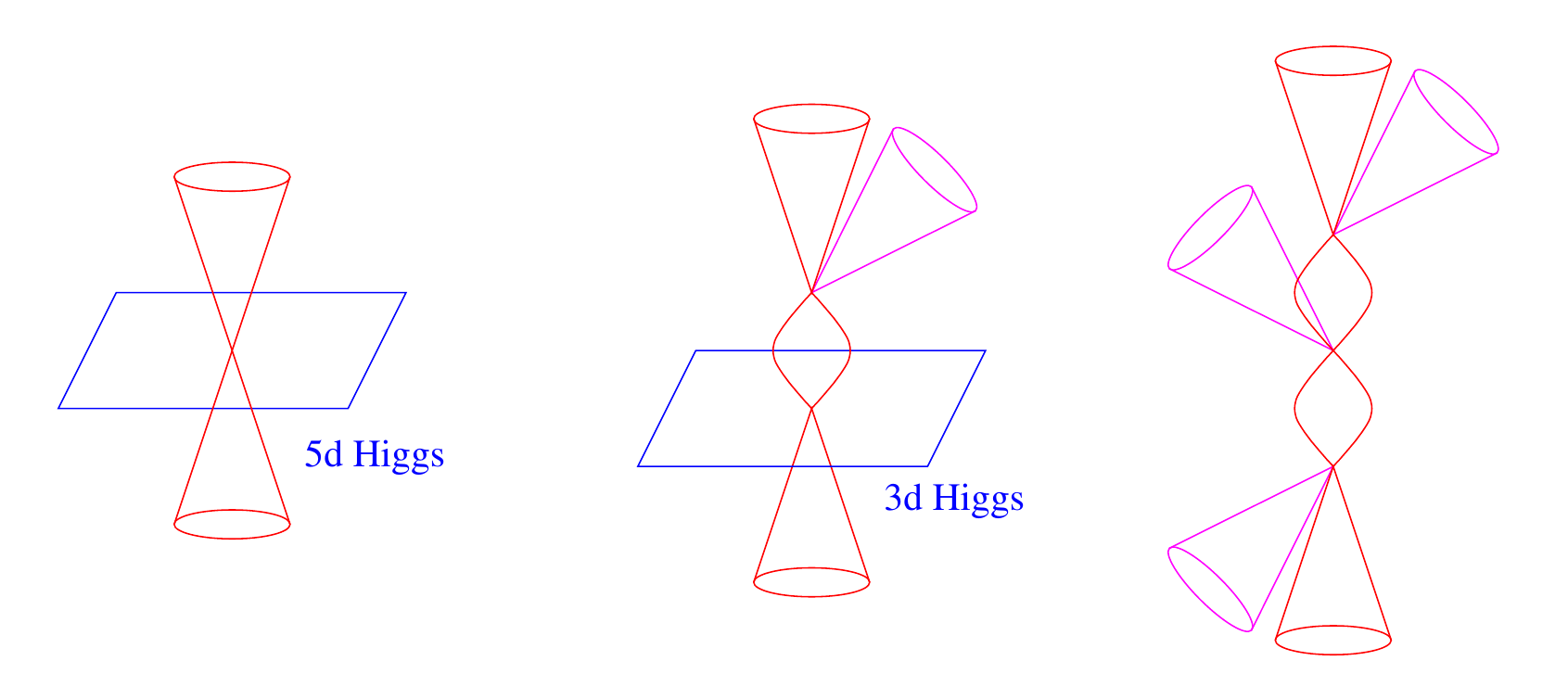}}
\caption{The moduli space of ${\cal N}=2$ $N_c=1$ $N_f=3$ theory with
  three equal real quark masses, two equal real quark masses, and all
  distinct real quark masses.
\label{figh}}
\end{figure}

It is straightforward to extend the analysis for $N_f=1,2$ for any
$N_f$.  When $N_f=3$, for example, with no mass deformations, one
expects a 1-dimensional Coulomb branch and a 5-dimensional Higgs
branch.  The branch structure for $N_f=3$ as the real masses of the 3
flavors of quarks are varied is illustrated in Figure \ref{figh}.  As
shown in the figure, for each flavor that we add (with a generic real
mass), the moduli space develops an additional compact branch and an
additional noncompact branch. Some further remarks on the Coulomb
branch for general $N_f$ appear in Appendix \ref{appA}.

\subsubsection{Complex Mass Deformation}

It is simple to deform the previous analysis by adding a complex mass
(of the electric theory.)  We do the analysis in the magnetic theory,
where it appears as a complex FI term.  In the brane picture this
means we displace one of the NS5$^\ensuremath{\prime}$-branes in the
$X_{7,8}$ directions.  We can implement this by changing equations
(\ref{Nc1Nf2YII}) and (\ref{Nc1Nf2YIII}) to
\beq
&&\mathcal{Y}_{II} = B_1 A_1=A_2 B_2 - \zeta_c\ , \\
&&\mathcal{Y}_{III}  =  B_2 A_2 -\zeta_c\ .
\eeq
We are forced to set $b=c=d=0$.  So the equations are
\beq
A_2 B_2 &=& \zeta_c\ ,\\
A_1 B_1 &=& 0\ ,\\
aA_1 = a B_1 &=& 0\ .
\eeq
We can use the complex gauge transformation to remove one degree of
freedom in $A_2, B_2$.  We are left with three branches of moduli
space, corresponding to $a\neq0, A_1=B_1=0$, or $a=0, A_1=0, B_1 \neq
0$ or $a=0, A_1 \neq 0, B_1 = 0$. It is natural to identify these
three branches with the branches we obtained from directly analyzing
the case $N_c=1$, $N_f=1$.

\subsection{$\mathcal{N}=2$ $U(1)$ Theory with Hidden Parameters}
\label{sec:hidden}

Some interesting phenomena can arise for brane configurations with
arbitrary arrangements of 1/4 BPS 5-branes which are hard to
understand from the point of view of 3$d$ field theory.  In
particular, as the branes are reordered in the $y$ direction, the low
energy theory undergoes phase transitions which can change the
geometry of the moduli space (in particular the dimensions of the
branches can change.)

The $y$ positions of the 5-branes should correspond to deformations by
irrelevant operators from the point of view of the three-dimensional
theory.  However, because they change the vacuum structure of the
theory, they are dangerously irrelevant.  In \cite{Hanany:1996ie}
these 5-brane positions were called ``hidden parameters'' of the 3$d$
theory, although of course they are not hidden in the 4$d$ defect
theory.

We will consider two Abelian examples which are related by varying the
hidden parameters.  Actually, these systems were considered previously
in \cite{Aharony:1997ju}.  There the moduli spaces were constructed by
making educated guesses for an effective superpotential.  We will
reproduce their results for the moduli spaces, but the point we wish
to emphasize is that the moduli space is {\it derived} rather than
guessed.  Because our method is systematic, it can be generalized to
more complicated systems where it may be hard to guess an appropriate
effective superpotential.

\subsubsection{D5---1D3---NS5---1D3---NS5$^\ensuremath{\prime}$---1D3---D5$^\ensuremath{\prime}$}
\label{sec451}

\begin{figure}
\centerline{\includegraphics[scale=0.8]{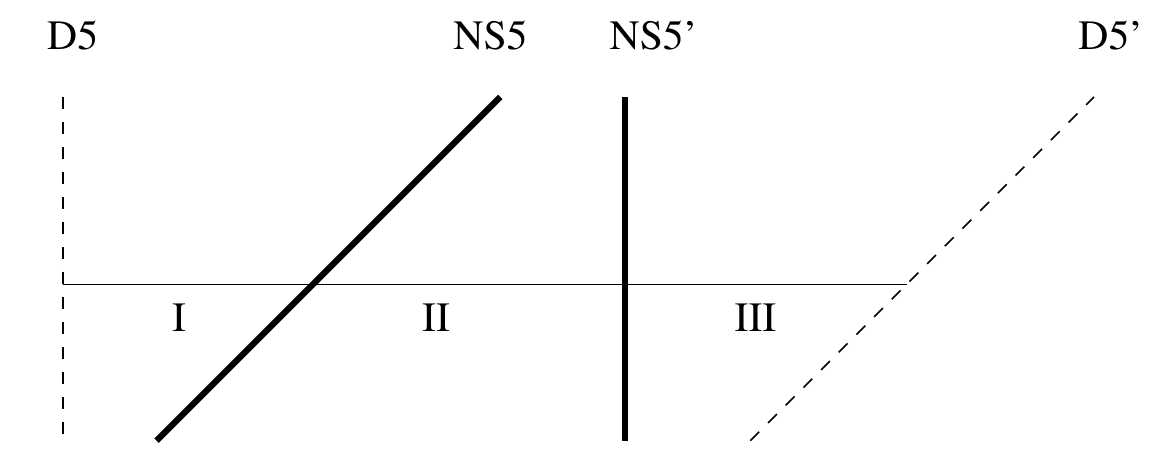}}
\caption{The brane configuration D5---1D3(I)---NS5---1D3(II)---NS5$^\ensuremath{\prime}$---1D3(III)---D5$^\ensuremath{\prime}$.
\label{d5ns5ns5pd5p}}
\end{figure}

Let us consider the brane configuration shown in Figure \ref{d5ns5ns5pd5p}.  

The scalars may be taken to be
\beq
\mathcal{X}_I &=& a\ ,\\
\mathcal{Y}_I &=& 0\ ,\\
\mathcal{X}_{II} &=& b\ ,\\
\mathcal{Y}_{II} &=& c\ ,\\
\mathcal{X}_{III} &=& 0\ ,\\
\mathcal{Y}_{III} & = & d\ .
\eeq

The interfaces have matter fields $A_1, B_1$ at the NS5 and $A_2, B_2$
at the NS5$^\ensuremath{\prime}$.  They obey the equations
\beq
&&\; \mathcal{X}_I= A_1 B_1\ , \quad \mathcal{X}_{II}= B_1 A_1\qquad \Longrightarrow \qquad b=a\ ,\\
&& \mathcal{Y}_{II} = A_2 B_2\ ,  \quad \mathcal{Y}_{III} = B_2 A_2 \qquad \Longrightarrow  \qquad d=c\ ,\\
&& (\mathcal{Y}_I - \mathcal{Y}_{II})A_1 = 0\ ,\\
&& (\mathcal{Y}_I - \mathcal{Y}_{II})B_1 = 0\ ,\\
&& (\mathcal{X}_{II} - \mathcal{X}_{III})A_2 = 0\ ,\\
&& (\mathcal{X}_{II} - \mathcal{X}_{III})B_2 = 0\ .
\eeq
The last four equations can be simplified to 
\beq
&& A_2 a = B_2 a = 0\ ,\\
&& A_1 c = B_1 c = 0\ .
\eeq
We also have
\beq
iX_{6,I} &=& A_1 A_1^{\dag}- B_1^{\dag}B_1\ ,\\
iX_{6,II} &=& A_1^{\dag} A_1 - B_1 B_1^{\dag}\ ,\\
iX_{6,II} &=& A_2 A_2^{\dag}- B_2^{\dag}B_2-\zeta_r\ ,\\
iX_{6,III} &=& A_2^{\dag} A_2 - B_2 B_2^{\dag}-\zeta_r\ .
\eeq

We can distinguish several branches.  First, there are two branches
with the complex structure of $\mathbb{C}$:
\begin{itemize}
\item[{\bf i}] $a=A_1 B_1\neq 0$ which forces $A_2=B_2=0$ and $c=0$. The relative magnitudes of $A_1, B_1$ are fixed by $|A_1|^2-|B_1|^2 = -\zeta_r$.
\item[{\bf ii}] $ c = A_2 B_2\neq 0$ which forces $A_1=B_1 =0$ and $a = 0$.   The relative magnitudes of $A_1, B_1$ are fixed by $|A_2|^2-|B_2|^2 = \zeta_r$.
\end{itemize}
We also have four branches with $a=c=0$:
\begin{itemize}
\item[{\bf iii}] $A_1=0, B_2=0$ and $A_2 \neq 0, B_1\neq 0$.  This requires $\zeta_r = |B_1|^2+|A_2|^2>0$.
\item[{\bf iv}] $A_1 = 0, A_2 = 0$ and $B_1\neq 0$, $B_2 \neq 0$. This requires $\zeta_r =|B_1|^2-|B_2|^2$.
\item[{\bf v}] $B_1 = 0, B_2 = 0$ and $A_1 \neq 0, A_2 \neq 0$.  This requires $\zeta_r =|A_2|^2-|A_1|^2$.
\item[{\bf vi}] $B_1 = 0, A_2 = 0$ and $A_1 \neq 0, B_2 \neq 0$.  This requires $\zeta_r =-|A_1|^2-|B_2|^2<0$.
\end{itemize}
So there are a total of six 1-dimensional branches on which the gauge
symmetry is completely broken.  Four of the branches exist for any
value of $\zeta_r$, and they consist of semistable points.  The third
and sixth of the bulleted branches are unstable.

If we allow some gauge symmetry to be unbroken, we can set all the
$A_i=0$ and $B_i=0$.  This also sets $a=b=c=d=0$.  Ignoring quantum
corrections, we have one classical dimension left in the moduli space
from $\mathcal{Z}_{II}$. This branch only exists when $\zeta_r = 0$.
Presumably for nonzero $\zeta_r$ it is transmuted into the third and
sixth of the above branches, which only exist when $\zeta_r>0$ or
$\zeta_r<0$.

We see that for any value of $\zeta_r$, we have 5 one-dimensional
branches of moduli space.

We can also analyze this system in the S-dual, or equivalently after
having done Hanany-Witten transitions.  The brane configuration is
NS5---1D3(I)---D5---1D3(II)---D5$^\ensuremath{\prime}$---1D3(III)---NS5$^\ensuremath{\prime}$.

\begin{figure}
\centerline{\includegraphics[scale=0.8]{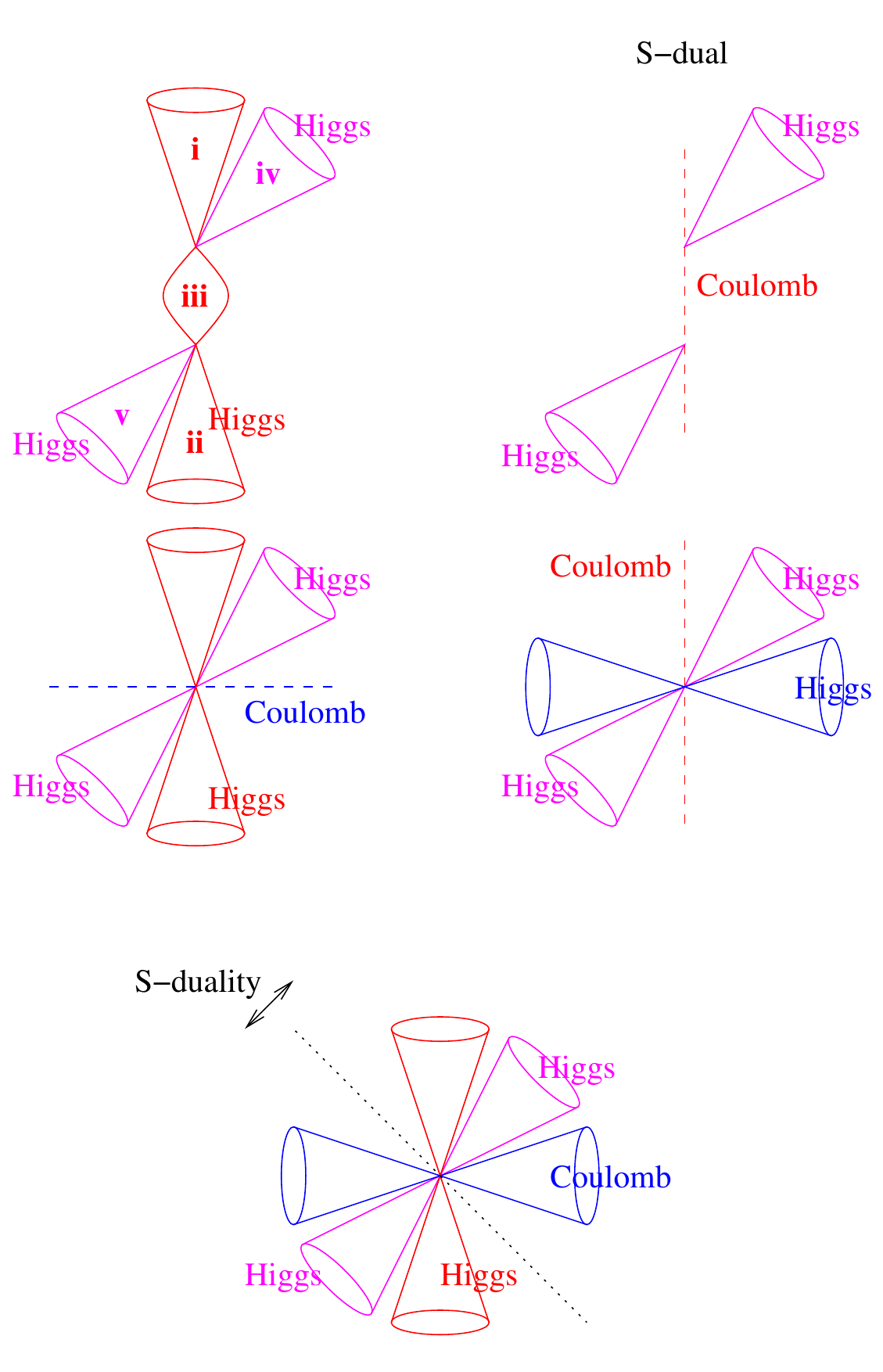}}
\caption{The branch structure of configuration 
D5---1D3(I)---NS5---1D3(II)---NS5$^\ensuremath{\prime}$---1D3(III)---D5$^\ensuremath{\prime}$ with FI deformation, and its S-dual also with FI deformation. In this example, two of the Higgs branches are stable and survive the $\zeta_r \rightarrow 0$ limit.
\label{figi}}
\end{figure}

The branch structure is illustrated in Figure \ref{figi}. Note that
even in the $\zeta_r \rightarrow 0$ limit, the Higgs branches {\bf i}
and {\bf ii} do not disappear in this case.

\subsubsection{D5---1D3---NS5$^\ensuremath{\prime}$---1D3---NS5---1D3---D5$^\ensuremath{\prime}$}
\label{sec452}

\begin{figure}
\centerline{\includegraphics[scale=0.8]{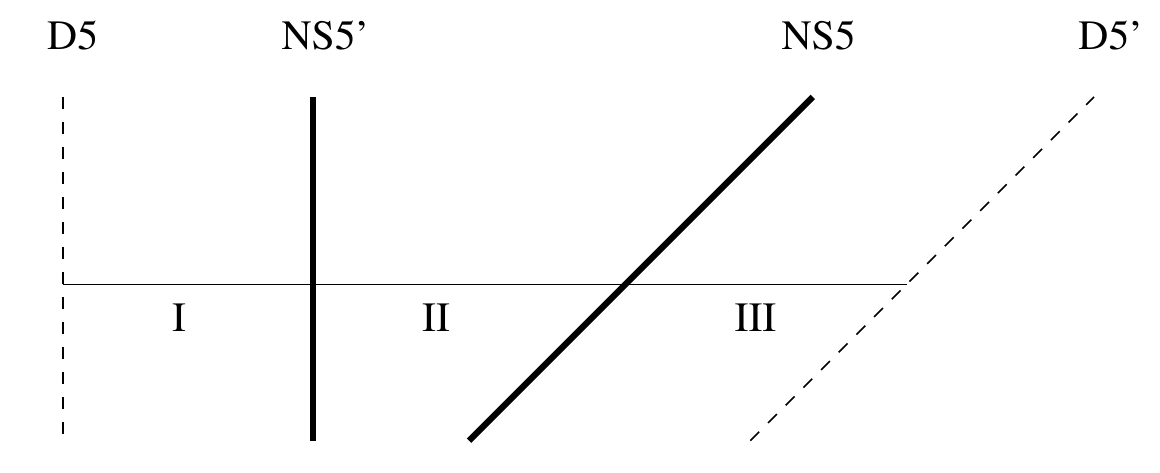}}
\caption{The brane configuration D5---1D3(I)---NS5$^\ensuremath{\prime}$---1D3(II)---NS5---1D3(III)---D5$^\ensuremath{\prime}$.   It differs from the configuration in Figure \ref{d5ns5ns5pd5p} by the ordering of the NS5 and NS5$^\ensuremath{\prime}$.
\label{d5ns5pns5d5p}}
\end{figure}

Now let us consider the system where we have interchanged the position
in $y$ of the two NS branes, corresponding to a change of what were
called ``hidden parameters'' by \cite{Hanany:1996ie}.  The brane
configuration is shown in Figure \ref{d5ns5pns5d5p}.  As the 5-brane positions are varied,
the moduli space changes qualitatively, signaling that there is a
phase transition when the NS-brane ordering is changed.

We can choose the starting ansatz
\beq
\mathcal{X}_I &=& a\ ,\\
\mathcal{Y}_I &=& 0\ ,\\
\mathcal{X}_{II} &=& b\ ,\\
\mathcal{Y}_{II} &=& c\ ,\\
\mathcal{X}_{III} &=& 0\ ,\\
\mathcal{Y}_{III} & = & d\ .
\eeq

We introduce $A_1,B_1$ at the NS5$^\ensuremath{\prime}$ and $A_2, B_2$
at the NS5.  The interface conditions are
\beq
&&\;\; \mathcal{Y}_{I}= A_1 B_1 = \mathcal{Y}_{II} \qquad \Longrightarrow \qquad  A_1 B_1 =c=0\ ,\\
&& \mathcal{X}_{II} = A_2 B_2 = \mathcal{X}_{III} \qquad \Longrightarrow \qquad A_2 B_2 =  b=0\ ,
\eeq
and also
\beq
&& (\mathcal{X}_I-\mathcal{X}_{II} ) A_1 =  (\mathcal{X}_I-\mathcal{X}_{II} ) B_1 = 0\ ,\\
&&(\mathcal{Y}_{II} -\mathcal{Y}_{III} ) A_2 = (\mathcal{Y}_{II} -\mathcal{Y}_{III} ) B_2 = 0\ ,
\eeq
which imply
\beq
&& A_1 a = B_1 a = 0\ ,\\
&& A_2 d = B_2 d = 0\ .
\eeq
We also have
\beq
iX_{6,I} &=& A_1 A_1^{\dag}- B_1^{\dag}B_1\ ,\\
iX_{6,II} &=& A_1^{\dag} A_1 - B_1 B_1^{\dag}\ ,\\
iX_{6,II} &=& A_2 A_2^{\dag}- B_2^{\dag}B_2-\zeta_r\ ,\\
iX_{6,III} &=& A_2^{\dag} A_2 - B_2 B_2^{\dag}-\zeta_r\ .
\eeq

Now we classify the branches of moduli space.  First, suppose that
either $a \neq 0$ or $d \neq 0$:
\begin{itemize}
\item[{\bf i}] $a \neq 0$  forces $A_1 = B_1 = 0$.  We also have $A_2 B_2 =  b=0$ so either $A_2$ or  $B_2$ is zero.  Which one is nonvanishing is controlled by the sign of the real FI term, $|A_2|^2-|B_2|^2 = \zeta_r$.
\item[{\bf ii}] $d\neq0$ gives a similar one-dimensional branch with $A_1, B_1$ nonzero and $|B_1|^2-|A_1|^2 = \zeta_r$.
\end{itemize}
For either sign of $\zeta_r$, this gives rise to two branches of
moduli space.  These branches are unstable, and do not exist for
$\zeta_r = 0$.

In addition to these, there are some 1$d$ branches with $a=d=0$:
\begin{itemize}
\item[{\bf viii}] $A_1=0, B_2=0$ and $A_2 \neq 0, B_1\neq 0$.   This requires $\zeta_r =-|B_1|^2-|A_2|^2<0$.  This branch is unstable; if it is made physical by activating $\zeta_r$, it has the complex structure of $\mathbb{P}^1$.
\item[{\bf vii}] $A_2 = 0, B_1 = 0$ and $A_1 \neq 0, B_2 \neq 0$.  This requires $\zeta_r =|A_1|^2+|B_2|^2>0$, and this branch is also unstable.
\item[{\bf iv}] $A_1 = A_2 = 0$ and $B_1 \neq 0, B_2 \neq 0$.  On this branch $|B_1|^2 - |B_2|^2 = \zeta_r$.
\item[{\bf v}] $ B_1 = B_2 = 0$ and $A_1 \neq 0, A_2\neq 0$.  On this branch $|A_1|^2 - |A_2|^2 = -\zeta_r$.
\end{itemize}
The last two of these branches are stable.

If we have $\zeta_r = 0$, we have solutions where all the $A_i,B_j$
vanish and there is unbroken gauge symmetry.  This moduli space is
classically 3-dimensional, parameterized by $a$, $d$, and
$\mathcal{Z}$. In fact, this space looks essentially like what was
illustrated in Figure \ref{figg}.

This example shows something important -- the moduli space is
qualitatively different than when the NS5 and
NS5$^\ensuremath{\prime}$-branes are interchanged.  This appears to be
a phase transition from varying what Hanany and Witten (and Aharony
and Hanany) called the ``hidden parameters.''

We may also deform this system by a real mass by displacing the
D5$^\ensuremath{\prime}$ brane in the $X_9$ direction.  To make this
explicit, we reintroduce the coordinate $\mathcal{Z}$ (recall the
discussion in Section \ref{sec:BPSeq}) and set
\beq
\mathcal{Z}_{I} &=& 0\ ,\\
\mathcal{Z}_{II} &=& f\ ,\\
\mathcal{Z}_{III} &=& m' \ ,
\eeq
where $m'$ is fixed.  Then we have two choices, $f=0$ or $f=m'$, for
which $A_1=B_1=0$ or $A_2=B_2=0$, respectively.  We see that the real
mass lifts all but the first two bulleted branches (which are only
present if $\zeta_r \neq 0$) and the $3d$ branch which has an unbroken
$U(1)$ gauge symmetry.

\begin{figure}
\centerline{\includegraphics[scale=0.8]{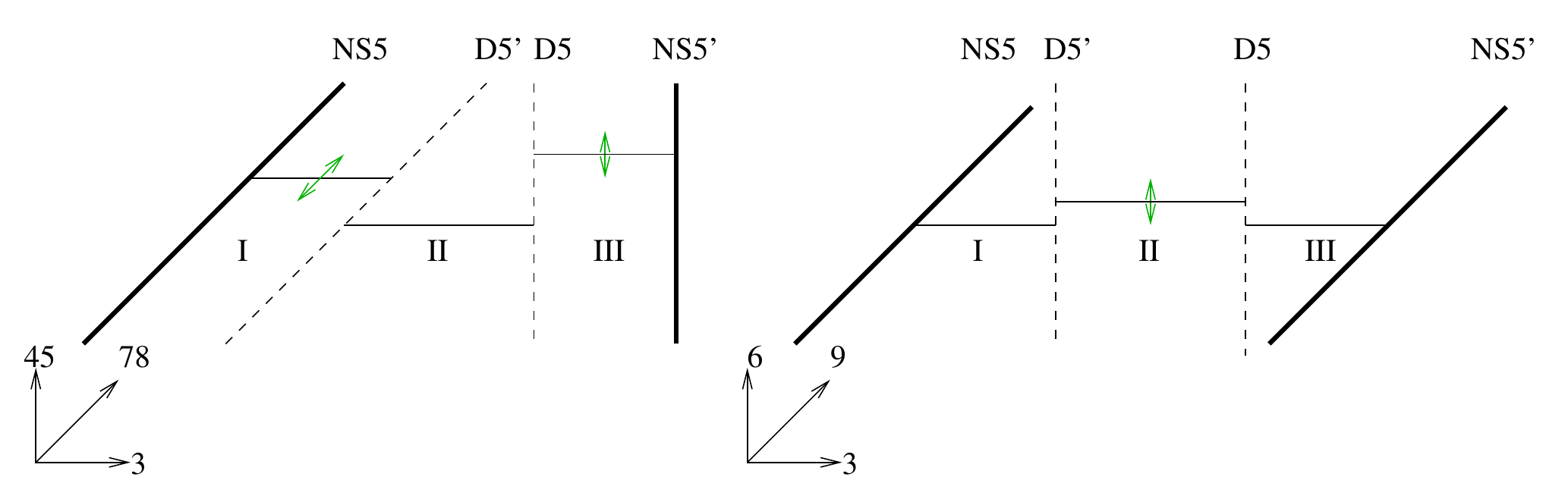}}
\caption{The brane representation of the 3 complex dimensional moduli space of the NS5---1D3(I)---D5$^\ensuremath{\prime}$---1D3(II)---D5---1D3(III)---NS5$^\ensuremath{\prime}$ system.
\label{Sd5ns5pns5d5p}}
\end{figure}

The 3-dimensional branch we found has unbroken gauge symmetry, so we
should analyze it in the S-dual configuration
NS5---1D3(I)---D5$^\ensuremath{\prime}$---1D3(II)---D5---1D3(III)---NS5$^\ensuremath{\prime}$
where the gauge symmetry will be broken. Starting with an ansatz
satisfying the boundary conditions at the endpoints,
\beq
\mathcal{X}_I &=& 0\ ,\\
\mathcal{Y}_I &=& a\ ,\\
\mathcal{X}_{II} &=& b\ ,\\
\mathcal{Y}_{II} &=& c\ ,\\
\mathcal{X}_{III} &=& d\ ,\\
\mathcal{Y}_{III} & = & 0\ ,
\eeq
and imposing the junction conditions at the D5$^\ensuremath{\prime}$
and D5, we obtain $b=c=0$ and
\beq
Q_1 \tilde{Q}_1 &=& a \ ,\\
Q_2 \tilde{Q}_2 &= & -d\ ,\\
|Q_1|^2 + |Q_2|^2 -|\tilde{ Q}_1|^2 - |\tilde{Q}_2|^2 &= & 0\ .
\eeq
Because $a$ and $d$ are free to vary, we see that there is a branch
with broken gauge symmetry where the $Q$ and $\tilde Q$ have only a
D-term constraint.  The resulting 3-dimensional space is the conifold.
There is also a branch with all the $Q, \tilde Q = 0$ with a $U(1)$
gauge symmetry.  From the brane perspective the three dimensional
moduli space can be visualized as illustrated in Figure
\ref{Sd5ns5pns5d5p}.

Note that in the original S-dual frame with unbroken gauge symmetry,
the gauge symmetry is $U(1)$.  Naively we might have thought that this
branch should have a direct product structure, with a 1-dimensional
Coulomb part and a 2-dimensional Higgs part.  However, the computation
using the S-dual shows that the Coulomb and Higgs parts merge to form
a 3-dimensional space which is not a direct product and instead has
the complex structure of the conifold.

\subsection{Vortices and Skyrmions in Defect Theories}

The three-dimensional $\mathcal{N}=2$ field theories can have an
interesting array of nonperturbative solutions, whose existence can
serve as order parameters for the vacuum phases of a given theory.  In
particular, for $U(1)$ theory with $N_f$ flavors, at various points on
the Higgs branch, it is possible to find ANO \cite{Abrikosov:1956sx,
  NO} vortex solutions which are BPS.  This was emphasized by
\cite{Aharony:1997bx}.  With a real FI parameter turned on, the
vortices have the asymptotic (in large $r$) behavior
\beq
Q \sim \sqrt{\zeta_r} \, e^{\pm i\phi}\ , \qquad A_{\phi} \sim \pm \frac{1}{r}\ .
\eeq
The full $r$-dependence of the vortex solutions can be found
numerically. In the case of ${\cal N}=2$, $N_c=1$, and $N_f=2$
illustrated in Figure \ref{figg}, these vortex solutions can appear in
all the Higgs branches, namely {\bf i}, {\bf ii}, {\bf v}, and {\bf
  vi} of column (d), as well as the S-duals of {\bf i} and {\bf ii} in
column (c). They could also appear in the $3d$ Higgs branch in column
(a).

It was argued in \cite{Aharony:1997bx} that the vortices are the
mirror duals of the (Coulomb branch) monopole operators.  In addition
to the vortex solutions, on compact Higgs branch such as {\bf vii} in
column (d) of Figure \ref{figg}, one expects to find skyrmions
\cite{Intriligator:2013lca}.  The skyrmion solutions are
nontopological solitons \cite{Vachaspati:1991dz,Hindmarsh:1992yy}.

We expect that similar nonperturbative solutions exist for our
3+1-dimensional defect theories -- after all, they must appear in the
limit where the defect theory becomes a 2+1-dimensional theory.
However, there is a difference between the classic analysis and our
defect systems, where the matter fields which acquire VEVs are only
supported at the interfaces, while the gauge fields propagate in the
3+1-dimensional bulk.  It would be interesting to construct these
solutions explicitly.  Similar systems have been considered in the
context of high-$T_c$ superconductivity, where the electromagnetic
fields are allowed to propagate in the bulk of some material with a
coupling to matter fields localized on planar defects.  In this
situation, vortex-like solutions called ``pancake" vortices have been
constructed; see \cite{Clem} for a brief review with references to the
earlier literature.

Such nonperturbative solutions, as understood from the point of view
of the theory in 3+1 dimensions, might serve as a probe of the
transition discussed in Section \ref{sec:hidden}, where some branches
of moduli space change dimension as the NS5-brane and
NS5$^\ensuremath{\prime}$-brane cross in $y$.  The transition is
desingularized by turning on FI terms (displacing the NS branes in
$X_6$.)  In this picture one expects to find vortices on the Higgs
branch and skyrmions in the S-dual.  As the FI terms are taken to
zero, the vortices condense and a Coulomb branch opens up.  In the
example of Section \ref{sec451}, the Coulomb branch is 1-dimensional,
while in Section \ref{sec452} it is a 3-dimensional mixed
Coulomb-Higgs branch which emerges.  This mixed branch should also
emerge from some kind of vortex condensation.  It would be interesting
if this difference were reflected in some kind of a change of the
soliton solutions as the 5-brane positions are varied in the $y$
coordinate.

\section{3$d$ $\mathcal{N}=2$ $U(2)$ Gauge Theories}
\label{sec:u2}

We now turn to an analysis of brane configurations which are related
to field theories with $U(2)$ gauge symmetry.  From the field theory
point of view, new phenomena appear (compared to the Abelian case)
because the the non-Abelian theories can have instantons
\cite{Affleck:1982as}.  The instantons generate corrections to the
superpotential which can change the complex structure of moduli space,
in some cases lifting the vacuum completely.

In this section, we will see how these instanton effects are encoded
in the 3+1-dimensional defect theory by using S-duality.  In the
magnetic formulation of a particular defect theory, the effect we need
to include is the existence of Nahm poles at D5-brane boundaries.
Because of the Nahm poles, the scalar fields of the bulk ${\cal N} =4$
theory do not necessarily commute; this intrinsically non-Abelian
behavior is what makes the naive geometric brane analysis invalid.

\subsection{$\mathcal{N}=2$, $N_c=N_f=2$}

The $\mathcal N =2$ $U(2)$ gauge theory with $N_f=2$ is a rather rich
system, and serves as a good illustration of the strengths and
limitations of our methods.  The cases with $N_f =1$ and $N_f =0$ can
be extracted from the two-flavor analysis by adding complex mass
deformations.

First, let us recall some expectations from the direct field theory
analysis for $N_c =2$, $N_f=2$.  When no mass deformations are
present, there should be a Coulomb branch with $U(1)\times U(1)$ gauge
symmetry, a Higgs branch with the gauge symmetry fully broken, and
some mixed branches with $U(1)$ gauge symmetry.  We expect from the
instanton-based computations that the Coulomb branch is 2 complex
dimensional.  We also have a Higgs branch which can be computed
classically in the electric theory; its dimension is $2 N_f N_c
-N_c^2=4$.\footnote{This Higgs branch is parameterized by four gauge
  invariant mesons, $Q_i \tilde{Q}_j$ with no constraints relating
  them.  Had the gauge group been $SU(2)$ rather than $U(2)$, we would
  have had to include two baryonic operators, $B=Q_1 Q_2$ and
  $\tilde{B}=\tilde{Q}_1 \tilde{Q}_2$, with the constraint $\det(M) -B
  \tilde{B} = 0$, leaving a 5-dimensional moduli space.}

There should also be mixed branches, which are 4 complex dimensional.
To see this, note that we can have expectation values for the 2
flavors of the form
\beq
Q_1 = \left( \begin{array}{c} q_1 \\ 0 \end{array} \right), \qquad Q_2 = \left( \begin{array}{c} q_2 \\ 0 \end{array} \right),\qquad \tilde{Q}_1^T = \left( \begin{array}{c} \tilde{q}_1 \\ 0 \end{array} \right), \qquad \tilde{Q}_2^T = \left( \begin{array}{c} \tilde{q}_2 \\ 0 \end{array} \right).
\eeq
These expectation values break the gauge symmetry from $U(2)$ to
$U(1)$. There are four complex degrees of freedom, one of which can be
eliminated by a gauge transformation.  In addition to the three Higgs
moduli, we expect a one-dimensional Coulomb branch because of the
unbroken $U(1)$ gauge symmetry, so the total dimension of the mixed
branch is 4.  For $U(1)$ theory the Coulomb branch actually consists
of two separate branches, and correspondingly we expect to see two
mixed branches in this case.

These features are summarized by an effective superpotential (obtained
in \cite{Aharony:1997bx} using considerations of holomorphy and
symmetry)
\beq
W = v_+ v_- \det(M)\ .
\label{Nc2Nf2W}
\eeq
In this expression $M$ is a $2\times2$ meson matrix while $v_{\pm}$
are monopole operators.  On the Coulomb branch, both $v_{\pm} \neq 0$
while $M_{ij} =0$, while on the Higgs branch $v_{\pm} =0$ and $M$ is
unconstrained.  On the mixed branches only one of $v_{\pm}$ is
nonzero, while $\det M =0$.

\subsubsection{Analysis in the Electric Theory}
\label{nc2nf2electric}

The brane construction which defines the $U(2)$ theory with two
flavors is
NS5---2D3(I)---NS5$^\ensuremath{\prime}$---2D3(II)---D5$^\ensuremath{\prime}$---1D3(III)---D5$^\ensuremath{\prime}$,
as shown in Figure \ref{figj}.a.  We label the three regions as
\beq
0 \;< &y& <\; y_1\ , \qquad {\rm (Region\; I)}\nonumber\\
y_1 \;< &y& <\; y_2\ , \qquad {\rm (Region\; II)}\label{y3regions}\\
y_2 \;< &y& <\; y_3\ . \qquad {\rm (Region\; III)}\nonumber
\eeq
In this system, there is enough complex gauge symmetry to set $\A=0$
in all three regions, which also makes the scalar fields piecewise
constant.  This choice does not completely fix the gauge; there is a
residual rigid $U(2)_{\mathbb C}$ gauge transformation in region I.

The analysis begins with the NS5 boundary condition at $y=0$ and then
proceeds from the left to the right.  With all the mass deformations
turned off, the NS5 boundary puts no constraint on $\Y_I$ but sets
\beq
\X_I = 0\ .
\label{nc2nf2electricXI}
\eeq

At the I--II interface ($y=y_1$), we have bifundamentals $A$ and $B$
which are $2\times 2$ matrices satisfying
\beq
\Y_I = AB, \qquad \Y_{II} = BA\ ,
\eeq
but these equations are trivial because the $\Y$ fields are otherwise
unconstrained.  There are also real moment map equations for $X_6$:
\beq
A A^{\dag} - B^{\dag} B &=& 0\ ,\\
A^{\dag} A- B B^{\dag}  &=& {\rm any}\ .
\eeq
For the moment, we can suppress these equations by using the
complexified gauge symmetry.

We also need to find $\X_{II}$ to satisfy the equations $\X_I A = A
\X_{II}$, etc.  But these are automatically satisfied because of
(\ref{nc2nf2electricXI}) combined with the fact that the
D5$^\ensuremath{\prime}$ ordinary Dirichlet boundary conditions on the
right set \beq \X_{II} = 0\ .  \eeq Moreover, the ordinary Dirichlet
boundary conditions put no constraint on $\Y_{II}$.

So the moduli space is given by $A$ and $B$ with no constraints other
than that we must mod out by the $U(2)$ gauge symmetry; given $A$ and
$B$, $\Y_{I, II}$ are determined and give no additional moduli.  This
means there are 8 degrees of freedom with 4 gauge symmetries, so the
Higgs branch is 4-dimensional.

In this analysis we have assumed that $A$ and $B$ are generic (so
that, for example, a $G_\mathbb{C}$ transformation can set $A=
\mathbb{I}$.)  However, if they have some vanishing eigenvalues, then
$\Y_I$ will also have vanishing eigenvalues.  When this happens there
can be unbroken gauge symmetry and we should restore the field
$\mathcal{Z}$, for which we have
\beq
\mathcal{Z}_I A = A\mathcal{Z}_{II}\ , \qquad B\mathcal{Z}_I = \mathcal{Z}_{II} B\ .
\eeq

The ordinary Coulomb branch arises when $A=B=0$ (which satisfies the
real equations.)  Then $\mathcal{Z}_I$ gives two moduli (because it is
arbitrary but we can diagonalize it by a gauge transformation) and we
have a 2$d$ Coulomb branch.

To see the mixed branch, we want to assume that $\mathcal{Z}_I$ has
one zero eigenvalue.  Then we can do a gauge transformation to
\beq
\mathcal{Z}_I = \left( \begin{array}{cc} 0 & 0 \\ 0 & v \end{array} \right)\ ,
\eeq
which breaks the gauge symmetry to $U(1) \times U(1)$.  To further
break the gauge symmetry to just $U(1)$, we take
\beq
A = \left( \begin{array}{cc} a_1 & a_2 \\ 0 & 0 \end{array} \right), \qquad B = \left( \begin{array}{cc} b_1 & 0 \\ b_2 & 0 \end{array}\right)\ .
\eeq
Modding out by a complex $U(1)$ leaves us with the three-dimensional
conifold.  So the classical mixed branch moduli space is
$\mathbb{C}^*$ times the conifold.

We can also consider deforming by a real FI parameter.  All the
equations are the same except that one of the real equations is
modified to
\beq
A A^{\dag} - B^{\dag} B &=&  \left( \begin{array}{cc} \zeta_r & 0 \\ 0 & \zeta_r \end{array} \right)\ .
\eeq
This completely lifts both the pure Coulomb branch and the mixed
branch, but does not affect the pure Higgs branch, for which there was
no constraint on $A$ and $B$ anyway.

\subsubsection{Analysis in the S-dual}
\label{nc2nf2magnetic}

The Coulomb branch as described in Section \ref{nc2nf2electric}
preserves some unbroken gauge symmetry, and is therefore subject to
instanton corrections.  We can study the quantum-corrected Coulomb
branch by considering the S-dual.

\begin{figure}
\centerline{\includegraphics[width=\hsize]{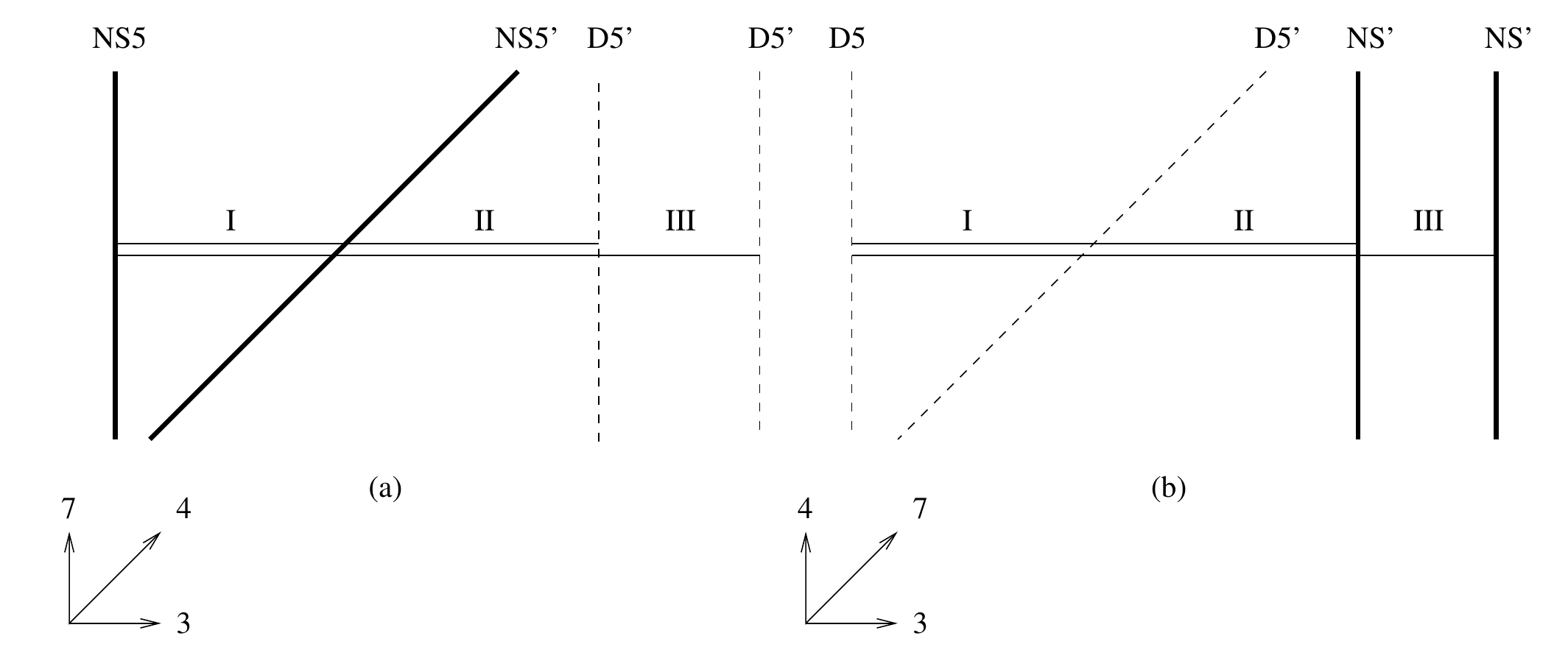}}
\caption{The defect system for 3$d$ $\mathcal{N}=2$ SYM with $N_c =2$, $N_f=2$.  It is shown in the defining electric representation (a) and the S-dual magnetic representation (b).
\label{figj}}
\end{figure}

The S-dual brane configuration is
D5---2D3(I)---D5$^\ensuremath{\prime}$---2D3(II)---NS5$^\ensuremath{\prime}$---1D3(III)---NS5$^\ensuremath{\prime}$,
divided into regions I, II, III as indicated.  The configuration is
shown pictorially in Figure \ref{figj}.b.

Because of the Nahm pole at the D5-brane boundary, we cannot set
$\A=0$.  Instead, we need to make a choice of gauge consistent with
the Nahm pole singularity.  In region I, we have
\beq
\mathcal{X}_I &=& \left( \begin{array}{cc} a & f_1(y) \\ bf_1(y)^{-1} & a \end{array} \right)\ , \label{pole2x2} \\
\mathcal{Y}_I &=& \left( \begin{array}{cc} 0 & 0 \\ 0 & 0 \end{array} \right)\ ,\\
\mathcal{A}_I &=& \left( \begin{array}{cc} f_2(y) & 0\\ 0 & - f_2(y) \end{array} \right) \ .
\eeq
We use $G_{\mathbb C}$ to pick the form where $f_1(y)$ and $f_2(y)$
are a particular choice of Euler top functions \cite{Weinberg:2006rq}.
We can choose them so that
\beq 
f_1(y) & = & {i {D}e^{i \phi}\over 2}  \left(\mbox{ns}_\kappa(D y) +  \mbox{ds}_\kappa(Dy)\right)\\
f_2(y) &=& -{i \over 2}  {f_1'(y) \over f_1(y)} =  {i D \over 2} \mbox{cs}_\kappa(D y)
\eeq
with real parameters $\kappa$, $D$, and $\phi$.  The parameter
$\kappa$ is fixed at the I--II interface by requiring $f_1'(y_1)=0$
which implies $f_2(y_1)=0$ which conveniently sets $\A_I(y_1)=0$. The
complex parameter $b$ is related to $D$ and $\phi$ via
\be b = - \kappa (De^{i \phi})^2 \ . \ee
With these choices, ${\cal X}$ has eigenvalues $a \pm \sqrt{b}$.
Then, without any loss of generality, we can further set $f_1(y_1)=1$
using complex gauge transformation to simplify the expressions which
follow.  We can then choose the gauge $\A=0$ in regions II and III, so
all the fields are piecewise constant in these regions.

The point of the preceding complicated analysis is that at $y=y_1$ we
are free to simply set
\beq
\mathcal{X}_{I}(y_1) = \mathcal{X}_{II}(y_1)= \left( \begin{array}{cc} a & 1 \\ b & a \end{array} \right) \ .
\eeq
To determine $\Y$ in region II, we introduce the matter fields at the
I--II interface, $\tilde{Q}$ and $Q$.  Because
\beq
\mathcal{X}_{II}(y_1) Q = \tilde{Q}\mathcal{X}_{II}(y_1) =0\ ,
\eeq
we have to either set ${\rm det}(\mathcal{X}_I ) = 0$ or else we have
$Q=\tilde{Q}=0$.  It is useful to distinguish three separate cases:
\begin{enumerate}
\item $Q=\tilde{Q}=0$, so there is no constraint on $\X$.  Then we
  have $\X$ with 2 distinct non-zero eigenvalues at generic points of
  moduli space.  We can do a rigid $G_{\mathbb{C}}$ transformation to
  simplify the analysis:
\beq
\X_{II} = \left( \begin{array}{cc} \mu_1 & 0 \\ 0 & \mu_2 \end{array} \right) \ ,
\eeq
unless the two eigenvalues are equal, in which case we can only put
the matrix in Jordan normal form:
\beq
\X_{II} = \left( \begin{array}{cc} a & 1 \\ 0 & a \end{array} \right).
\eeq
\item At least one of $\tilde{Q}$ or $Q$ is nonzero so ${\rm
  det}(\mathcal{X}_I ) = 0$.  The eigenvalues of $\X$ are $2a$ and
  $0$, and suppose $a \neq 0$.  Then we may do a rigid
  $G_{\mathbb{C}}$ transformation to
\beq
\X_{II} = \left( \begin{array}{cc} 2a & 0 \\ 0 & 0 \end{array} \right).
\eeq
\item At least one of $\tilde{Q}$ or $Q$ is nonzero and both
  eigenvalues of $\X$ vanish.  Then we put $\X$ in Jordan normal form:
\beq
\X_{II} = \left( \begin{array}{cc} 0 & 1 \\ 0 & 0 \end{array} \right).
\eeq
\end{enumerate}
The way to think about the $G_{\mathbb{C}}$ transformation is that we
do a rigid transformation in both regions I and II; this changes the
form of $\A$ in region I so it is not solution-generating; it simply
rewrites every solution in a more convenient form for doing the linear
algebra calculations.  The reason why we are free to do this rotation
is that the 4$d$ $U(2)$ gauge theory has an NS5-like boundary
condition on the right.  It should also be apparent that the only
nontrivial effect of the Nahm pole is to force $\X_{II}$ to take
Jordan normal form if two of the eigenvalues are equal.

Now let us analyze each case, beginning with case 1.  Because
$Q=\tilde{Q}=0$, we have $\Y_{II}=0$.  Using
\beq
\Y_{II} &=& AB\ , \\
\Y_{III} &=& BA \; =\;  0  \ ,
\eeq
we see that $\Y_{III}=0$ as well, and that either $A=0$ or $B=0$.

In the absence of real mass deformations, stability requires that in
fact both $A=0$ and $B=0$.  Then the gauge symmetry in region III
between the NS5$^\ensuremath{\prime}$-branes is unbroken and we do not
strictly trust our classical solution.  Nevertheless, if we compute
the associated moduli space we find 4 complex degrees of freedom: 2
corresponding to $a$ and $b$ (or equivalently, the two eigenvalues of
$\X_{II}$), and two corresponding to $\X_{III}$ and
$\mathcal{Z}_{III}$ which are unconstrained if the interface matter is
trivial.

Next, we proceed to case 2.  The relations $\mathcal{X}_{II} Q =
\tilde{Q}\mathcal{X}_{II} =0$ imply that
\beq
Q = \left( \begin{array}{c} 0 \\ v_+ \end{array} \right) \;, \qquad \tilde{Q} =\left( 0 \;\; v_-  \right)\ .
\eeq
This implies
\beq
\Y_{II} = Q \tilde{Q} = \left( \begin{array}{cc} 0 & 0\\ 0 &v_+ v_-  \end{array}\right) = AB\ ,
\eeq
but because
\beq
\mathcal{Y}_{III} = BA = 0\ ,
\eeq
we are forced to set 
\beq
v_+ v_- = 0\ .
\eeq
This implies that at least one of $A,B$ vanishes, and stability then
requires that we have both $A=0$ and $B=0$.

The remaining analysis is similar to that of case 1, except that we
can have either $v_+\neq 0$ or $v_- \neq 0$.  As in case 1, because
both bifundamentals vanish, $\X_{III}$ and $\mathcal{Z}_{III}$ are
free to vary.  So we have two 4-dimensional branches, parameterized by
$a$, the nonzero $v_{\pm}$, $\X_{III}$, and $\mathcal{Z}_{III}$, again
with an unbroken $U(1)$ gauge symmetry.

Last but not least, we turn to case 3, where both eigenvalues of
$\X_{II}$ vanish.  In this case we can have both $Q,
\tilde{Q}$ nonvanishing, and of the form
\beq
Q = \left( \begin{array}{c} v_+ \\ 0 \end{array} \right) \;, \qquad \tilde{Q} =\left( 0 \;\; v_-  \right)\ .
\eeq
Then we have 
\beq
\Y_{II} = Q \tilde{Q} = \left( \begin{array}{cc} 0 & v_+ v_- \\ 0 & 0  \end{array}\right) = AB \ ,
\eeq
which is compatible with $\Y_{III} = BA = 0$.  The solution requires
$A= p Q, B= p^{-1} \tilde{Q}$ and the parameter $p$ can be thought of
as being fixed by $G_{\mathbb C}$.  So we are left with a
two-dimensional branch, which is evidently semistable (and, as one can
check, it exists for any value of the real FI parameter $\zeta_r$.)
The expectation values of $A$ and $B$ break the $U(1)$ gauge symmetry
in region III, so for this branch the classical analysis should be
reliable.  It is natural to identify this as the dual of the Coulomb
branch in the original (electric) gauge theory.

Let us summarize our findings for $N_f=N_c=2$ with no mass or FI
deformations.  We have a four-dimensional branch in case 1 which we
identify as the Higgs branch of the electric theory.  This branch has
unbroken gauge symmetry in the magnetic formulation, but it can be
computed reliably in the electric formulation.  In case 2, there are
two mixed Coulomb-Higgs branches which are 4-dimensional; for the
mixed branches, there is unbroken gauge symmetry in both S-duality
frames, so our analysis is not fully trustworthy.  Finally, in case 3
there is a single 2-dimensional branch which we identify as the
Coulomb branch; on this branch the analysis in the magnetic frame is
reliable but not the electric analysis.

The counting of the branches and their dimensions match what one would
have inferred from the superpotential (\ref{Nc2Nf2W}).  Our analysis
might not seem terribly impressive, because the dimensions of the
branches of moduli space are simply the classical ones, but we will now
proceed to some more nontrivial examples by adding mass deformations.

\subsubsection{Complex Mass Deformation to $\mathcal{N}=2, N_c=2, N_f=1$\label{sec:deform}} 

We may add a complex mass (of the electric theory) by moving one of
the NS5$^\ensuremath{\prime}$ branes in the 78-directions.  This will,
for example, set
\beq
\mathcal{Y}_{III} = BA = c' \ , \label{yiii}
\eeq
with $c' \neq 0$.  

This criterion lifts the moduli spaces in case 1 and 3, both of which
require $\Y_{III}=0$.  Case 2 survives, with the constraints
\beq
a &\neq& 0\ , \label{aneq0}\\
v_+ v_- &=& c' \ .  \label{vpvmc}
\eeq
We see that we no longer have one of the $v_{\pm}$ vanishing, so the
solution space has characteristics of both cases 2 and 3.  The branch
has the same complex structure as $(\mathbb C^*)^2$.  Note that we
could have changed the normalization of $v_{\pm}$ by a function of $a$
without changing the complex structure.

Note also that because we can solve for $B, A$ as $B \sim \tilde{Q}, A
\sim Q$, with both $B$ and $A$ nonvanishing, this branch is
semistable.

The complex structure is independent of the value parameter $c'$ (as
long as it is nonzero), so we should be able to take the limit $c'
\rightarrow \infty$ which corresponds to completely removing one of
the NS5$^\ensuremath{\prime}$-branes.

To compare with the field theory, we should consider a superpotential
with a mass deformation
\beq
W = v_+ v_- \det(M) + \mu M_{22}\ ,
\eeq
and one can show that the complex structure from this superpotential
is also $(\mathbb C^*)^2$.  In the classical field theory analysis,
there is a 2-dimensional Coulomb branch and a 2-dimensional mixed
Higgs-Coulomb branch, but the quantum-corrected analysis has only a
single 2-dimensional branch.  This moduli space has characteristics of
the Coulomb branch (both monopole operators have nonzero expectation
values) and of the Higgs or mixed branches (the mesons have
expectation values), so one can think of it as a quantum mechanical
merging of the Coulomb and mixed branches.

\subsubsection{Real Mass Deformations}

As in the Abelian theories, it is also interesting to consider real
mass deformations.  In the S-dual formulation, the (electric) real
masses appear as real FI terms.  In this situation, the defect
analysis is especially useful -- real masses break holomorphy, so the
standard field theory arguments based on superpotentials are
inapplicable.  Moreover, in the S-dual formulation, the FI terms
generically break the gauge symmetry completely, which is precisely
the condition that we want to guarantee that our analysis is reliable.

When the real masses take generic values, a brane analysis described
in \cite{Elitzur:1997hc,Giveon:1998sr} claims that for $N_c = N_f =2$, there should
be six branches of moduli space, each of which is 2-dimensional.  The
counting is given by the following picture.  In the magnetic
description, one considers 2 D3-branes suspended between a D5 and a
D5$^\ensuremath{\prime}$ brane.  In between the D5 branes one places 2
NS5 (or NS5$^\ensuremath{\prime}$) branes, displaced in the
$X_6$-direction which is common to the D5 and
D5$^\ensuremath{\prime}$.  This creates five slots in which the D3
branes can be inserted.  One then adds the constraint that repulsive
forces between the D3-branes prevent 2 D3's from being placed in the
same slot or in adjacent slots.  Then, according to this recipe, there
are six distinct allowed configurations for the D3-branes. For general
values of $N_c$ and $N_f$, this counting argument gives
\be \left(\begin{array}{c} 2 N_f - N_c + 2 \\ N_c \end{array}\right)\label{modulicomponents}\ee
branches of moduli space.

We will now show that the Nahm computation reproduces this counting,
although the way our method accounts for the six branches is slightly
different from what is described in the literature.  We do the
analysis in the S-dual, and revisit the three cases in turn.  The FI
terms modify the real equations and relax the stability conditions.

In case 1, we obtained the constraint that either $A=0$ or $B=0$.
Then the stability condition forced us to set both $A=B=0$, but with
an FI term we are no longer required to do so.  Instead, we have the
real equation.
\beq
A A^{\dag} - B^{\dag} B = \zeta_r\ ,
\label{Nc2Nf2FI}
\eeq
and $A=0$ requires $\zeta_r < 0$ while $B=0$ requires $\zeta_r > 0$.
Suppose that $\zeta_r > 0$ so that $B=0$.  Then we must have
\beq
A = \left( \begin{array}{c} v \\ 0 \end{array} \right)  \qquad { \rm or}\qquad A = \left( \begin{array}{c} 0  \\ v \end{array} \right)
\ ,
\label{AwithFI}
\eeq
and the relation $\X_{II} A = A \X_{III}$ implies that for each choice
we have either $\X_{III}= \lambda_1$ or $\lambda_2$.  It might seem
that these are two distinct branches, but really they are not -- we
can move between the two choices continuously by varying $a$ and $b$
and then doing a discrete gauge transformation to interchange the
choices of $A$.  (When the two eigenvalues are equal there is only one
choice of $A$.)  So we find that in case 1 there is one 2$d$ branch of
moduli space if $\zeta_r \neq 0$, which we can think of as the remnant
of the $4d$ branch with unbroken gauge symmetry which exists for
$\zeta_r =0$.

In case 2, the analysis with $\zeta_r \neq 0$ is similar to case 1,
except that the two forms of $A$ (or $B$) in (\ref{AwithFI}) are
distinct.  Combining this with the choice of whether $v_+$ or $v_-$ is
nonzero, we see that there are four 2-dimensional branches if $\zeta_r
\neq 0$.

In case 3, the analysis with $\zeta_r =0$ is unchanged when $\zeta_r
\neq 0$, so this case contributes a single 2-dimensional branch.  So
to summarize, when $\zeta_r$ is nonzero, we see six 2-dimensional
branches, one from case 1, four from case 2, and one from case 3.

\begin{figure}
\centerline{\includegraphics[width=\hsize]{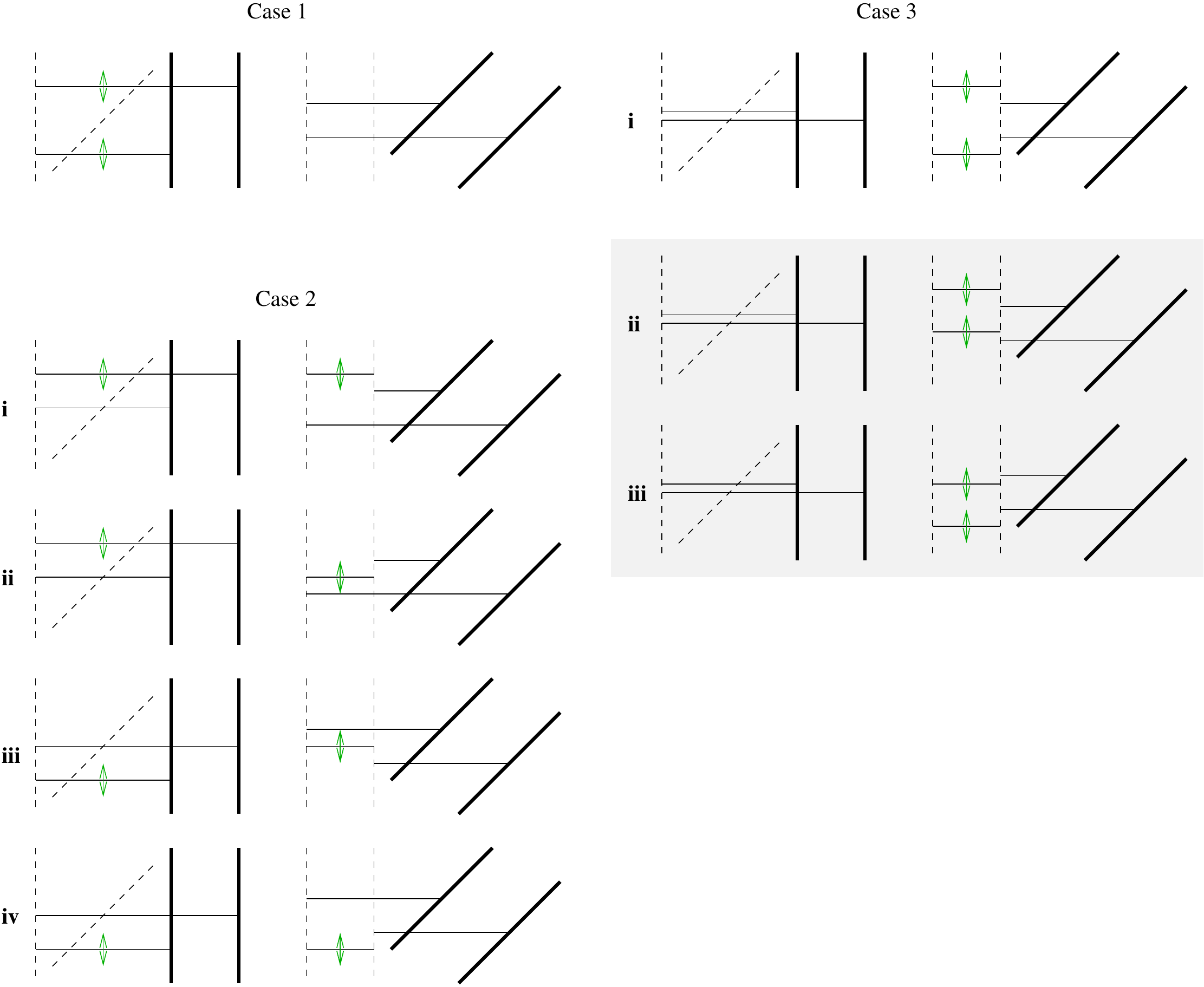}}
\caption{Brane configurations corresponding to one complex branch in
  case 1, four complex branches in case 2, and one complex branch in
  case 3. Each green arrow corresponds to one complex dimension of
  moduli. The branches 3.{\bf ii} and 3.{\bf iii} do not appear
  explicitly in our analysis. However, they can be understood as
  having merged with branches 2.{\bf iii} and 2.{\bf ii} respectively
  through the mechanism of Coulomb-Higgs merging, and therefore do not count 
  as separate branches. This is exactly what is needed
  since in the counting of (\ref{modulicomponents}), branches 2.{\bf
    ii} and 2.{\bf iii} were not included but 3.{\bf ii} and 3.{\bf iii}
  were.  That such a merging occurs is easy to see, by
  moving one of the NS5-branes in 3.{\bf ii} and 2.{\bf iii} to
  effectively reduce the system to $N_c=2$ and $N_f=1$. 
\label{fign2nc2nf2brane}}
\end{figure}

For the sake of comparison, it is useful to revisit the naive
geometric brane analysis. The brane configurations corresponding to
the branches of the moduli space in cases 1, 2, and 3 for $\zeta_r>0$
are shown in Figure \ref{fign2nc2nf2brane}. In the figure, the Higgs
branch moduli correspond to unbroken D3-branes, while the Coulomb-type
moduli (shown by green arrows) correspond to broken D3-branes.  In the
counting argument of \cite{Elitzur:1997hc,Giveon:1998sr}, the
configurations 2.{\bf ii} and 2.{\bf iii} have D3-branes in adjacent
slots, and so do not give rise to branches of moduli space.  On the
other hand, in the S-dual Nahm analysis, we retained the cases 2.{\bf
  ii} and 2.{\bf iii} but excluded the diagrams from case 3, namely
3.{\bf ii} and 3.{\bf iii}.

The difference between the two ways of accounting for the branches of
moduli space is partly an artifact of the way we organized the
algebraic computation, but more importantly, is a consequence of the
quantum merging of branches.  For example, in 2.{\bf ii} and 3.{\bf
  iii} of Figure \ref{fign2nc2nf2brane}, we could move one of the
NS5$^\ensuremath{\prime}$-branes to infinity, effectively reducing the
system to $N_c=2$ $N_f=1$.  The classical moduli spaces of 2.{\bf ii}
and 3.{\bf iii} (or 2.{\bf iii} and 3.{\bf ii}) then merge quantum
mechanically, following the analysis of Section \ref{sec:deform}.

The larger point is that the geometric brane drawing, with D3-branes
depicted as straight lines, is misleading because it does not
accurately capture the non-Abelian nature of the defect theory.  This
is not just a matter of nomenclature; from the S-dual Nahm analysis we
can determine the complex structure of the mass-deformed moduli space,
including the loci of intersection of the branches.  In general these
loci as computed by the Nahm analysis will be different from what one
might infer from the brane diagram.

\subsubsection{Complex Mass Deformation to $\mathcal{N}=2, N_c=2, N_f=0$}

Next, suppose we turn on {\em two} complex masses by moving both
NS5$^\ensuremath{\prime}$-branes (in the magnetic picture.)  Then we
have
\beq
\Y_{II}= \left( \begin{array}{cc} c' & 0 \\ 0 & d' \end{array} \right)\ ,
\eeq
but we also have to satisfy $\Y_{II} = Q \tilde{Q}$, and it is easy to
check that this is impossible if both $c'$ and $d'$ are nonzero.

We could have obtained the same result by attempting to solve the
system D5---2D3---D5$^\ensuremath{\prime}$ with poles at both the D5
and D5$^\ensuremath{\prime}$.  A careful analysis of the generalized
Nahm equations shows that no solution exists.

This recovers the classic result that pure $\mathcal{N}=2$ $U(2)$
gauge theory in 3 dimensions has no supersymmetric vacuum
\cite{Affleck:1982as}.

\subsection{$\mathcal{N}=2, N_c=2, N_f=1$\label{sec:merge}}

It is also possible to do the $U(2)$ with $N_f=1$ analysis directly,
in the brane ordering
D5---2D3(I)---NS5$^\ensuremath{\prime}$---2D3(II)---D5$^\ensuremath{\prime}$
(it is much more cumbersome to analyze this system if
we do a Hanany-Witten transition to put the NS5$^\ensuremath{\prime}$
to the right of the D5$^\ensuremath{\prime}$.)

We allow the gauge transformations to be discontinuous at the
NS5$^\ensuremath{\prime}$, which we place at $y=0$, and we place the
D5 and D5$^\ensuremath{\prime}$ at $ y= \pm 1$.  Then we can set
\beq
\A_I &=& \left( \begin{array}{cc} \frac{1}{2(y+1)} & 0 \\ 0 & - \frac{1}{2(y+1)}  \end{array} \right)\ ,\\
\A_{II} &=& \left( \begin{array}{cc} \frac{1}{2(1-y)} & 0 \\ 0 & - \frac{1}{2(1-y)}  \end{array} \right)\ ,
\eeq
and we will have a solution of the Nahm equations, with
\beq
\Y_I &=& 0\ ,\\
\X_{II} &=& 0\ ,
\eeq
and
\beq
\X_I &=& \left(\begin{array}{cc} a & 1/(y+1) \\ b (y+1)  & a \end{array} \right)\ ,\\
\Y_{II} &=& \left(\begin{array}{cc} c & 1/(1-y) \\ d(1-y)  & c \end{array} \right)\ .
\eeq

In addition to these equations, we also have to satisfy $\Y_I = AB$,
$\Y_{II}= BA$.  This forces the characteristic polynomial of $\Y_{II}$
to vanish, so we have to set $c = d= 0$.

We also have to satisfy $\X_{I}A = 0$ and $B\X_I=0$.  This requires
that at least one eigenvalue of $\X_I$ vanishes, so we set $b = a^2.$

Solving for $A, B$ we find
\beq
A &=& \left(\begin{array}{cc} 0 & s \\ 0   & -sa\end{array} \right)\ ,\\
B &=& \left(\begin{array}{cc}  -a t & t \\ 0   & 0\end{array} \right)\ ,
\eeq
along with the constraint 
\beq
-2a s t = 1\ .
\eeq
This is the quantum deformed moduli space with the complex structure
$(\mathbb{C}^* )^2$.

\subsection{$\mathcal{N}=2, N_c=2, N_f=3$}

We can apply our method to any number of flavors.  For example we can
analyze the situation with $N_f=3$.  The algebra becomes rather
complicated, so we will simply state the results which an energetic
reader will be able to confirm.

The branches of moduli space fall into three parts.  There is an
8-dimensional part which is the ordinary Higgs branch and which can be
computed classically, a 6-dimensional mixed branch with $U(1)$ gauge
symmetry, and a 2-dimensional Coulomb branch which can be computed
classically in the S-dual.

\begin{figure}
\centerline{\includegraphics[scale=0.8]{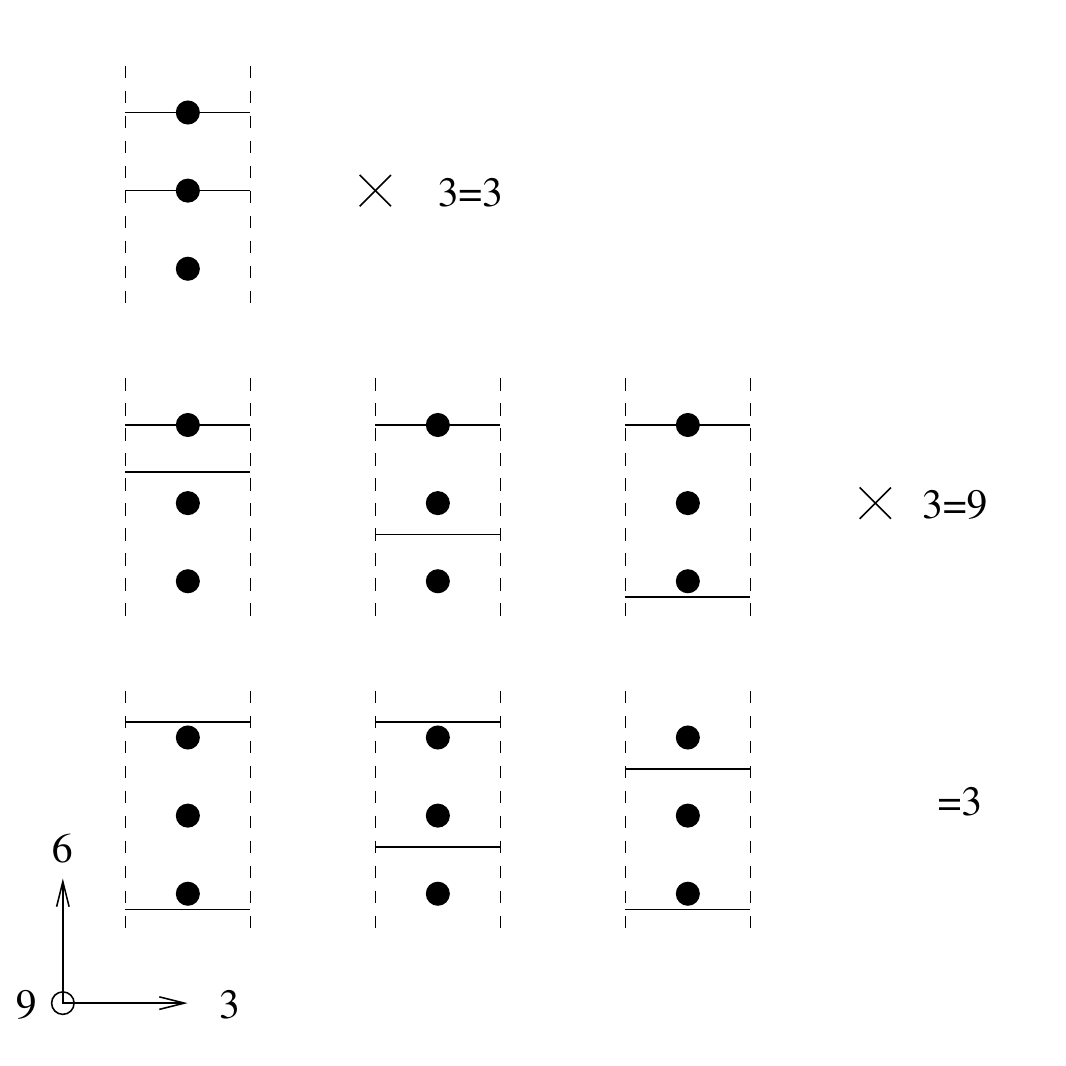}}
\caption{Enumeration of moduli space branches for the ${\cal N}=2$ $N_c=2$ $N_f=3$ theory with real masses form the S-dual perspective.  \label{n2nc2nf3brane}}
\end{figure}

With generic real masses turned on, we find 15 distinct 2-dimensional
branches, consistent with the old brane counting 
  (\ref{modulicomponents}) for $N_c=2$ and $N_f=3$. An abbreviated
version of Figure \ref{fign2nc2nf2brane} for this setup accounting for
the 15 branches is illustrated in Figure \ref{n2nc2nf3brane}.

%------------------------------------------------
\section{$U(3)$ Examples}
\label{sec:u3}

In this section, we will extend our analysis to the case where the
gauge group is $U(3)$. The main novelty of theories with $N_c > 2$ is
the fact that purely Coulomb branches are completely lifted by
non-perturbative effects. There are, however, mixed branches with some
Coulomb components.

\subsection{$\mathcal{N}=2$, $N_c=3$, $N_f=3$}

We continue our analysis with the simplest example for a $U(3)$ gauge
group.

Let us take a quick look at the classical moduli space of the field
theory (in the electric description.)  There is a classical Coulomb
branch with $U(1)^3$ gauge symmetry which is 3-dimensional and a
classical (exact) Higgs branch which is $2 N_f N_c -N_c^2=$
9-dimensional.  There are also mixed branches with varying amounts of
gauge symmetry.  For example, we can choose quark expectation values
which leave a $U(1)$ gauge symmetry.  This mixed branch has a product
structure with a Higgs part of dimension dim$(U(3)/U(1))=8$ and a
1-dimensional Coulomb part, so it is 9-dimensional.  There is another
branch for which the quarks leave a $U(2)$ gauge symmetry unbroken,
which is further broken to $U(1)\times U(1)$ on a partial Coulomb
branch.  The dimension of the Coulomb part is 2 and the Higgs part has
dimension dim$(U(3)/U(2))=5$.  This mixed branch is 7-dimensional.

Let us compare this with the superpotential proposed by Aharony
et.al.\cite{Aharony:1997bx}:
\beq
W = v_+ v_- \det(M) \ ,
\eeq
where $M$ is an $N_f \times N_f$ matrix.  The F-term equations have
three solution branches:
\begin{itemize}
\item $v_{+} = v_- = 0$, $M$ is unfixed -- which is 9-dimensional.
\item $v_+=0$ with $v_- \neq 0$ (or vice versa) with $\det(M)=0$ -- which is 9-dimensional
\item $v_{\pm} \neq 0$ with $\delta \det(M)=0$ -- which is 7-dimensional.
\end{itemize}
Comparing to the classical moduli space, the simplest picture is that
the effective superpotential is capturing the Higgs and mixed branches
of the moduli space but not the pure Coulomb branch.  We will see that
the Nahm computation matches this field theory analysis, with the
mixed branches but not the 3$d$ Coulomb branch with $U(1)^3$ gauge
symmetry.

Working in the magnetic formulation, the brane configuration is
D5---3D3(I)---D5$^\ensuremath{\prime}$---3D3(II)---NS5$^\ensuremath{\prime}$---2D3(III)---NS5$^\ensuremath{\prime}$---1D3(IV)---NS5$^\ensuremath{\prime}$.
The I--II interface supports fundamental quarks $Q, \tilde{Q}$.  At
the II--III interface, the NS5$^\ensuremath{\prime}$ supports
bifundamental fields $A_1, B_1$ and at the III--IV interface, the
NS5$^\ensuremath{\prime}$ supports bifundamentals $A_2, B_2$.  In the
following analysis we disregard all the unstable branches.

As in the previous section, we can do a $G_C$ transformation in
regions II, III, IV to make $\X$ constant and to put it in Jordan
normal form.  Then we distinguish three cases by the form of the
eigenvalues of $\X_{II}$.  The first case is when all three
eigenvalues are distinct and nonzero.  Then we have to set
$\tilde{Q}=Q=0$, and $\Y=0$ everywhere.  Neglecting unstable branches,
we are forced to set $A_1,B_1,A_2,B_2$ all to zero.  The resulting
moduli space has unbroken gauge symmetry, so we don't trust the
geometry, but we can compute the dimension of the moduli space anyway.
There are 3 moduli corresponding to the eigenvalues of $\X_{II}$.  In
addition to these, between the NS5$^\ensuremath{\prime}$-branes there
are three D3-brane segments, each of which contributes 2 complex
moduli corresponding to their positions in $\X$ and $\mathcal{Z}$.  So
there are a total of 9 complex moduli.  It is natural to identify this
as the Higgs branch of the electric theory.

The second case is when two eigenvalues are nonzero and one vanishes.
In this case either $Q$ or $\tilde{Q}$ is nonzero.  Again, $\Y=0$
everywhere so there are 6 moduli from the D3 segments between the
NS5$^\ensuremath{\prime}$-branes.  Comparing with the previous case,
we lose one modulus because one of the eigenvalues of $\X_{II}$ was
fixed to zero, but we gain one from the length of $Q$ or $\tilde{Q}$
(whichever is nonvanishing.)  So the total moduli space again is
9-dimensional, with two branches.  This appears to correspond to the
mixed branch of the electric theory with $U(1)$ gauge symmetry.

Next we consider the case where two eigenvalues of $\X_{II}$ vanish,
and take the $G_C$ gauge $\A=0$ in regions II, III, and IV.  Then we
can write
\beq
\X_{II} = \left( \begin{array}{ccc} a & 0 & 0 \\ 0 & 0 & 1\\ 0 & 0 & 0 \end{array} \right)  \ ,
\eeq
for which we have
\beq
Q^T &=& (0, v_+, 0) \  , \\
\tilde{Q} &=& (0,0,v_-) \ ,
\eeq
and
\beq
\Y_{II}  = \left( \begin{array}{ccc} 0 & 0 & 0 \\ 0 & 0 & v_+ v_- \\ 0 & 0 & 0 \end{array} \right) = A_1 B_1 \ .
\eeq
The bifundamental fields also have to satisfy 
\beq
\X_{II} A_1 &=& A_1 \X_{III} \ , \\
B_1 \X_{II} &=& \X_{III} B_1 \ .
\eeq

We identify two independent solutions of these conditions which break
all the gauge symmetry:
\beq
A_1 = \left( \begin{array}{cc} 0 & 0 \\ 0 & v_+ \\ 0 & 0  \end{array} \right) \; , \qquad B_1 = \left( \begin{array}{ccc} 0 & v_- & 0 \\ 0& 0 & v_- \end{array} \right) \ ,
\eeq
and 
\beq
A_1 = \left( \begin{array}{cc} 0 & 0 \\ v_+ & 0 \\ 0 & v_+  \end{array} \right) \; , \qquad B_1 = \left( \begin{array}{ccc} 0 & 0 & v_- \\ 0& 0 & 0 \end{array} \right)\ ,
\eeq
with (in both cases)
\beq
\X_{III} &=& \left( \begin{array}{cc} 0 & 1\\ 0 &  0 \end{array} \right) \ ,\\
\Y_{III} &=& \left( \begin{array}{cc} 0 & v_+ v_- \\ 0 &  0 \end{array} \right) \ . 
\eeq
The rest of the analysis is identical to case 3 of $N_c=2, N_f=2$.  We
see that we have two 3-dimensional branches on which the gauge
symmetry is completely broken.  However, we need to be a bit careful
about the action of $G_C$.  We have $U(2)$ complex gauge
transformations which act as $A_1 \rightarrow A_1 g^{-1}$, $B_1
\rightarrow g B_1$, $A_2 \rightarrow g A_2$, $B_2 \rightarrow B_2
g^{-1}$, with
\beq
g = \left( \begin{array}{cc} z_1 & 0 \\ 0 & z_2 \end{array}\right) \ . 
\eeq

In evaluating the gauge quotient, we need to consider not just the
regular gauge transformations but also the {\it closure} of the gauge
orbit.  It is not hard to see that the closure of the gauge orbit can
map these points to
\beq
A_1 = \left( \begin{array}{cc} 0 & 0 \\ 0 & v_+ \\ 0 & 0  \end{array} \right) \; , \qquad B_1 = \left( \begin{array}{ccc} 0 & 0 & 0 \\ 0& 0 & v_- \end{array} \right) \ ,
\eeq
where the gauge symmetry is broken.  At these points, new moduli
appear.  We have
\beq
\X_{III} &=& \left( \begin{array}{cc} r & 0\\ 0 &  0 \end{array} \right) \ ,\\
\Y_{III} &=& \left( \begin{array}{cc} 0 & 0 \\ 0 &  0 \end{array} \right) \ ,
\eeq
where $r$ is unfixed.  Now, at the III--IV interface condition at
least one of $A_2,B_2$ must vanish.  If one of the two vanishes, we
have an unstable branch; if both vanish then the $U(1)$ gauge symmetry
in region IV is unbroken.  It is the latter case which is of interest
to us.  We see that there is an unbroken $U(1)$ symmetry in region III
and an unbroken $U(1)$ in region IV.  Associated with each unbroken
$U(1)$ are two complex moduli, from $\X$ and $\mathcal{Z}$.  The total
dimension of this branch of moduli space is therefore 7 (2 from
$v_{\pm}$, 1 from $a$, and 4 from the scalars between the
NS5$^\ensuremath{\prime}$-branes.)  This appears to correspond to the
7-dimensional mixed branch.

\begin{figure}
\centerline{\includegraphics[scale=0.8]{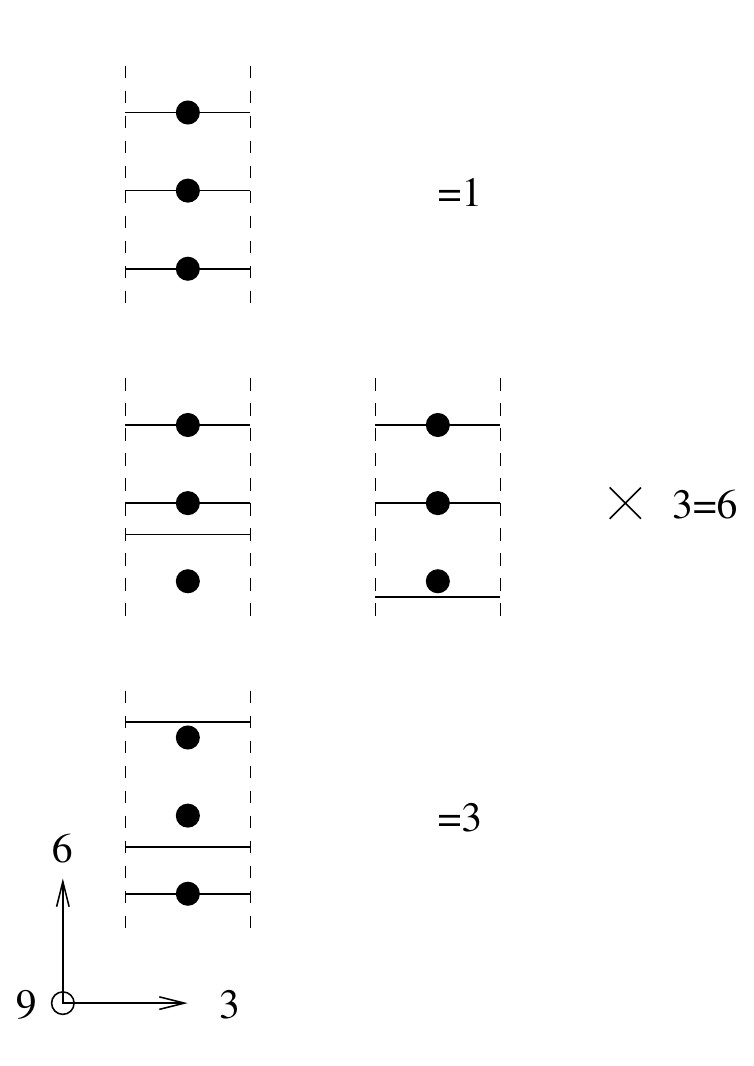}}
\caption{Enumeration of moduli space branches for the ${\cal N}=2$ $N_c=3$ $N_f=3$ theory with real masses form the S-dual perspective.  \label{n2nc3nf3brane}}
\end{figure}

If we turn on real masses (which are FI terms in the magnetic
formulation we are analyzing), when we solve the real equation we will
find that we have 3 branches from case 3.  This is because we can
solve the real equation for all three forms of $A_1, B_1$ (up to
$G_C$) with distinct solutions. The brane picture associated with each
of these branches are illustrated in Figure \ref{n2nc3nf3brane}.

The total counting of the number of branches in this case
  comes out to
\be 1 +6+3=10 \ , \ee
which again is consistent with (\ref{modulicomponents}) for $N_c=3$
and $N_f=3$.

\subsection{$\mathcal{N}=2, N_c=3, N_f =2$ by Complex Mass Deformation}

Analogously to the $N_c=2$, $N_f=1$ case, we can extract the moduli
space of $N_c=3$, $N_f=2$ by adding a complex mass deformation.  This
means in particular that we should set $\Y_{IV}=c' \neq 0$.

Classically, we would have the following expectations.  There is a
pure Higgs branch with dimension 3, and a Coulomb branch which is 3
dimensional. The classical geometry of the Coulomb branch is
$\mathbb{C}^3/\mathcal{W}$ where $\mathcal{W}$ is the $SU(3)$ Weyl
chamber. We also have a mixed branch with an unbroken $U(1)$ gauge
symmetry and 3 Higgs moduli; this mixed branch is 4 dimensional.
There is no separate mixed branch with $U(1) \times U(1)$ gauge
symmetry as it is part of the $U(1)^3$ Coulomb branch.

Quantum mechanically we expect to see merging of the Higgs and Coulomb
branches.

As in the case for $N_c=2$, $N_f=1$, the unlifted solutions appear
when there is only one vanishing eigenvalue of $\X_{II}$, so we can
write
\beq
\X_{II} = \left( \begin{array}{ccc} a & 0 & 0 \\ 0 & b & 0\\ 0 & 0 & 0 \end{array} \right)  \ ,
\eeq
and with $Q^T = (0,0,v_+)$, $\tilde{Q} = (0,0,v_-)$, we have
\beq
\Y_{II} = \left( \begin{array}{ccc} 0 & 0 & 0 \\ 0 & 0 & 0\\ 0 & 0 & v_+ v_- \end{array} \right) \ . 
\eeq

If we require that $\Y_{IV}=c' \neq 0$, then we cannot eliminate the
zero eigenvalue of $\X$ until we reach the last
NS5$^\ensuremath{\prime}$.  So we have to have
\beq
\X_{III} &=& \left( \begin{array}{cc} r & 0\\ 0 &  0 \end{array} \right) \ , \\
\Y_{III} &=& \left( \begin{array}{cc} 0 & 0 \\ 0 &  v_+ v_-  \end{array} \right)\ ,
\eeq
and we need to find $A_1, B_1$ to implement this.  Up to $G_{\mathbb
  C}$, we can write
\beq
A_1 &=& \left( \begin{array}{cc} 0 & 0 \\ 0 & 0 \\ 0 & v_+\end{array}\right) \ , \label{u33case3A1}\\
B_1 &=&  \left( \begin{array}{ccc} 0 & 0 & 0 \\0 & 0 & v_- \end{array} \right) \ ,\label{u33case3B1}
\eeq
and
\beq
A_2 &=& \left( \begin{array}{c}  0 \\  v_+\end{array}\right) \ ,\label{u33case3A2}\\
B_2 &=&  \left( \begin{array}{cc} 0 &  v_- \end{array} \right).\label{u33case3B2}
\eeq
Because of the structure of the bifundamentals, we have a free
parameter $r$ which appears in $\X_{III}$.  We are also (classically)
allowed to turn on $X_9$ and $\varphi$, that is
\beq
\mathcal{Z}_{III} &=& \left( \begin{array}{cc} s& 0\\ 0 &  0 \end{array} \right).
\eeq

We have a five-dimensional merged branch, parameterized by nonzero $a,
b$, arbitrary $r,s$, and $v_+ v_- = c'$.  However there is an unbroken
$U(1)$ gauge symmetry and the moduli space will be subject to quantum
corrections.

Note that because of the form of $\Y_{II}$, it can have at most one
nonzero eigenvalue.  This means that there are no solutions with two
nonzero complex masses turned on.  So the $\mathcal{N}=2$ theory with  $N_c=3$ and $N_f=1$ has no
supersymmetric vacuum.

%%%%%%%%%%%%%%%%%%%%%%%%%%%%%%%%%%%%%

\section{3$d$ $\mathcal{N}=2$ $U(N_c)$ Gauge Theories}\label{sec:un}

In this section, we will briefly describe how our analysis for $N_c=2$
and $N_c=3$ generalizes to arbitrary $N_c$.

\subsection{$N_f=N_c$}

From the $U(2)$ with $N_f=2$ and $U(3)$ with $N_f=3$ examples we can
attempt to see the pattern in the Nahm analysis so that we can
generalize to all $N_c= N_f\equiv N$.

The brane configuration is D5---$N$
D3(I)---D5$^\ensuremath{\prime}$---$N$ D3(II)---$T[SU(N),
  \textrm{NS5}^\ensuremath{\prime}]$.  We have flavors $Q$, $\tilde Q$
at the D5$^\ensuremath{\prime}$-brane.  There is a Nahm pole in $\X$
in region I. The coupling to a $T[SU(N)]$ boundary theory on the right
with NS5$^\ensuremath{\prime}$-branes fixes the eigenvalues of
$\Y_{II}$ to all be zero.  We can choose a gauge where $\X$ at the
I--II interface is
\beq
\left( \begin{array}{cccccc} a_1 & 1 & &  &&  \\ a_2 & a_1 & 1 & & &\\ a_3& a_2 & a_1 & 1 & &\\ \ldots &\ldots  &\ldots &\ldots &\ldots & \\ \ldots
&\ldots &a_3 &a_2&a_1& 1\\ a_N& \ldots& a_4& a_3&a_2 &a_1 \end{array}\right)
\label{XIform} \ . 
\eeq
Moreover, we have enough gauge freedom to make $\X$ and $\Y$ constant
in regions II, III, etc.  Note that there are $N$ parameters which are
free to vary.

In region II we can organize the computation by the eigenvalue
structure of $\X$; this involves choosing a different gauge than the
one used in (\ref{XIform}):
\begin{enumerate}
\item All the eigenvalues of $\X_{II}$ are distinct and nonzero:
\beq
\X_{II} = \left( \begin{array}{ccccc} \mu_1 &&&&\\ &\mu_2&&&\\ &&\mu_3&&\\ &&&\ldots&\\ &&&&\mu_N\end{array} \right) \ .
\eeq
This choice forces 
\beq
Q,\tilde Q &=& 0 \ ,\\
\Y_{II}&=& 0 \ .
\eeq
\item One eigenvalue of $\X_{II}$ is zero; the rest are distinct and nonzero
\beq
\X_{II} = \left( \begin{array}{ccccc} \mu_1 & &&&\\ &\mu_2&&&\\ &&\ldots&&\\ &&&\mu_{N-1}&\\ &&&&0\end{array} \right) \ .
\eeq
This forces
\beq
Q^T &=& (0,\ldots,0, v_+) \ , \\
\tilde Q &=& (0,\ldots, 0, v_-) \ , \\
\Y_{II} &=& {\rm diag}(0,\ldots, 0, v_+ v_-) \ . \label{YIIcase2}
\eeq
\item Two eigenvalues of $\X_{II}$ vanish and the rest are distinct and nonzero
\beq
\X_{II} = \left( \begin{array}{ccccc} \mu_1 & &&&\\ &\ldots&&&\\ &&\mu_{N-2}&&\\ &&&0&1\\ &&&0&0\end{array} \right) \ .
\eeq
Then we have
\beq
Q^T &=& (0,\ldots,0, v_+,0) \ , \\
\tilde Q &=& (0,\ldots, 0, 0, v_-) \ ,
\eeq
and 
\beq
\Y_{II} &=&  \left( \begin{array}{ccccc} 0 & &&&\\ &0&&&\\ &&\ldots&&\\ &&&0&v_+ v_-\\ &&&&0\end{array} \right) \ . \label{YIIcase3}
\eeq
\end{enumerate}
The other possibilities (for example, when more than two eigenvalues
vanish, or when two eigenvalues coincide but don't vanish) can be
smoothly related to these three cases, which are distinguished by the
form of the fundamental quarks $Q, \tilde Q$.  So it suffices to
consider these three cases (other possibilities do not give distinct
branches of moduli space.)

Now we proceed to analyze the dimension of moduli space for these
three cases.  This calculation is not strictly reliable because there
will be some unbroken gauge symmetry, but we can do it anyway and see
what we get.

In case 1, we have $\Y_{II}=0$.  Stability requires that all the $A_i,
B_i$ supported at the NS5$^\ensuremath{\prime}$-branes vanish.  Then
we have $N(N-1)/2$ brane segments between the
NS5$^\ensuremath{\prime}$-branes which are free to move.  Each brane
segment contributes 2 complex moduli.  In addition we have $N$ moduli
from the eigenvalues of $\X$.  So the total dimension of moduli space
is $N^2$.

In case 2, because the eigenvalues of $\Y_{II}$ have to vanish (from
the coupling to $T[SU(N)]$), we must have $v_+ v_-=0$.  There is no
gauge symmetry to fix $v_{\pm}$.  We see that there are two branches
of moduli space where either $v_+$ or $v_-$ is zero.  Because
$\Y_{II}$ vanishes completely, we again have to set all the $A_i, B_i$
to zero.  So (as in case 1) we get $N(N-1)$ moduli from the brane
segments which are free to move.  We also have $N-1$ eigenvalues of
$\X$, and one modulus from $v_{\pm}$.  The total dimension is $N^2$.
Note however that this is a distinct branch of moduli space from case
1 because the fundamentals $Q,\tilde{Q}$ are not both zero.

In case 3, $\Y_{II}$ does not vanish even though all its eigenvalues
are zero, so it is no longer necessary for all the bifundamentals of
$T[SU(N)]$ to vanish.  Instead we can set $\Y_{II}= A_1 B_1$ with
\beq
A_1 &=& \left( \begin{array}{cccc} 0&\ldots&0&0 \\ 0 &\ldots &0 &0\\ \ldots &\ldots&\ldots&\ldots \\0&\ldots&0& v_+\\ 0&\dots&0&0\end{array}\right) \ ,
\eeq
\beq
B_1 &=& \left( \begin{array}{ccccc} 0&\ldots&\ldots&\ldots& 0\\ \ldots&&&&\ldots\\0&&&&0 \\0 &\ldots &0&0& v_-\end{array} \right) \ . 
\eeq
Note that $B_1A_1=0$ so that $\Y_{III}=0$.  This forces all the
remaining $A_i, B_i$ to vanish.  I believe this is the only solution
for the bifundamentals consistent with stability, up to gauge
transformations.

Counting brane segments which are free to move between the
NS5$^\ensuremath{\prime}$-branes, we see that there are $N^2-N-2$
associated complex moduli.  In addition to this, we have 2 moduli from
$v_{\pm}$ and $N-2$ eigenvalues of $\X_{II}$.  The total dimension of
this branch of moduli space is $N^2-2$.

These three branches can be compared to our field theory expectations.
We expect an $N_c^2$-dimensional Higgs branch.  There should also be a
$U(1)$ mixed branch with Coulomb dimension 1 and Higgs dimension
dim($U(N)/U(1)$) = $N^2-1$ so the total dimension is $N^2$.  For $N\ge
2$ there is a mixed branch with $U(1)^2$ gauge symmetry with Coulomb
dimension 2 and Higgs dimension dim($U(N)/U(2)$) = $N^2-4$, so the
total dimension is $N^2-2$.

Crucially, these are the only three cases.  The pure Coulomb branch
(for $N_c > 2$) and most of the mixed branches, which are present
classically, do not appear in the S-dual Nahm analysis.  These
additional branches appear to be lifted quantum mechanically, in
accord with the superpotential of \cite{Aharony:1997bx}.

\subsection{$N_f<N_c$}

We can extract the dimensions of moduli space for $N_f < N_c$ by
giving complex masses to the quarks.  In the algebraic analysis this
corresponds to fixing some of the eigenvalues of $\Y_{II}$ to be
nonzero.

For $N_f = N_c-1$, we can start from $N_f=N_c$ and deform by a single
complex mass.  We see that cases 1 and 3 are immediately excluded,
leaving only case 2.  There are $N_c-1$ moduli contained in $\X_{II}$
and 1 modulus in $\Y_{II}$ (and the quarks $Q, \tilde Q$.)  In
addition there are $(N_c-1)(N_c-2)$ moduli from the D3-brane segments
between the NS5$^\ensuremath{\prime}$-branes which are allowed to move when the
bifundamentals vanish.  So the total dimension of moduli space is
$N_c^2 -2N_c +2$.

For $N_f \le N_c -2$, however, we see immediately that there are no
supersymmetric solutions, because $\Y_{II}$ in the $N_f = N_c$
analysis can have at most one nonzero eigenvalue.

\subsection{$N_f = N_c +1$}

The brane configuration is D5---$N$
D3(I)---D5$^\ensuremath{\prime}$---$N+1$ D3(II)---$T[SU(N+1),
  \textrm{NS5}^\ensuremath{\prime}]$.  Again we have $\X_I$ given by
(\ref{XIform}).  However, moving into region II we have
\beq
\X_{II} = \left(\begin{array}{cc} \X_I & 0 \\ 0 & 0 \end{array} \right) \ , \\
\Y_{II} = \left(\begin{array}{cc} 0 & Q \\ \tilde Q & t \end{array} \right) \ , 
\eeq
and the commutator $\lbrack \X, \Y \rbrack = 0$ implies $ \X_{II} Q =
\tilde Q \X_{II} = 0$.

Again, we divide the analysis into separate cases depending on the
eigenvalue structure of $\X$, doing appropriate $GL(N, \bC)$ rotations
to simplify the form of the fields to one of the following three
possibilities:
\begin{enumerate}
\item All the eigenvalues of $\X_{I}$ are distinct and nonzero, so 
\beq
\X_{II} = \left( \begin{array}{ccccc} \mu_1 & &&&\\ &\mu_2&&&\\ &&\ldots&&\\ &&&\mu_{N_c}&\\ &&&&0\end{array} \right) \ . 
\eeq
This forces $Q, \tilde Q =0$.
\item One eigenvalue of $\X_I$ zero and the rest are distinct and
  nonzero.  Then $\X_{II}$ has the form
\beq
\X_{II} = \left( \begin{array}{ccccc} \mu_1 & &&&\\ &\ldots&&&\\ &&\mu_{N_c-1}&&\\ &&&0&\\ &&&&0\end{array} \right) \ . 
\eeq
Then $\Y_{II}$ is
\beq
\Y_{II} = \left( \begin{array}{ccccc} 0 & &&&\\ &\ldots&&&\\ &&0&&\\ &&&0&v_+\\ &&&v_-&t\end{array} \right) \ .
\eeq
\item Two eigenvalues of $\X_I$ vanish
\beq
\X_{II} = \left( \begin{array}{cccccc} \mu_1 & &&&&\\ &\ldots &&&&\\ & &\mu_{N_c-2}&&&\\&&&0&1&0\\ &&&&0&0\\ &&&&&0\end{array} \right)  \ ,
\eeq
\beq
\Y_{II} = \left( \begin{array}{ccccc} 0 & &&&\\ &\ldots&&&\\ &&0&0&v_+\\ &&0&0&0\\ &&0&v_-&t\end{array} \right)\ . 
\eeq
\end{enumerate}

Case 1 here is much like Case 1 for $N_c=N_f$.  We are forced to set
$\Y_{II}=0$ and then all the $A_i, B_i$ vanish (for stability.)  The
number of moduli from the D3 segments between
NS5$^\ensuremath{\prime}$ branes is $N_f(N_f -1)$.  In addition there
are $N_c = N_f-1$ moduli from $\X$.  So the total dimension of moduli
space is $N_f^2-1$.

Case 2 is also similar to the analogous situation for $N_c=N_f$.  The
coupling to $T(SU(N_f))$ demands that the characteristic polynomial of
$\Y_{II}$ vanishes.  This sets $t=0$ and $v_+ v_-=0$.  However unlike
the $N_c=N_f$ analysis, it is no longer the case that $\Y_{II}$
vanishes completely, as one of $v_+$ or $v_-$ is nonzero.  Counting
the number of brane segments which are free to move, the corresponding
number of complex moduli is $(N_f+1)(N_f -2)$.  In addition to this,
we have $N_f-1$ moduli from $\X$ and $v_{\pm}$.  The total dimension
of this branch is $N_f^2-3$.

In case 3, vanishing of the characteristic polynomial of $\Y$ implies
$t=0$ but places no constraint on $v_{\pm}$.  We find that up to
$G_{\mathbb C}$, the first bifundamentals are
\beq
A_1 &=& \left( \begin{array}{cccc} 0&\ldots&0&0 \\
\ldots &\ldots&\ldots&\ldots \\
0&\ldots&0& v_+\\  
0 &\ldots &0 &0 \\ 
0&\dots&v_-&0\end{array}\right)\ , \\
B_1 &=& \left( \begin{array}{ccccc} 0&\ldots&\ldots&\ldots& 0\\ \ldots&&&&\ldots\\0&&&1&0 \\0 &\ldots &0&0& 1\end{array} \right) \ . 
\eeq
This also implies
\beq
\Y_{III} = \left( \begin{array}{cccc} 0 & &&\\ &\ldots&&\\ &&0&0\\ &&v_-&0\end{array} \right) \ . 
\eeq
There are $N_f(N_f-3)$ complex moduli from the D3 segments between
NS5$^\ensuremath{\prime}$-branes.  In addition there are two moduli
from $v_{\pm}$ and $N_f-3$ moduli from $\X$.  So the total dimension
of this branch of moduli space is $N_f^2-2N_f -1$.

It is easy to verify that the dimensions of these branches match the
classical expectations for the branches with no gauge symmetry, with
$U(1)$ gauge symmetry, and with $U(1)^2$ gauge symmetry.  However the
branches with more gauge symmetry do not arise in the Nahm analysis;
presumably they are lifted by quantum effects.

%%%%%%%%%%%%%%%%%%%%%%%%%%%%%%%%%%

\section{Examples of Quantum Merging with One NS5}\label{sec:merging}

One of the interesting phenomena in 3$d$ $\mathcal{N}=2$ theories is
the quantum merging of Higgs and Coulomb branches.  The prototypical
example where this occurs is in the $U(2)$ theory with one flavor.  In
that theory, classically there is a 2-dimensional Coulomb branch with
$U(1)^2$ gauge symmetry and a 2-dimensional mixed Higgs-Coulomb branch
with $U(1)$ gauge symmetry (in this case there is no pure Higgs
branch.) The two branches intersect on a complex line. Quantum
mechanically the Coulomb and mixed branches merge and there is only
one 2$d$ branch which is smooth.  This merging can be understood by
analyzing the field theory but is not manifest from considering the
brane cartoons.  In Section \ref{sec:u2} we described the quantum
merging from the point of view of the 3+1-dimensional defect
realization.

We would like to see more examples of this quantum merging, but the
field theory analysis is more difficult when the theories become more
complicated, as the effective Lagrangians on moduli space are less
constrained by symmetry as the field content becomes richer.  However
it turns out that the Nahm analysis is well-suited for constructing
some of these examples.

The strategy for finding theories with quantum merging is to look for
theories where the instanton effects are strongest; we might expect
this to occur when the number of flavors is as small as possible
without breaking supersymmetry.  In the case with only a $U(2)$ gauge
group, the pure $\mathcal{N}=2$ theory broke supersymmetry because of
instanton effects.  However, by adding one flavor, the theory became
supersymmetric with a quantum-merged moduli space.  So what we will do
is to look for other theories which break supersymmetry but which can
be saved by adding a single flavor.

\begin{figure}
\centerline{\includegraphics[scale=0.8]{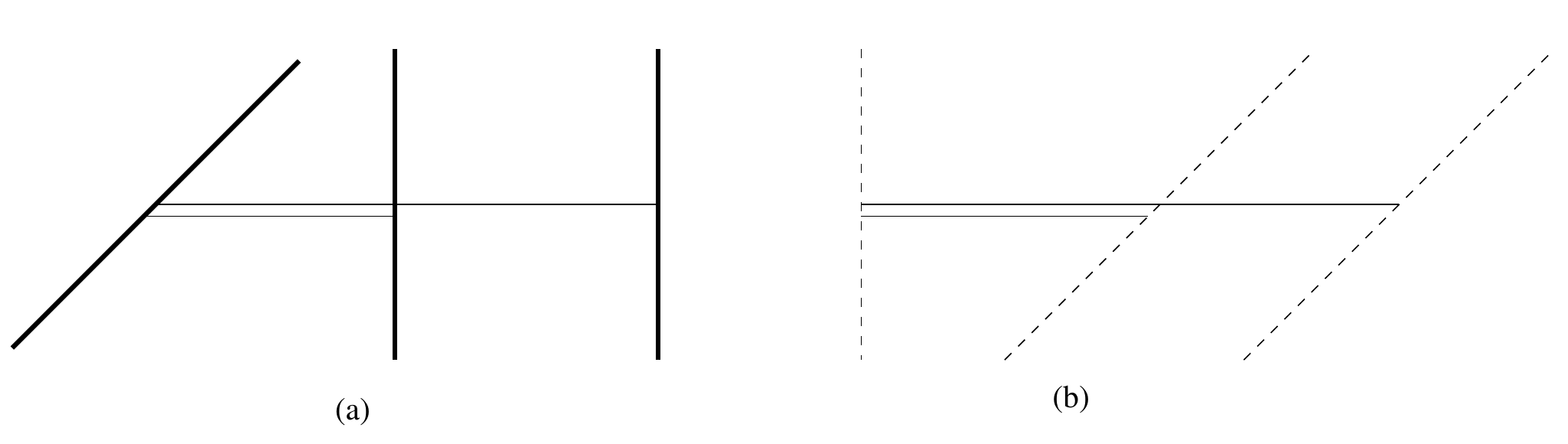}}
\caption{$U(2)\times U(1)$ quiver theory (a) and its S-dual, (b).
\label{u2u1}}
\end{figure}

Consider the $U(2)\times U(1)$ quiver theory defined by the brane
configuration
NS5---2D3---NS5$^\ensuremath{\prime}$---1D3---NS5$^\ensuremath{\prime}$,
shown in Figure \ref{u2u1}.a.  From analyzing the S-dual,
D5---2D3---D5$^\ensuremath{\prime}$---1D3---D5$^\ensuremath{\prime}$,
shown in Figure \ref{u2u1}.b, it is clear that this theory has no
supersymmetric vacuum.  We can think of it in terms of a Nahm pole on
the left and D5$^\ensuremath{\prime}$ ordinary Dirichlet boundary
conditions on the right.  The Nahm pole from the D5 requires that $\X$
is nonvanishing but the D5$^\ensuremath{\prime}$ boundary conditions
set $\X=0$.  So it is impossible to solve the Nahm equations and
supersymmetry is broken.

We can add a flavor to the $U(2)\times U(1)$ theory by adding either a
D5 or D5$^\ensuremath{\prime}$ to make the configuration
NS5---2D3---D5/D5$^\ensuremath{\prime}$---2D3---NS5$^\ensuremath{\prime}$---1D3---NS5$^\ensuremath{\prime}$.
The S-dual is
D5---2D3---NS5/NS5$^\ensuremath{\prime}$---2D3---D5$^\ensuremath{\prime}$---1D3---D5$^\ensuremath{\prime}$.
The S-dual has no gauge symmetry in the 3$d$ limit and so we expect
the Nahm computation to determine the complex structure reliably.  We
will analyze the situation for both types of flavors.

\begin{figure}
\centerline{\includegraphics[scale=0.8]{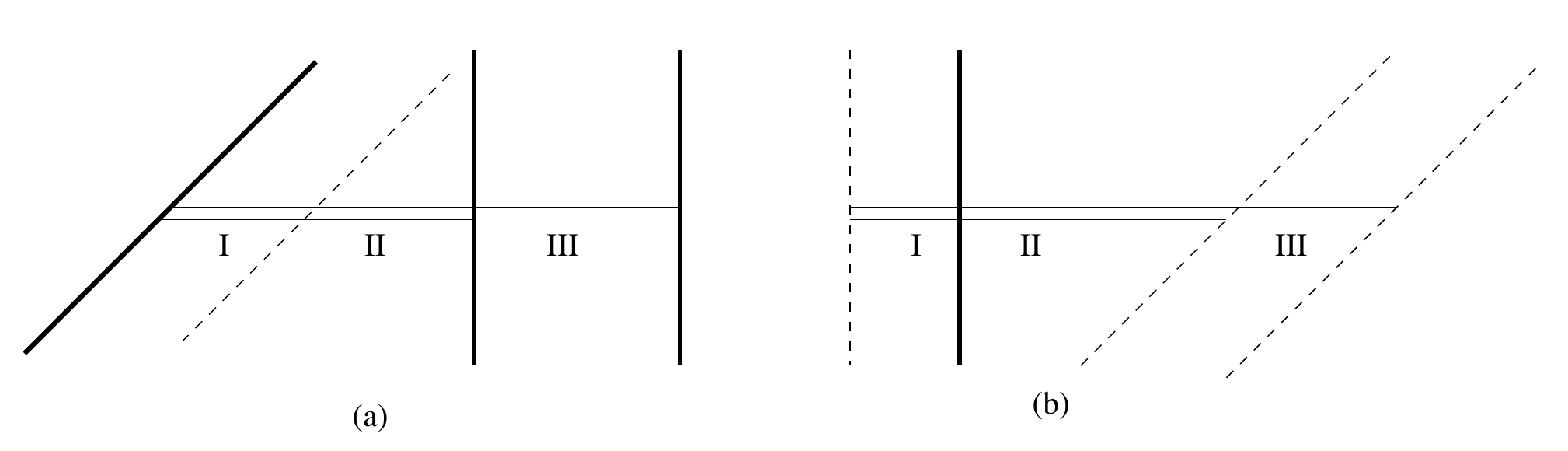}}
\caption{$U(1)\times U(2)$ quiver theory with one flavor added to the $U(2)$ gauge group (a) and its S-dual, (b).
\label{u2u1nf1}}
\end{figure}

\subsection{D5---2D3---NS5$^\mprime$---2D3---D5$^\mprime$---1D3---D5$^\mprime$}

Let us consider
D5---2D3(I)---NS5$^\mprime$---2D3(II)---D5$^\mprime$---1D3---D5$^\mprime$.
It suffices to consider regions I and II.  At the location of the
NS5$^\ensuremath{\prime}$, we can take the form of $\X$ generated by
the Nahm pole as
\beq
\X_I = \left( \begin{array}{cc} a & 1 \\ b & a \end{array}\right).
\eeq
Because of the ordinary Dirichlet boundary condition on the right, we have 
\beq
\X_{II} =0\ .
\eeq
We also have to satisfy
\beq
\Y_{I}= AB &=& 0\ , \\
\Y_{II} = BA &=& {\rm any}\ ,
\eeq
and
\beq
\X_I A = A\X_{II} &=& 0\ ,\\
B \X_I = \X_{II} B &=& 0\ .
\eeq

One solution branch is given by setting the bifundamentals $A=B=0$.
Then $a, b$ are free parameters.  We have a 2-dimensional moduli
space.  This is natural to interpret as being parameterized by the
D3-brane segments between the D5 and NS5$^\ensuremath{\prime}$ while
everything else is fixed.  So this is a sort of Higgs branch (of
self-dual type.)

If $A$ or $B$ is not strictly zero, then we have to set $b=a^2$ to
find solutions.  The solutions for the bifundamentals are most easily
written as an outer product
\beq
A &=& \left( \begin{array}{c} 1\\-a\end{array} \right) \left( \begin{array}{cc} x_1 & x_2 \end{array} \right)\ ,\\
B &=& \left( \begin{array}{c} x_3\\x_4\end{array} \right) \left( \begin{array}{cc} -a & 1\end{array} \right)\ ,
\eeq
with the constraint
\beq
x_1 x_3 + x_2 x_4 = 0\ .
\eeq
This branch of moduli space is $\mathbb C$ (parameterized by $a$)
times the conifold.  The dimension of moduli space comes out to what
one expects based on counting degrees of freedom in a brane diagram,
but the fact that the complex structure is that of a conifold would
have required additional data not contained in the brane diagram. The
Nahm analysis, however, contains this data. Perhaps it is also
possible to arrive at the same conclusion from non-perturbative
consideration of the field theory in the electric formalism.

\subsection{D5---2D3---NS5---2D3---D5$^\mprime$---1D3---D5$^\mprime$}

Next we consider the other type of flavor:
D5---2D3(I)---NS5---2D3(II)---D5$^\ensuremath{\prime}$---1D3---D5$^\ensuremath{\prime}$.
Again we have
\beq
\X_I &=& \left( \begin{array}{cc} a & 1 \\ b & a \end{array}\right)\ ,\\
\X_{II} &=& 0\ ,
\eeq
but now because of the NS5 brane we have
\beq
\X_I &=& AB\ ,\\
\X_{II} &=& BA\ .
\eeq
These actually require the trace and determinant of $\X_I$ to vanish,
so we have to set $a=b=0$.  This fixes the form of $A, B$:
\beq
A &=& \left( \begin{array}{cc} x_1 & x_2 \\ 0 & 0 \end{array}\right)\ ,\\
B &=& \left( \begin{array}{cc}0 & x_3 \\ 0 & x_4 \end{array} \right)\ ,
\eeq
with the constraint
\beq
x_1 x_3 + x_2 x_4 = 1\ ,
\eeq
parameterizing a deformed conifold.

We still have to consider the $\Y$ equations, $\Y_I =0$, $A
\Y_{II}=\Y_I A=0$ and $\Y_{II} B=B\Y_{I}=0$.  These actually fix the
form of $\Y_{II}$ to
\beq
\Y_{II}=p \left( \begin{array}{c} x_3\\x_4\end{array} \right) \left( \begin{array}{cc} x_1 & x_2\end{array} \right)\ ,
\eeq
with a free coefficient $p$.

So the total moduli space appears to be $\mathbb C$ times the deformed
conifold.  There is only one branch, which we might be able to
interpret as a quantum-merged branch of moduli space. Once again, the
form of the complex structure would have been impossible to infer from inspection of
the brane diagram.

\subsection{$U(1)^2\times U(2)$ Example}

Another interesting quiver theory is defined by the brane
configuration
NS5---1D3---NS5---2D3---NS5$^\ensuremath{\prime}$---1D3---NS5$^\ensuremath{\prime}$.
We see that it has gauge group $U(1)^2 \times U(2)$.

\begin{figure}
\centerline{\includegraphics[scale=0.8]{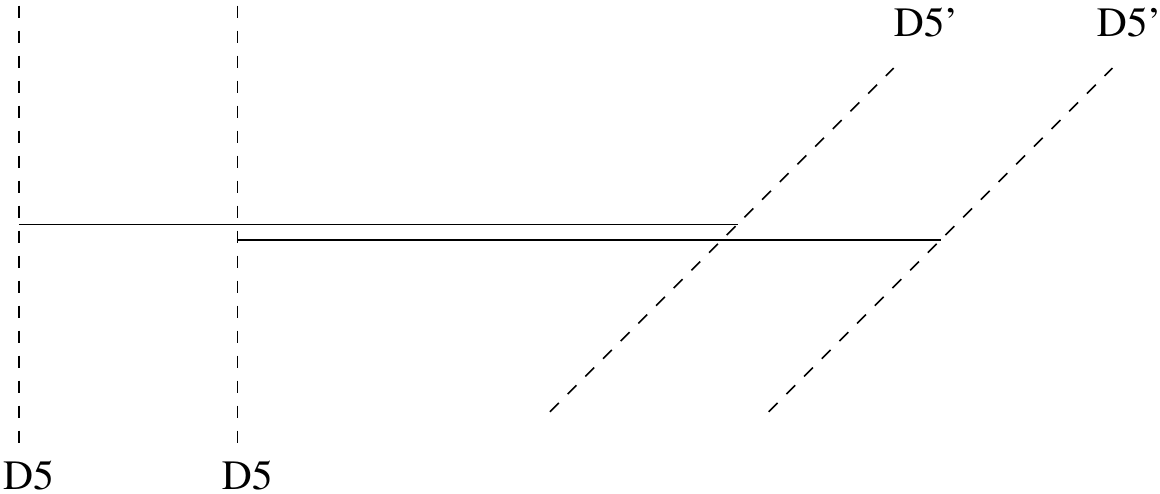}}
\caption{The S-dual of a  $U(1) \times U(2)\times U(1)$ quiver theory.  The brane configuration can be understood as $U(2)$ gauge theory with ordinary Dirichlet boundary conditions of D5 type on the left and D5$^\ensuremath{\prime}$ type on the right.
\label{u1u2u1}}
\end{figure}

It is best to analyze this theory in the S-dual frame where the gauge
symmetry is broken.  The S-dual is
D5---1D3---D5---2D3---D5$^\ensuremath{\prime}$---1D3---D5$^\ensuremath{\prime}$
or in other words, it is a D5 ordinary Dirichlet boundary condition
for $U(2)$ gauge theory on the left and a D5$^\ensuremath{\prime}$
ordinary Dirichlet boundary condition on the right.  The brane
configuration is pictured in Figure \ref{u1u2u1}.  The Nahm equations
force $\X=\Y=0$ in all regions.  The moduli space is still not quite
trivial because we have to fix the gauge carefully.  We are not
allowed to fix $\A=0$ but only $\A=$ constant.  This leaves a
four-dimensional moduli space. 
% (We will see how to compute this
%cleanly after adding an NS5-brane.)
%

\begin{figure}
\centerline{\includegraphics[scale=0.8]{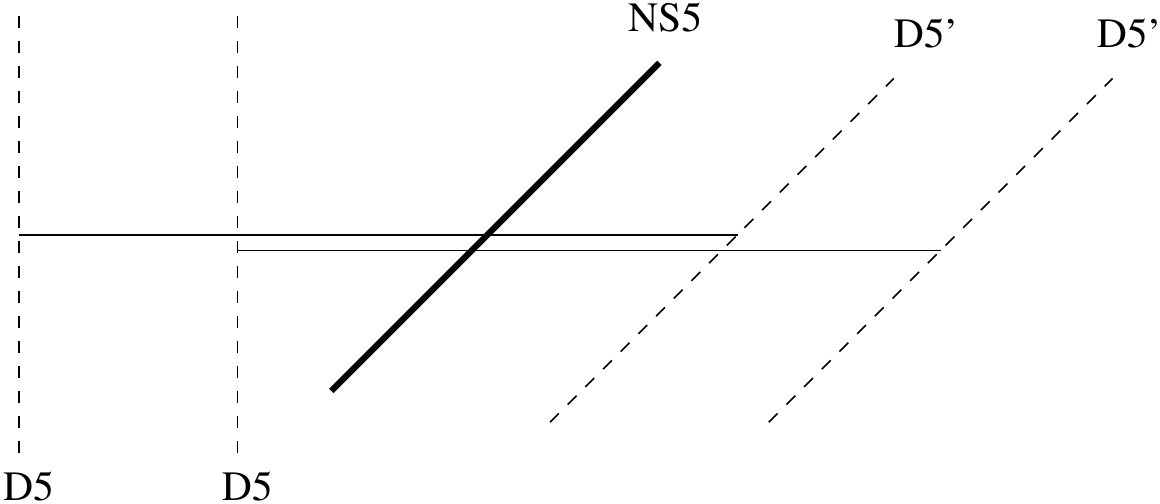}}
\caption{The S-dual of a  $U(1) \times U(2)\times U(1)$ quiver theory with one flavor added to the $U(2)$ gauge group.
\label{u1u2u1nf1}}
\end{figure}

Now suppose we add a flavor by adding an NS5-brane to the S-dual, as
in Figure \ref{u1u2u1nf1}.  So now we have D5 ordinary Dirichlet on
the left and D5$^\ensuremath{\prime}$ ordinary Dirichlet on the right.
These boundary conditions set
\beq
\X_L &=& {\rm any}\ ,\\
\X_R &=& 0\ ,\\
\Y_L &=& 0\ ,\\
\Y_R &=& {\rm any}\ .
\eeq

The additional constraints we need to impose are
\beq
\X_L &=& AB\ ,\\
\X_R &=& BA=0\ ,\\
A \Y_R &=& 0\ ,\\
\Y_R B &=& 0\ .
\eeq
If we write
\beq
A &=& \left(\begin{array}{cc} a_1 & a_2 \\ a_3 & a_4 \end{array} \right)\ ,\\
B &=& \left(\begin{array}{cc} b_1 & b_2 \\ b_3 & b_4 \end{array} \right)\ ,
\eeq
the constraints from $BA=0$ are
\beq
a_1 b_1 + a_3 b_2 &=&0\ ,\\
a_2 b_1 + a_2 b_4 &=&0\ ,\\
a_3 b_1 + a_4 b_3 &=& 0\ ,\\
a_3 b_2 + a_4 b_4 &=& 0\ .
\eeq
It appears that there are four constraints, but actually only three of
them are independent.  In particular, when $A, B$ are essentially
generic, they can be written as
\beq
a_1 a_4 - a_2 a_3 &=&0\ ,\\
b_1 b_4 - b_2 b_3 &=& 0\ ,\\
a_1 b_1 + a_3 b_2 &=& 0\ .
\eeq
The constraints degenerate, of course, if either $A=0$ or $B=0$.

When $A\neq 0$ and $B\neq 0$, we have $\det A=0$ and $\det B = 0$ so
they both have a single zero eigenvector.  By taking the outer product
of these eigenvectors we can construct $\Y_R$ up to an overall
normalization.  So $\Y_R$ contributes one complex modulus.  From $A$
and $B$ we get five complex moduli.  Note that if we start by thinking
about $B$ as parameterizing a conifold, then $A$ contains two moduli
that are completely unconstrained.  So the moduli space for this
branch is the conifold times ${\mathbb C}^3$.

Next, suppose that $A=0$ and $B$ is completely generic. This fixes
$\Y_R=0$ and we have a four-dimensional branch of moduli space which
is just $\mathbb{C}^4$.  There is a second copy of this branch from
interchanging $A$ and $B$.

Or, if $A=0$ and $B$ has vanishing determinant (but is otherwise
generic) then $\Y_R$ contributes two moduli.  So this branch is the
conifold times $\mathbb{C}^2$.  This branch also has a second copy.

Finally, if $A=0$ and $B=0$ then $\Y_R$ is completely free and there
is a $\mathbb{C}^4$ branch of moduli space.
We stress that the complex structure as determined by this
analysis should be quantum exact, on all of the branches.  

Suppose we give a complex mass to the quarks.  In the S-dual picture
this comes from adding a complex FI term from moving the NS5 in the 45
directions.  We need to change the moment map to
\beq
\X_L &=& AB-m\mathbb I\ ,\\
\X_R &=& BA- m\mathbb I=0\ .
\eeq
This sets $B= A^{-1}$.  Because $A, B$ are invertible, their
determinants are nonzero and they have no vanishing eigenvalues.  This
forces $\Y=0$.

Of course, adding a complex mass returns us to the unflavored case, so
we just recovered the result that the moduli space is $4d$.  Because
the defining equation is $a_1 a_4 - a_2 a_3 \neq 0$ the moduli space
is the group manifold $GL(2, \mathbb{C})$.

\section{Discussion}

In this article, we applied the junction and boundary conditions of
${\cal N}=4$ SYM in 3+1 dimensions preserving 1/4 of the
supersymmetries (which was worked out in \cite{Hashimoto:2014vpa}) to
analyze the moduli space of a defect/impurity system on an
interval. Such a theory naturally flows in the IR to an ${\cal N}=2$ SYM in
2+1 dimensions. The main input is the boundary condition and the
generalized Nahm equations reviewed in section \ref{sec2} and Appendix
\ref{appB}. The main output is the mapping out of the moduli space of
examples with varying degrees of sophistication in Sections
\ref{sec:abelian}--\ref{sec:merging}.

By analyzing several examples in detail, we illustrated how one can
reliably extract the complex structure of the moduli space,
from the classical
analysis for the branch of moduli spaces where the gauge group is
completely broken. On branches where some gauge symmetry remain
unbroken, the classical analysis is subject to quantum
corrections. Nonetheless, in many cases, one can reliably extract the
quantum corrected moduli space classically by analyzing the S-dual
defect/impurity system.

Using these techniques, we are able to reproduce all of the known
moduli space structures for $U(1)$, and $U(2)$ theories with ${\cal
  N}=2$ supersymmetry in 2+1 dimensions
\cite{deBoer:1997ka,deBoer:1997kr,Aharony:1997bx}. As one might expect
with fewer supersymmetries, the structure of the moduli space for the
${\cal N}=2$ models is far more intricate than their ${\cal N}=4$
counterparts. We also found that some issues such as stability
\cite{Mumford,Mumford:1977} play a critical role in mapping out the
moduli space of ${\cal N}=2$ theories.

Despite these subtleties, the analysis of the boundary/impurity system
eventually boils down to an algebraic analysis of solutions to the
generalized Nahm equations with appropriate boundary/junction
conditions. Combined with S-duality, this formalism appears to know
about many subtle dynamical effects including the $s$-rule, splitting
of the Coulomb branch into multiple branches, merging of various
branches, etc. Some effects, such as that of instanton generated
superpotential lifting certain branches of the moduli space, arise in
non-Abelian examples. These effects can be accounted for classically
in the S-dual frame, allowing one to infer the at least part of the
structure of the exact moduli space with relative ease.

Perhaps the most striking observation is that intricate dynamics of
non-perturbative effects such as generation of superpotentials,
quantum merging, and the $s$-rule on the electric side is reproduced
faithfully through the intricacy of non-commutativity of non-abelian
field configurations.

The power of this approach over direct analysis in 2+1 dimensions is
that in a number of examples, no guess-work regarding the form of the
superpotential or the mirror description was necessary. From the point
of view of the defect/impurity theory on an interval, the theory as
well as its S-dual is defined microscopically. Using this
construction, one can specify, in the UV, a system which may not admit
a Lagrangian description in the IR, and map out the moduli space.

One can think of the program being employed in this work in the same
broad class as the approach to study field theories in 3+1 dimensions
using M-theory \cite{Witten:1997sc,Witten:1997ep}. In both of these
approaches, the field theory of interest is embedded in some UV
framework, which enable one to apply broader set of analytical
tools. The regime of validity and reliability of these tools is always
an issue in interpreting the analysis for the original field theory
system, but for certain classes of observables and features, one can
take advantage of non-renormalization theorems. The generalized Nahm
equations and the interface/junction conditions described in
\cite{Hashimoto:2014vpa} are  playing roles analogous to
the holomorphic curves characterizing the M5-brane worldvolume in the
approach of \cite{Witten:1997sc,Witten:1997ep}. One advantage of the
generalized Nahm analysis is the fact that the UV completion is field
theoretic and does not rely on the full machinery of string theory or
M-theory.

These impurity systems, of course, are conveniently engineered as zero
slope limit of brane constructions.  We have focused mainly on the
case where the ${\cal N}=2$ system are constructed only using the NS5,
D5, NS5$^\ensuremath{\prime}$, and D5$^\ensuremath{\prime}$
branes. The formalism can also be generalized to the $(p,q)$ 5-brane
boundary/junction conditions, a problem to which we hope to return in the future.

%%%%%%%%%%%%%%%%%%%%%%%%%%%%%%%%%%%%%%%%%
\section*{Acknowledgments}
%%%%%%%%%%%%%%%%%%%%%%%%%%%%%%%%%%%%%%%%%%

We would like to thank P.~Argyres, T.~Clark, D.~Gaiotto,
K.~Intriligator, S.~Khlebnikov, M.~Kruczenski, G.~Michalogiorgakis,
N.~Seiberg, B.~Willett, E.~Witten, D.~Xie and K.~Yonekura for discussion.

This work is supported in part by the DOE grant DE-FG02-95ER40896
(A.~H.), by DOE grant DE-FG02-91ER40681 (P.~O.) and partly by
Princeton Center for Theoretical Science, by Institute for Advanced
Study and by WPI program, MEXT, Japan (M.~Y.).

This work was initiated at the SPOCK meeting (University of
Cincinnati, 2011) and we would like to thank the organizers for the
meeting.  M.~Y. would like to thank Aspen Center for Physics, Newton
Institute (Cambridge), Simons Center (Stony Brook), YITP (Kyoto), and
KITP (UCSB) for hospitality where part of this work has been
performed.  The content of this paper was presented at presentations
by A.~H. (Michigan, Dec.\ 2012, KITP Apr.\ 2014, Princeton
June.\ 2014), P.~O. (Purdue, Nov.\ 2013, Madison Apr.\ 2014) and
M.~Y. (IAS, Apr.\ 2013, Mar.\ 2014; Kavli IPMU, Aug.\ 2013; Perimeter,
Jan.\ 2014; KITP, Feb.\ 2014), and we thank the audiences of these
talks for feedback.

\appendix

\section{Coulomb Branch of ${\cal N}=4,2, N_c=1$ Theories with $N_f$ Flavors \label{appA}}

In this appendix, we review the complex geometry of the Coulomb branch
of ${\cal N}=4$ and ${\cal N}=2$ theories with general $N_f$. 

\subsection{Coulomb Branch of $\mathcal{N}=4, N_c=1, N_f$ Theory}

Let us revisit an old result, the Coulomb branch moduli space for
$\mathcal{N}=$ $U(1)$ gauge theory with $N_f$ flavors.  Because we are
interested in the Coulomb branch, we work in the S-dual frame where
the brane construction is D5---NS5---NS5---...---NS5---D5.

In this $\mathcal{N}=4$ example, we can set $ \Y= \mathcal{Z}=0$ in
all regions.  As for $\mathcal{X}$, we have a sequence of relations at
each NS5 interface,
\beq
A_1 B_1 = A_2 B_2 = \ldots = A_{N_f} B_{N_f} \equiv x \ ,
\eeq
when all the complex masses (of the electric theory) vanish.  With the
complex mass deformations turned on, we have instead
\beq
A_1 B_1 -m_1 = A_2 B_2 - m_2  = \ldots = A_{N_f} B_{N_f} - m_{N_f} \equiv x\ .
\eeq

To analyze the moduli space, it is convenient to define the gauge
invariant quantities
\beq
a &\equiv&\prod_i A_i \ ,\\
b &\equiv& \prod_i B_i\ ,
\eeq
which satisfy the constraint equations
\beq
a b = \prod (x+m_i)\ .
\eeq
We see that we have $N_f -1 $ complex parameters which act as
deformations of the complex structure, and one ``center of mass'' of
the complex deformations which acts trivially.  The moduli space is
evidently $\mathbb{C}^2/\bZ_{N_f}$ with deformations induced by the
complex masses.

\subsection{Coulomb Branch of $\mathcal{N}=2, N_c=1, N_f$ Theory 
}

To repeat the preceding computation for $U(1)$ $\mathcal{N}=2$ theory,
we need to rotate one of the D5-branes to D5$^\ensuremath{\prime}$, so
that the brane configuration is
D5---NS5---NS5---...---NS5---D5$^\ensuremath{\prime}$.  The analysis
is the same as before, except that we have to set
\beq
x=0\ .
\eeq
The result is that the complex structure of the Coulomb branch is
\beq
ab = \prod(m_i)\ .
\eeq
If any of the complex masses vanishes, we are simply left with
\beq
ab = 0\ ,
\eeq
which is a $1d$ moduli space with two branches.  If all the complex
masses are nonzero, we have instead $ab = {\rm const}$ which is simply
a cylinder.  This recovers the result that the Coulomb branch of
$U(1)$ $\mathcal{N}=2$ is $\mathbb{C} \oplus \mathbb{C}$ except when
$N_f=0$, for which it is $\mathbb{R} \times S^1$.

\section{Review of 1/4 BPS Boundary Conditions \label{appB}}

In this section, we review the supersymmetry conditions for these
defect systems. The derivations of these equations as well as detailed
explanations may be found in our earlier paper.  The spirit of the
analysis follows the treatment of Gaiotto and Witten
\cite{Gaiotto:2008sa,Gaiotto:2008ak}.

There are two especially important classes of boundary conditions for
this paper, which one can think of as corresponding to some number of
D3-branes either ending on or intersecting a D5-brane or an NS5-brane.

At a D5-like interface with $N$ D3-branes on each side, the interface
conditions are
\beq \Delta \mathcal{X}(y_0) &=& Q \tilde{Q} \ ,\label{dX} \\
i\Delta X_6 &=& Q Q^{\dag}-\tilde{Q}^{\dag} \tilde{Q}\ , \label{dX6} \\
{\cal Y}  Q &=& 0\ , \label{YQtilde} \\
\tilde Q {\cal Y} &  =& 0 \ , \label{QY}
\eeq
where the fields $Q$ ($\tilde{Q}$) are complex scalars which transform
in the fundamental (anti fundamental) representation of $U(N)$.  These
boundary conditions are derived from the effective theory of D3-branes
intersecting a D5-brane, as discussed in
\cite{Kapustin:1998pb,Tsimpis:1998zh}.  They are equivalent to the
standard jumping equations used to find monopole solutions; see
\cite{Bielawski} for a review.  For a D5$^\ensuremath{\prime}$-brane
interface, the analogous conditions are obtained by exchanging $\X$
and $\Y$:
\beq \Delta \mathcal{Y}(y_0) &=& Q \tilde{Q}\ , \label{dY} \\
i\Delta X_6 &=& Q Q^{\dag}-\tilde{Q}^{\dag} \tilde{Q} \ , \\
{\cal X}   Q &=& 0\ , \label{XQtilde} \\
\tilde Q {\cal X} &  =& 0\ .  \label{QX}
\eeq
These conditions can be generalized by displacing the D5-brane in the
$X_{7,8,9}$ directions (or displacing D5$^\ensuremath{\prime}$-branes
in the $X_{4,5,9}$ directions.)  This will amount to shifting the
scalars in the equations (\ref{YQtilde}, \ref{QY}, \ref{XQtilde},
\ref{QX}) by appropriate constants proportional to the identity.  In
the field theory these deformations correspond to turning on masses
for the quarks.

When the gauge groups on the two sides of the D5 interface are
unequal, the conditions are a little more subtle, see
\cite{Hashimoto:2014vpa} for details. It is possible to understand
them by starting with the case where there are equal gauge groups on
the two sides of the interface and then taking expectation values of
the interface fields $Q$, $\tilde Q$ to partially Higgs the gauge
group on one side of the interface.

For an NS5 oriented along 012789 with $N$ D3 ending from the left and
$M$ D3 ending from the right, we impose the junction conditions
\beq
{\cal X}_L &=& AB -\zeta_c \mathbb I_L \ ,\\
{\cal X}_R &=& BA -\zeta_c \mathbb I_R \ ,\\
i X_{6,L} &=& A A^\dag -  B^\dag B -\zeta_r \mathbb I_L \ ,\\
i X_{6,R} & = & A^\dag A - B B^\dag -\zeta_r \mathbb I_R\ ,\\
\mathcal{Y}_L A &=& A \mathcal{Y}_R \ ,\\
B\mathcal{Y}_L &=& \mathcal{Y}_R B \ . 
\eeq
We can also generalize these conditions for the case of
NS5$^\ensuremath{\prime}$ brane junction oriented along 012459 by
exchanging ${\cal X}$ and ${\cal Y}$.
\beq
{\cal Y}_L &=& AB-\zeta_c \mathbb I_L\ ,\\
{\cal Y}_R &=& BA -\zeta_c \mathbb I_R\ ,\\
i X_{6,L} &=& A A^\dag -  B^\dag B -\zeta_r \mathbb I_L \ , \\
i X_{6,R} & = & A^\dag A - B B^\dag -\zeta_r \mathbb I_R\ ,\\
\mathcal{X}_L A &=& A \mathcal{X}_R\ , \\
B\mathcal{X}_L &=& \mathcal{X}_R B \ . 
\eeq
We have included FI deformations $\zeta_c$ and $\zeta_r$.  The complex
FI term $\zeta_c$ corresponds to changing the positions of the
NS5(NS5$^\ensuremath{\prime}$)-branes in the 45(78) directions, while
the real FI term comes from moving the NS branes in the $x^6$
direction.

In addition to the simple D5 and NS5 interfaces, one can build more complicated composite boundaries from multiple 5-branes.
Details about such composite boundaries may be found in \cite{Hashimoto:2014vpa,Gaiotto:2008sa,Gaiotto:2008ak}.

\bibliography{boundary}\bibliographystyle{utphys}

\end{document}